\documentclass[%
 aip,
 amsmath,amssymb,
 reprint,%
]{revtex4-1}

\usepackage{graphicx}
\usepackage{dcolumn}
\usepackage{bm}

\usepackage[utf8]{inputenc}
\usepackage[T1]{fontenc}
\usepackage{mathptmx}
\usepackage{xcolor}
\usepackage{afterpage}
\usepackage{booktabs}
\usepackage{xr}

\makeatletter
\newcommand*{\addFileDependency}[1]{
  \typeout{(#1)}
  \@addtofilelist{#1}
  \IfFileExists{#1}{}{\typeout{No file #1.}}
}
\makeatother

\newcommand*{\myexternaldocument}[1]{%
    \externaldocument{#1}%
    \addFileDependency{#1.tex}%
    \addFileDependency{#1.aux}%
}
\myexternaldocument{SM_12}

\begin{document}
\preprint{AIP/123-QED}

\title{Identifying the leading dynamics of ubiquitin: a comparison between the tICA and the LE4PD slow fluctuations in amino acids' position\cite{CreditLine}}

\author{E. R. Beyerle}
 \affiliation{Institute for Fundamental Science and Department of Chemistry and Biochemistry, University of Oregon, Eugene, Oregon 97403, USA}
\author{M. G. Guenza}%
 \email{mguenza@uoregon.edu.}
\affiliation{Institute for Fundamental Science and Department of Chemistry and Biochemistry, University of Oregon, Eugene, Oregon 97403, USA}

\date{\today}

\begin{abstract}
Molecular Dynamics (MD) simulations of proteins implicitly contain the information connecting the atomistic molecular structure and proteins' biologically relevant motion, where large-scale fluctuations are deemed to guide folding and function. In the complex multiscale processes described by MD trajectories it is difficult to identify, separate, and study those large-scale fluctuations. This problem can be formulated as the need to identify a small number of collective variables that guide the slow kinetic processes. The most promising method among the ones used to study the slow, leading processes in proteins' dynamics, is the time-structure based or time-lagged independent component analysis, or tICA, which identifies the dominant components in a noisy signal. Recently, we developed an anisotropic Langevin approach for the dynamics of proteins, called the anisotropic Langevin Equation for Protein Dynamics or LE4PD-XYZ. This approach partitions the protein's MD dynamics into mostly uncorrelated, wavelength-dependent, diffusive modes. It associates to each mode a free-energy map, where one measures the spatial extension and the time evolution of the mode-dependent, slow dynamical fluctuations. Here, we compare the tICA modes' predictions with the collective LE4PD-XYZ modes. We observe that the two methods consistently identify the nature and extension of the slowest fluctuation processes. The tICA separates the leading processes in a smaller number of slow modes than the LE4PD does. The LE4PD provides time-dependent information at short times, and a formal connection to the physics of the kinetic processes that are missing in the pure statistical analysis of tICA.
\end{abstract}

\maketitle

\section{Introduction}
Large-scale fluctuations and global structural rearrangements play an essential role in the biological functions of biopolymers. DNA transcription and replication involve the self-assembly of large multiprotein complexes that spontaneously form through step-by-step processes where binding of proteins is facilitated by the molecular flexibility.\cite{Alberts2008} At the single molecule level, folding of the proteins to their most probable conformation involves large-scale molecular fluctuations and slow global structural rearrangements that are guided by cooperative dynamics.\cite{Socci1996, WU2008, Maisuradze2009, Hegger2007} 
Proteins' slow fluctuations may have important biological implication in the mechanism of protein binding and function.\cite{Kitao1999, Amadei1993, Sittel2018,Zhuravlev2010, Atilgan2001, Demirel1998, Tirion1996, Bahar1999}  Following the hypothesis of the Monod-Wyman-Changeux model, a protein's spontaneous fluctuations can lead the conformation selection mechanism where the binding partner selects the most favorable conformation among the ones made available by fluctuations.\cite{Monod1965,Boehr2009,Csermely,Kahler2020}  Thus, fluctuations in the unbound protein can signal the regions that are involved in protein binding and function.

Molecular dynamics (MD) simulations of proteins in solvent are a powerful method to identify fluctuations and investigate the role that the chemical structure, or primary sequence, of a protein play in multiscale dynamics. However, the information contained in the simulation trajectory is difficult to analyze because dynamical processes are often coupled on multiple length scales. Therefore, it is crucial to devise statistical procedures that conveniently separate the multidimensional trajectory of a simulation into a set of \textit{independent} dynamical processes that, when added together, form the observed data. These different contributions should be as independent as possible for one to be able to analyze and classify their dynamical response separately. Traditionally, this issue has been addressed by adopting statistical tools from signal processing to extract from a noisy response the most critical information, which is usually a slowly fluctuating signal or a collection of slowly fluctuating signals.

The most widely used analysis method for simulation trajectories is the principal component analysis or PCA method, based on the definition of a covariance matrix of the selected variables.\cite{Jolliffe2002} Correlation is a linear association measure, and uncorrelated processes are defined as having the cross terms in the covariance matrix equal to zero.
However, the goal is to isolate from the trajectory \textit{independent} fluctuations. Independent processes and uncorrelated processes are different from the mathematical point of view. Independent processes are defined as having a joint probability distribution that can be separated into a product of individual distributions.\cite{Hyvarinen2001} Independence includes not only linear (uncorrelated) but also nonlinear relationships.\cite{Goodfellow2016} In a nutshell, linear models such as PCA, in general, cannot decompose the dynamics into independent motions because those motions can be nonlinear. Independent, nonlinear, mode dynamics can be identified by an Independent Component Analysis (ICA).\cite{Hyvarinen2001}

In 2011, Naritomi and Fuchigami introduced to the field of protein dynamics the time-structure based independent component analysis or tICA, which is a specific type of ICA method.\cite{Naritomi2011} In tICA the dynamics is defined by a time-lagged covariance matrix and its limit at zero lag time, which is the conventional covariance matrix. By solving the generalized eigenvalue problem, the protein's dynamics are separated by tICA into modes that are uncorrelated at both zero lag time and at $\tau_{\text{tICA}}$.\cite{Klus2018} These constraints substitute for the stringent independence criteria normally required; i.e., the mode independence at $\tau_{\text{tICA}}$ substitutes for the independence of nonlinear zero-lag correlations.\cite{Hyvarinen2001, Molgedey1994} The time-structure based method was later revisited by Pande and coworkers, and with the name of `time-lagged independent component analysis' by Noe and coworkers.\cite{Perez-Hernandez2013, Schwantes2013} We will use the terms `time-structure based' and `time-lagged' interchangeably. When tICA is paired with a Markov State Model (MSM) of the kinetics of transition between modes, it accurately detects dominant slow modes of motion.\cite{Perez-Hernandez2013, Schwantes2013} The tICA modes identify the slowest dynamical decorrelation, and thus are considered to be the optimal linear coordinates to represent the slow dynamics:\cite{Tiwary2016} for example, the tICA modes have been used as variationally-optimal collective coordinates for enhanced sampling in metadynamics.\cite{Sultan2017,McCarty2017}

While tICA remains a rigorous statistical analysis of the multidimensional simulation trajectory, it still doesn't provide a physical interpretation of the slow dynamics or the connection between slow motions and protein's atomistic structure and interactions. Naritomi and Fuchigami partially addressed this issue by examining the slow dynamics of protein's domains and backbone using tICA. Still, an equation of motion that related structure and interaction potential to the dynamics was missing in their study.\cite{Naritomi2011,Naritomi2013}The degrees of freedom or `features' input to the tICA are chosen based on their ability to predict the slowest dynamics but are not necessarily connected to an equation of motion for describing the time evolution of the input coordinates. One determines the efficacy of the chosen features \emph{a posteriori} using cross-validation methods.\cite{Scherer2019, McGibbon2015} In this study, we relate the tICA formalism to our Langevin Equation for Protein Dynamics (LE4PD).\cite{ Beyerle2021a,Caballero-Manrique2007, Copperman2014, Copperman2015, Copperman2016} In the limit in which the tICA lag time is zero, the two are formally equivalent.\cite{Beyerle2021a} Furthermore, the LE4PD is a formal extension of the equation of motion for a macromolecule, obtained from the first-principles Liouville equation using Mori-Zwanzig projection operators.\cite{Zwanzig2001,Guenza1999,Schweizer1998,Lyubimov2011} Thus, albeit involved, there is a formal connection between the tICA fluctuations and the fundamental equation of motion for the dynamics of a protein.

The LE4PD formalism is based on the Rouse-Zimm equation for the dynamics of a polymer in solution,\cite{Doi1988,Bird1987, Zwanzig1988} that we extended to include physical characteristics that are specific to folded proteins. Typically i) inside the hydrophobic core of a protein, where atoms are not exposed to the solvent, the hydrodynamic interaction is screened, but atoms still experience friction; thus, the LE4PD adopts a residue-dependent friction coefficient calculated from the extent that each amino acid is solvent-exposed; ii) protein dynamics are non-linear and molecular rearrangements of the protein during fluctuations involve the crossing of energy barriers that play a major role in protein dynamics and folding. The LE4PD approximately accounts \emph{a posteriori} for the nonlinearities in the dynamics through the construction of free-energy landscapes for each mode and the rescaling of the timescale of barrier-crossing.\cite{Beyerle2021a,Copperman2014, Beyerle2019} 

Through normal mode diagonalization, the LE4PD separates the dynamics sampled in a long MD simulation, or in a set of short MD simulations, into a set of diffusive normal modes that are largely independent. These modes directly depend on real-space information, as the amino acid-specific local frictions, the water's viscosity, the potential of mean force, and the internal energy barriers are included. From the mode-dependent free energy landscape, one can identify the energy minima and the pathways that cross energy barriers, thus isolating the mode-dependent local fluctuations. Along the pathway, one can sample the protein conformation, thus depicting the conformational transitions during barrier crossing. A simple Kramers' rescaling, or applying a MSM analysis to the mode-dependent free energy surface (FES), provides the transition times of the mode-dependent fluctuations. Relaxation dynamics of time correlation functions predicted by LE4PD have been shown to be accurate when compared with experimental data of T$_1$, T$_2$, and NOE NMR relaxation,\cite{Copperman2014, Copperman2015} as well as to short-time Debye-Waller factors from X-ray scattering experiments.\cite{Caballero-Manrique2007} Recently, Beyerle et al. extended the LE4PD approach to describe \textit{anisotropic} fluctuations in the LE4PD-XYZ model.\cite{Beyerle2021a}

In this study, we focus on tICA, given that at zero lag time, the covariance matrix is formally consistent with the inverse of the LE4PD-XYZ force matrix. We compare the position, timescale, and pathway of slow fluctuations measured by tICA with the ones described by the LE4PD-XYZ. The question we aim to address is if a Langevin-mode decomposition can be as effective as tICA in isolating the leading dynamical processes from a protein MD trajectory.

We analyze an extensive, 1-$\mu$s long MD simulation of the protein ubiquitin in a solution of sodium chloride at physiological conditions.
Ubiquitin is a regulatory protein in eukaryotic cells, known for its role as a post-translational modifier of other proteins through mono- and poly-ubiquitination processes. It can bind to substrates either covalently\cite{Komander2009, Komander2012} or non-covalently.\cite{Lv2018} By identifying the slowest fluctuations in the isolated protein, we can locate important regions for ubiquitin's binding where a partner's selection can be guided by the thermodynamics and the kinetics of the different fluctuation processes.

This study addresses several relevant questions related to protein fluctuations and the tICA method: are the tICA's slow fluctuations similar to the ones that LE4PD-XYZ identifies? How are the results of this study dependent on the choice of the tICA lag time? Are there different but compatible procedures to correctly identify the best tICA lag time? Can a Langevin equation that adopts the tICA covariance matrix to build the intramolecular force matrix describe the correct dynamics of the system as measured by time correlation functions? How important is the role of hydrodynamics in tICA? How significant are the internal energy barriers that are only approximatively accounted for by tICA?

By comparing LE4PD-XYZ to tICA, this study formally connects the tICA method to a Langevin equation of motion. Along similar lines of thinking, Takano and coworkers recently proposed the Relaxation Mode Analysis (RMA), which is similar to tICA.\cite{Takano1995,Mitsutake2011,Mitsutake2015, Mitsutake2018} Both RMA and tICA maximize the time-dependent correlation matrix of the fluctuations at a given lag time, $\tau_{\text{tICA}}$, and at an initial time, $t_0$, while dynamics faster than $t_0$ is averaged out.\cite{Mitsutake2018} The difference between the two approaches is that RMA calculates the covariance matrix at a time $t_0\neq 0$, while tICA is a particular case of RMA, where $t_0=0$.\cite{Mitsutake2018} The RMA also has some similarities with our LE4PD approach as both accurately model with a Langevin equation of motion the slow dynamics of the protein, even if the details of the two dynamical equations are different.

The paper is structured as follows. Section \ref{LE4PD} briefly summarizes the anisotropic LE4PD-XYZ approach and the calculation of the LE4PD mode-dependent free energy surfaces. Section \ref{tICA} presents the tICA method while also proposing the calculation of a single-mode-dependent free energy map, which MSM analyzes. Section \ref{comparison} illustrates the comparison between LE4PD-XYZ's and tICA's slowest fluctuations. This Section also includes a biological interpretation of the observed fluctuations. Section \ref{wwohydro} discusses the compatibility of the FES for the slowest tICA mode and the slowest LE4PD-XYZ modes with and without hydrodynamic interactions. The calculation of time correlation functions with the two approaches is in Section \ref{TCFS}. Because the results of the tICA method depend on the tICA lag time selected, we analyze the dependence of the tICA single-mode FES on this parameter in Section \ref{lag time}. Finally, we assess the advantages and disadvantages of the proposed single-mode tICA method in comparison with the more traditional two-modes tICA analysis in Section\ref{2D}. A brief discussion with conclusions summarizes the findings of this study in Section \ref{conclusions}.

\section{The Anisotropic Langevin Equation for Protein Dynamics or LE4PD-XYZ}
\label{LE4PD}
In recent years, we have developed a coarse-grained model to describe protein fluctuations in the amino acid positions, called the Langevin Equation for Protein Dynamics (LE4PD).\cite{Caballero-Manrique2007, Copperman2014, Copperman2015, Copperman2016, Beyerle2019, Copperman2017}
The original LE4PD is isotropic and is presented in Section S6 of the Supplementary Material. Beyerle at al. have recently extended it to the related anisotropic formalism, called the LE4PD-XYZ method.\cite{Beyerle2021a} The anisotropic LE4PD directly connects the PCA fluctuations to an equation of motion that contains the covariance matrix in the amino acids positions.\cite{Beyerle2021a} We briefly review the LE4PD-XYZ model here, while we refer for more details on both LE4PD models to the aforementioned original manuscripts. 

The first step in developing the anisotropic LE4PD is to define as the leading variables the deviations of the position of the protein's alpha-carbons from their average values, $\Delta\vec{ R}_i(t)= \vec{ R}_i(t) -  \langle\vec{ R}_i(t)\rangle$.\cite{Beyerle2021a} 
Here, $\langle a(t)\rangle=\frac{1}{M}\sum\limits_{t=1}\limits^{M}a(t)$ denotes the usual static average calculated over a trajectory of length $M$ frames.

 Each component of the position vector fluctuation follows the anisotropic LE4PD equation of motion
\begin{equation}
\frac{d {\Delta R_i^{\alpha}}(t)}{dt}=-\frac{k_BT}{\overline{\zeta}}\sum\limits_{\beta,\gamma\in\{x,y,z\}}\sum\limits_{j=1}\limits^{N}\sum\limits_{k=1}\limits^{N} H_{ij}^{'\alpha\beta}A_{jk}^{'\beta\gamma}{\Delta R_k^{\gamma}}(t)+{\Delta v}_i^{\alpha} (t),
    \label{le4pdxyz}
\end{equation}
where $\alpha,\beta,\gamma \in \left\{x,y,z\right\}$. Furthermore, $k_B$ is the Boltzmann constant, T is the temperature in Kelvin, and $\vec{v}_i(t)$ is a stochastic velocity. The average residue friction coefficient is $\overline{\zeta}=\frac{1}{N}\sum_i\zeta_i$, where $\zeta_i$ is the friction coefficient of amino acid $i$. 
The matrix  $ H_{ij}^{'\alpha\beta}$ describes the hydrodynamic interaction between the $\alpha$ component of residue $i$ and the $\beta$ component of residue $j$,
 \begin{equation}
H^{\alpha\beta}_{ij} = \frac{\overline{\zeta}}{\zeta_{i}}\delta_{ij}\delta_{\alpha\beta}+(1-\delta_{ij})\delta_{\alpha\beta}\overline{r}_w\left\langle\frac{1}{r_{ij}}\right\rangle \ , 
 \label{Hmat}
 \end{equation}
with
$\langle\frac{1}{r_{ij}}\rangle$ is the average inverse distance between residues $i$ and $j$, and $\overline{r}_w=\frac{1}{N}\sum_ir_{w,i}$ is the average residue radius exposed to the solvent. 

In this equation, the dynamics is defined in a body-fixed system of coordinates, where both translation and rotation dynamics have been eliminated. The trajectory of the protein, analyzed to build the $\mathbf{H}'$ and $\mathbf{A}'$ matrices for example, is also in a body-fixed reference system, where translation and rotation are absent.\cite{Eckart1934,Sayvetz2004,Kneller2008,Chevrot2011}
The equation is solved by applying the fluctuation-dissipation condition, as described in our previous publication.\cite{Beyerle2021a}

The matrix $A_{jk}^{'\beta\gamma}$ describes the inverse of the covariance between the $\beta$ component of residue $j$ and the $\gamma$ component of residue $k$ as
\begin{equation}
 A^{\beta\gamma}_{jk}=\left( \left[ \mathbf{a} \otimes \mathbf{I} \right]^T
 \mathbf{U} \left[ \mathbf{a} \otimes  \mathbf{I}  \right] \right)_{jk}^{\beta\gamma },
 \label{Amat}
\end{equation}
where $\mathbf{U}^{-1}=\langle\Delta\vec{l}(t)\ \Delta\vec{l}(t)^T\rangle$ is the matrix of correlations of the bond fluctuations in Cartesian coordinates with $\Delta\vec{l}(t)=\left(\mathbf{a} \otimes \mathbf{I} \right)\Delta\vec{R}(t)$, $\Delta\vec{l}^{\alpha}_i(t)=\sum_{j}\left(\mathbf{a} \otimes \mathbf{I} \right)_{ij}\delta_{\alpha\beta}\Delta\vec{R}^{\beta}_j(t)$. $\mathbf{I}$ is the 3 $\times$ 3 identity matrix and $\mathbf{a}$ is the $N-1\times N$ matrix of the amino acid connectivity (with $ i=1, ..., N-1$ and $j=1, ..., N$),
\begin{equation}
a_{ij}=\left\{ \begin{array}{l} -1,\  j=i  \\ 1, \ \ j=i+1 \\ 0,  \ \ \text{otherwise} \end{array}\right. 
\end{equation}
Here, $\delta_{\alpha\beta}$ is the Kronecker delta, and the `$\otimes$' symbol denotes the Kronecker product.\cite{Horn1994} 

From the simulation trajectory, we calculate i) the average fluctuations in the $\alpha$-carbon positions, which enter through the $\mathbf{U}$ matrix the inverse of the covariance matrix, Eq. \ref{Amat}; ii) the average inverse distance between the residues, which enter the hydrodynamic interaction matrix, Eq. \ref{Hmat}; iii) the friction coefficient of each residue, $\zeta_i$ and the average residue radius exposed to the solvent, $\overline{r}_w$, which also enter Eq. \ref{Hmat}. The simulation trajectory is also used to test the quality of the theoretical predictions of time correlation functions in Section \ref{TCFS}.

More details on the anisotropic LE4PD model, and how it is formally related to the isotropic LE4PD, are given in \cite{Beyerle2021a}. Eq. \ref{le4pdxyz} is solved using the eigenvalue decomposition of the $\mathbf{H'A'}$ matrix product, $\mathbf{Q'}^{-1}\mathbf{H'A'Q'}=\Lambda'$, which gives the equation of motion for the evolution of the LE4PD-XYZ modes, $Delta{\vec \xi'}_a(t)$: 

\begin{equation}
\frac{d\Delta{\vec \xi'}_a(t)}{dt}=-\frac{k_BT}{\overline{\zeta}}\lambda'_a\Delta{\vec\xi'}_a(t) + \Delta{\vec v'}_a(t).\label{le4pdxyzxi}
\end{equation}
with $\sigma_a=k_BT\lambda_a/\overline{\zeta}$ the characteristic diffusive rate of mode $a$,\cite{deGennes1979} and $\Delta{\vec v'}_a(t)$ the stochastic velocity in mode coordinates.
\subsection{Building a free energy map in anisotropic coordinates and measuring fluctuations timescales}
\label{FES}
Using the decomposition of $\mathbf{Q'}$ for the anisotropic $\mathbf{H'A'}$ matrix, the mode coordinate $\xi'_a(t)$ of the anisotropic LE4PD can be separated into its $x-,y-,$ and $z-$ components as
\begin{align}
\vec\xi'_a(t) &= \sum\limits_{i=1}\limits^{3N} Q'^{-1}_{ai}\Delta\vec{R}_i(t) \notag\\
&= \sum\limits_{i=1}\limits^{3N} \left[\left(Q'^{-1}_{a,x} \otimes \hat{x}^T\right)_i + \left(Q'^{-1}_{a,y} \otimes \hat{y}^T\right)_i + \left(Q'^{-1}_{a,z} \otimes \hat{z}^T\right)_i\right]\Delta\vec{R}_i(t) \notag\\
&=\sum\limits_{i^{\prime}=1}\limits^{N} Q'^{-1}_{ai^{\prime},x}\Delta x_{i^{\prime}(t)} + Q'^{-1}_{ai^{\prime},y}\Delta y_{i^{\prime}}(t) + Q'^{-1}_{ai^{\prime},z}\Delta z_{i^{\prime}}(t)\notag\\
&=\xi'_{a,x}(t) + \xi'_{a,y}(t) + \xi'_{a,z}(t) \ , \label{xixyz}
\end{align}
and the spherical mode coordinates and free-energy surfaces can be defined as
\[\theta'_a (t) = \arccos\left({\xi'_{a,z} (t)}/{|\xi'_a (t)|} \right)\] 
\[\phi'_a (t) = \arctan\left({\xi'_{a,y} (t)}/{{\xi}'_{a,x} (t)} \right),\]
\begin{equation}
    F'(\theta'_a,\phi'_a)=-k_BT\ln\left[P'(\theta'_a,\phi'_a)\right]  \ ,
    \label{fesxyz}
\end{equation}
where the probability for the protein of adopting, in mode $a$, a conformation with angles $\theta'_a,\phi'_a$  is 
 \[P'(\theta'_a,\phi'_a)=\int P'\left(\vert \vec{\xi}'_a\vert\theta'_a,\phi'_a\right)d\vert\vec{\xi}'_a\vert.\]

\begin{figure*}[htb] 
\center
\includegraphics[width=1.4\columnwidth]{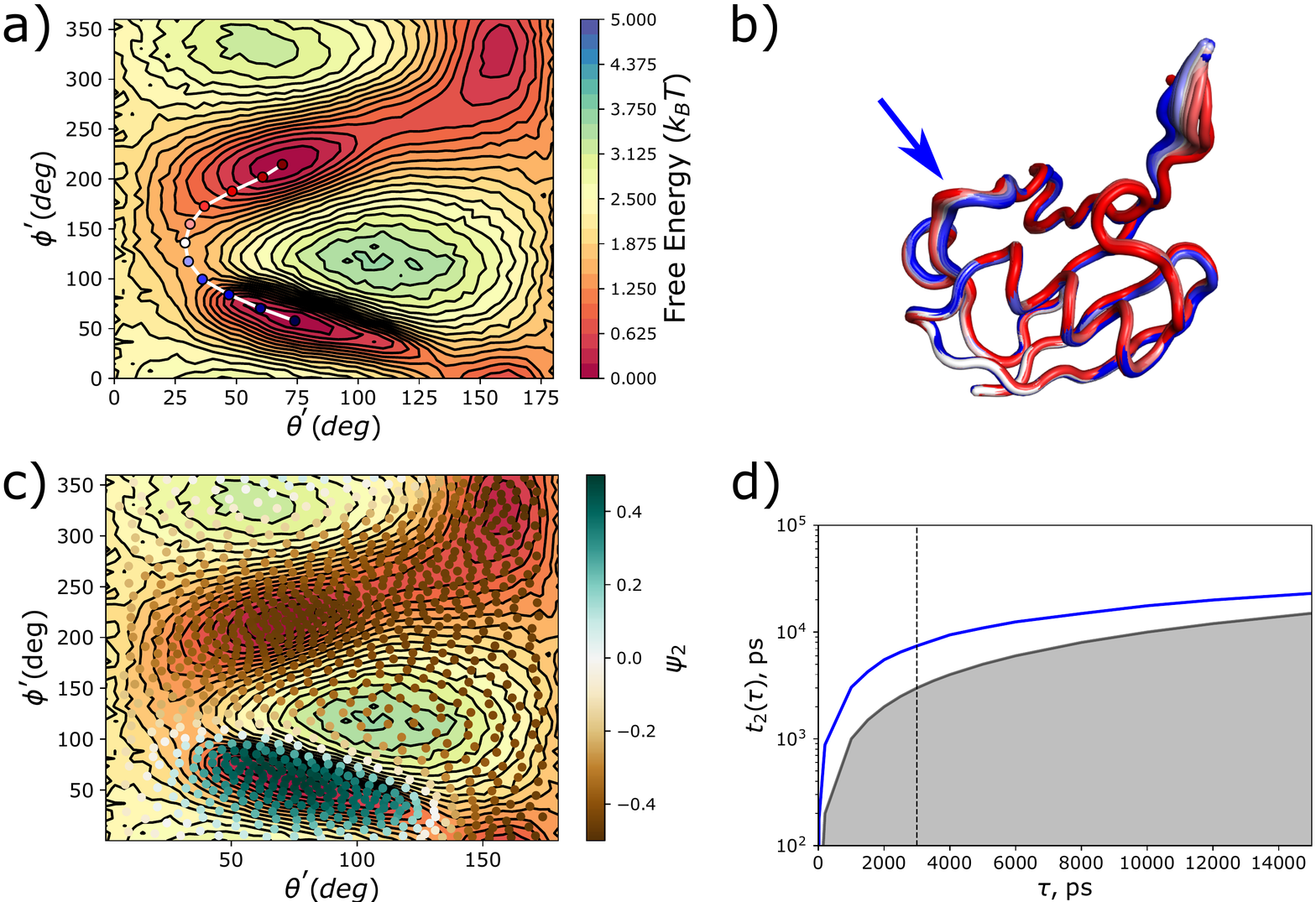}
\caption{Analysis of the seventh LE4PD-XYZ mode without hydrodynamics. Panel a) shows the free energy landscape of the seventh LE4PD-XYZ mode in the two spherical coordinate reference system. The pathway of crossing the energy barrier between the two minima is identified with a rubber band, using a variant of the string method.\cite{Beyerle2019} Panel b) shows ubiquitin's conformations that correspond to the pathway shown in panel a) with the red conformation referring to the energy minimum at the top of the map, and the blue conformation corresponding to the energy minimum at the bottom of the map. The blue arrow points to the region of ubiquitin experiencing the largest amplitude fluctuation for this mode corresponding to the $50 \text{s}$ loop. Panel c) displays the second eigenvector resulting from the diagonalization of the transition matrix defined in the Markov State Model (MSM) procedure for this mode, which identifies the two minima in the FES. The projection of $\psi_2$ onto the discrete states of the MSM has colors that correspond to the scaled-and-shifted value of $\psi_2$ at that discrete state, $\psi_2 = \frac{\psi_2-\min(\psi_2)}{\max(\psi_2)-\min(\psi_2)}-0.5$. Panel d) shows how the transition time for the second MSM eigenvector changes when we select a different lag time in the calculation of the MSM transition matrix.  The black, vertical line demarcates the lag time corresponding to the second eigenvector mapping the two minima, as reported in panel c).}
\label{tp1squareLE4PDXYZa}
\end{figure*}

The linear combination of all the anisotropic modes leads to structural and time-dependent properties, directly comparable to simulations or experimental data. From the anisotropic free-energy surfaces, we calculate fluctuations in the three spatial directions.
As an example, Figure \ref{tp1squareLE4PDXYZa} shows the LE4PD-XYZ analysis of a 1-$\mu$s long MD simulation of the protein Ubiquitin (for details on the simulation, see Section S8 in the Supplementary Material). The same trajectory will be analyzed with the tICA to provide a comparison between the two methods. Figure \ref{tp1squareLE4PDXYZa} shows in panel a) the FES in the mode coordinates for the seventh LE4PD-XYZ mode without hydrodynamics (mode seventh is the slowest one in this formalism). The FES displays two minima separated by a small energy barrier. The protein's conformations along the transition pathway between these two minima are displayed in panel b). Panels c) and d) report data from a MSM analysis of the mode trajectory (for details on the MSM, see Section S9 in the Supplementary Material). 

More specifically, Panel c) shows the projection of the second MSM eigenvector, $\psi_2$ onto the FES. The second eigenvector of the MSM transition matrix identifies the two deepest minima and the top of the energy barrier between them in an FES.\cite{Bowman2013} When the second MSM eigenvector matches the population in the two minima and at the top of the barrier, the MSM lag time identified by this process provides the transition time across the barrier.\cite{Beyerle2019} Panel d) shows the calculation of the transition time as a function of the MSM lag time. When a process fulfills the Chapman-Kolmogorov equation, the process follows markovian statistics, and the transition time becomes independent of the lag time, i.e., the transition time becomes constant.\cite{Bowman2013} Panel d) shows that the transition to markovian dynamics (when the blue line becomes flat) is reasonably close to the transition time calculated from the second MSM eigenvector (the vertical dashed line). Thus, the two procedures to evaluate the transition time give similar values for the seventh mode.  
Note that Figure \ref{tp1squareLE4PDXYZa} displays results for the LE4PD-XYZ theory without hydrodynamic interactions (see for a discussion Section \ref{wwohydro}). An identical calculation performed for the slowest LE4PD-XYZ mode with hydrodynamic interactions (mode six) gives similar free-energy maps and MSM analyses. This result suggests that the hydrodynamics interaction does not affect the dynamics of the slowest fluctuations. Note that the calculations performed take into account hydrodynamics and also include residue-dependent friction coefficients.

The procedure briefly presented here was proposed in our recent publications.\cite{Beyerle2019,Beyerle2021a} The same well-tested procedure will be applied to the analysis of the tICA modes in Section \ref{tICA}. 

\section{Time-lagged independent component analysis or \MakeLowercase{t}ICA}
\label{tICA}

The time-lagged independent component analysis is a method extensively used in the field of signal processing, information theory, artificial neural networks to identify hidden factors that are shared and underlie the observed multivariate data.\cite{Molgedey1994} This technique has been applied in several fields, including the analysis of protein dynamics to identify the prevalent large-scale motion inside a simulation trajectory. By introducing a time lag in the covariance matrix, one effectively includes the temporal dimension in the analysis of the leading fluctuations making it possible to model kinetic processes. The time-lagged ICA is an extension of the principal component analysis (PCA) method, where one takes care of isolating \text{the most slowly decorrelating dynamics} while including the time dependence of the data as an explicit variable in the analysis. The tICA method has been reviewed in several recent publications and will be only summarized here.\cite{Perez-Hernandez2013,Sultan2018, Naritomi2011, Naritomi2013,Noe2015, Scherer2015}

While tICA is a general approach that applies to any set of coordinates, here, we are interested in performing a tICA of the alpha-carbon trajectory of a protein with $N$ residues.  We define as tICA coordinates the  ${\Delta \mathbf{R}}(t)^T={\vec{R}}_1(t)-{\langle\vec{R}}_1(t)\rangle, {\vec{R}}_2(t)-{\langle\vec{R}}_2(t)\rangle, \ldots , {\vec{R}}_n(t)-{\langle\vec{R}}_n(t)\rangle $, where  $\Delta{\vec{R}}_i(t)=\vec{R}_i(t)-\langle\vec{R}_i(t)\rangle$ represents the fluctuations out of the equilibrium structure of the position of the space coordinates ${\vec{R}}_i(t)$, with ${\vec{R}}_i(t)=x_i(t),y_i(t), z_i(t)$  and $i=1, \ldots, N$ with $N$ the number of amino acids in the protein. The time-lagged covariance matrix is defined, for a lag time $\tau_{\text{tICA}}$, as 
\begin{equation}
 \mathbf{C}^r(\tau_{\text{tICA}})=\langle {\Delta\mathbf{R}}(t_0+\tau_{\text{tICA}})^T{\Delta\mathbf{R}}(t_0) \rangle_{\tau_{\text{tICA}}} \ ,
\label{CTau}
\end{equation}
and for $\tau_{\text{tICA}}=0$ the covariance matrix recovers the static, structural matrix that is used in PCA, as $\mathbf{C}^r(0)= \langle\Delta{\mathbf{R}}(t_0)^T \Delta{\mathbf{R}}(t_0) \rangle$. Here, $\langle a(t+\tau)b(t)\rangle_{\tau}=\frac{1}{M-\tau}\sum\limits_{t=1}\limits^{M-\tau}a(t+\tau)b(t)$ denotes an average over the time-lagged trajectory containing $M$ frames.

The tICA modes, or tICs, are found by solving the following generalized eigenvalue equation\cite{Hyvarinen2001, Perez-Hernandez2013}:
\begin{equation}
\mathbf{C}^r(\tau_{\text{tICA}}){\Omega}=\mathbf{C}^r(0){\Omega}\Lambda_{IC}(\tau_{\text{tICA}}),
    \label{gee}
\end{equation}
where ${\Omega}$ is the matrix of right eigenvectors of $\mathbf{C}^r(\tau_{\text{tICA}})$, and $\Lambda_{IC}(\tau_{\text{tICA}})$ is the diagonal matrix of the related eigenvalues. 

From the solution of the generalized eigenvalue problem, one has that the eigenvector matrix, ${\Omega}$, diagonalizes both $\mathbf{C}^r(\tau_{\text{tICA}})$ and $\mathbf{C}^r(0)$: 

\begin{align}
{\Omega}^T\mathbf{C}^r(\tau_{\text{tICA}}){\Omega}&=\Lambda_{IC}(\tau_{\text{tICA}}) \notag\\
{\Omega}^T\mathbf{C}^r(0){\Omega}&=\Lambda_{IC}^{\prime}(0)=\mathbf{I},
    \label{diagonalization}
\end{align}
where $\mathbf{I}$ is an identity matrix of the same dimensions as $\mathbf{C}^r(\tau_{\text{tICA}})$, and 
$\mathbf{C}^r(0)$. 
The tICA modes, $\mathbf{z}(t)$, are determined by transforming the input coordinates $\Delta\mathbf{R}(t)$ as $\mathbf{z}(t)={\Omega}^T\Delta\mathbf{R}(t)$. 

\subsection{Connection between the tICA and the LE4PD-XYZ}
Given the definition of the LE4PD-XYZ $\mathbf{A}$ matrix of Eq. \ref{Amat}, it is straightforward to show that 
\begin{equation}
\mathbf{A}=\lim_{\tau_{\text{tICA}}=0}\mathbf{C}^r(\tau_{\text{tICA}})^{-1} \,
\end{equation}
where the tICA matrix is defined in Eq. \ref{CTau}. A more detailed calculation of this relation is reported in Ref.\cite{Beyerle2021a}.

It follows that in the limit of zero lag time, the tICA eigenvectors are identical to the eigenvectors of the LE4PD-XYZ matrix, when hydrodynamics is discarded, i.e., the hydrodynamic matrix $\mathbf{H}=\mathbf{I}$. The tICA eigenvalues, instead, are equivalent to the inverse of the LE4PD-XYZ eigenvalues. For the hydrodynamic matrix to become equal to the identity matrix, one needs to assume that there are not solvent-mediated hydrodynamic interactions between amino acids, and that the friction coefficient of each amino acid is set equal to the average friction coefficient. Note that the formal equivalence between the tICA dynamics at lag time zero and the LE4PD-XYZ approach represented by Eq. \ref{le4pdxyzxi}, with $\mathbf{H}=\mathbf{I}$, implies that the fluctuations are harmonic and the internal energy barriers are also discarded. Thus, both the LE4PD-XYZ and the tICA dynamics should include in the equation of motion the correction due to the energy barriers calculated from the mode-dependent free energy landscapes and the hydrodynamic interaction.


\subsection{Converting to spherical coordinates creates a free-energy surface for each tICA mode}
\label{FESTICAa}

\begin{figure*}[htb] 
\center
\includegraphics[width=1.4\columnwidth]{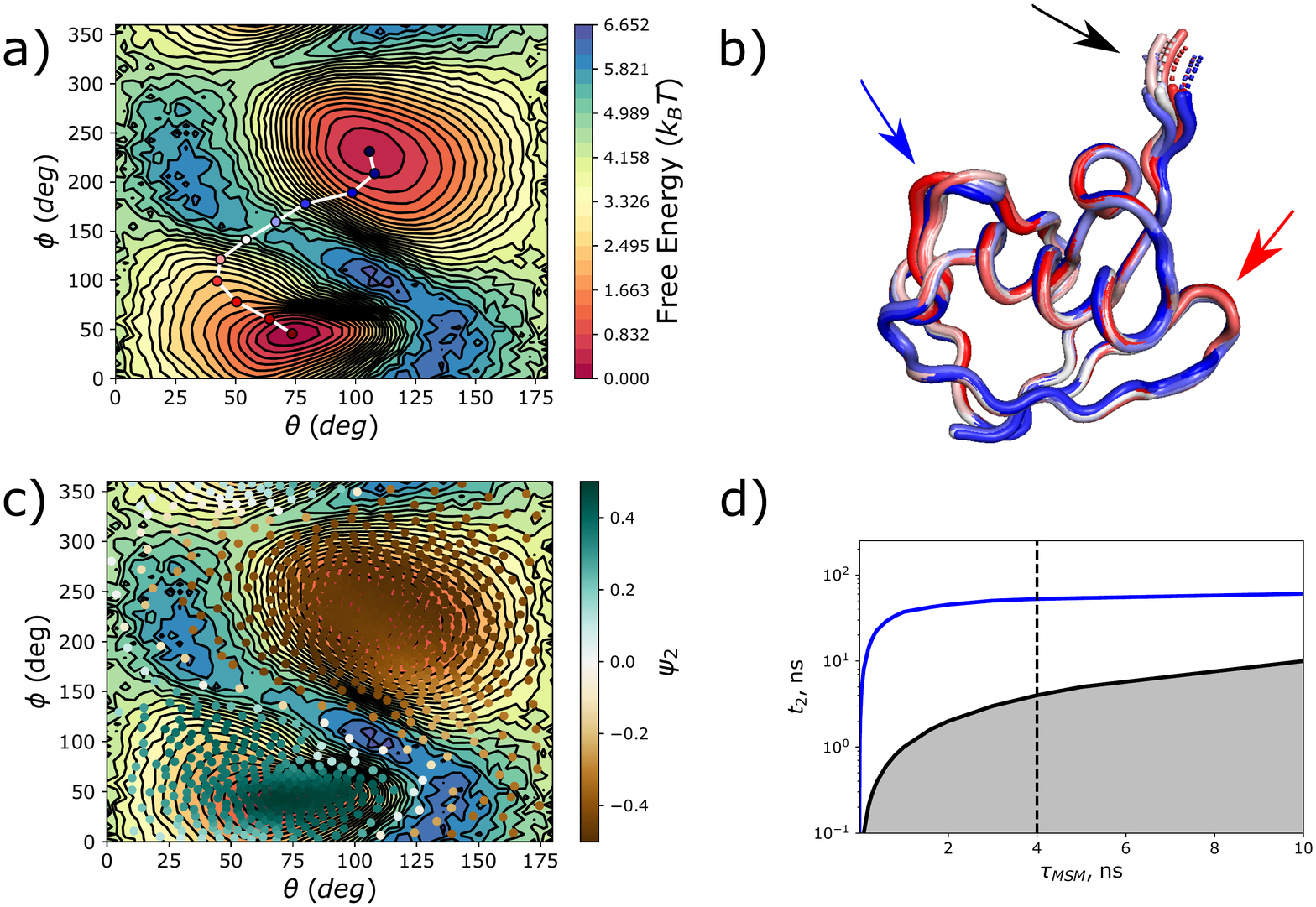}
\caption{Analysis of the free energy map of the first  tICA mode. a): Free-energy surface along the $\left(\theta_a, \phi_a\right)$ coordinates for the slowest tIC.  b): Structures of ubiquitin from the trajectory along the free-energy surface given in a). The colors of the structures correspond to the given colored marker along the transition pathway. Arrows point to the three regions of ubiquitin showing the largest amplitude fluctuations: the C-terminal tail (black arrow), the $50\text{ s}$ loop (blue arrow), and the Lys11 loop (red arrow). c): projection of $\psi_2$ onto the discrete states of the MSM; colors correspond to the scaled-and-shifted value of $\psi_2$ at that discrete state, $\psi_2 = \frac{\psi_2-\min(\psi_2)}{\max(\psi_2)-\min(\psi_2)}-0.5$. d): implied timescales of the MSM as a function of MSM lag time. The black vertical line demarcates the lag time selected when constructing the MSM,  $\tau_{MSM}=4 \ \text{ns}$.}
\label{tp1square}
\end{figure*}

\begin{figure*}[htb] 
\center
\includegraphics[width=1.4\columnwidth]{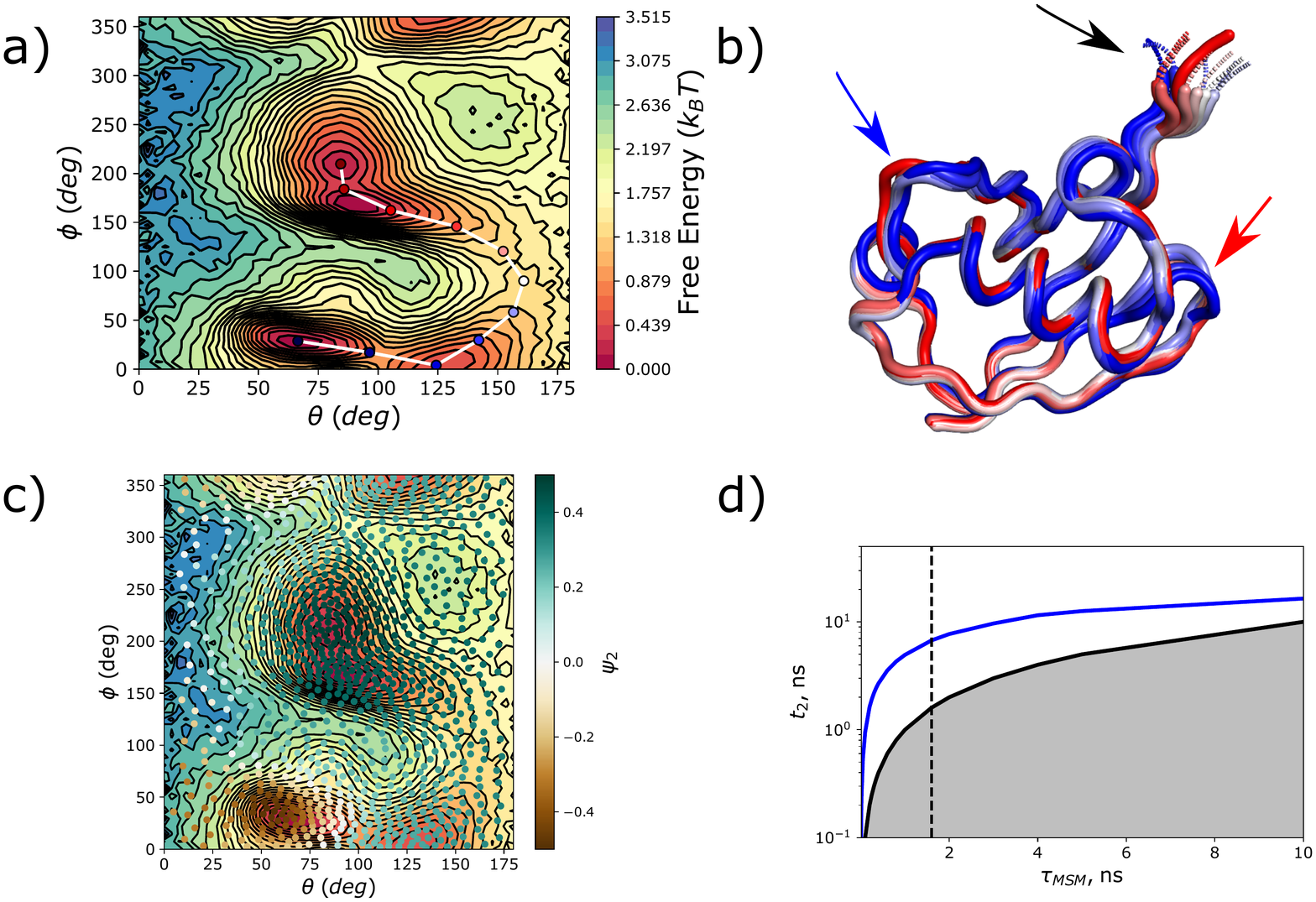}
\caption{Analysis of the free energy map for the second tICa mode. a): Free-energy surface along the $\left(\theta_a, \phi_a\right)$ coordinates for the slowest tIC.  b): Structures of ubiquitin from the trajectory along the free-energy surface given in a). The colors of the structures correspond to the given colored marker along the transition pathway. Movement along the pathway in a) correspond to fluctuations mostly in the Lys11 loop (red arrow) and C-terminal tail (black arrow) of ubiquitin, as well as smaller amplitude motions in the $50\text{ s}$ loop (blue arrow) c): projection of $\psi_2$ onto the discrete states of the MSM; colors correspond to the scaled-and-shifted value of $\psi_2$ at that discrete state, $\psi_2 = \frac{\psi_2-\min(\psi_2)}{\max(\psi_2)-\min(\psi_2)}-0.5$. d): implied timescales of the MSM as a function of MSM lag time. The black vertical line demarcates the lag time selected when constructing the MSM,  $\tau_{MSM}= 1.6 \ \text{ns}$.}
\label{tp2square}
\end{figure*}

In this section we show how a free energy landscape can be associated to each tICA mode. This step is important because the mode-specific free energy barriers allow one to study the details of conformation fluctuations, thus determining the pathway of transition and the height of the energy barriers in the fluctuations associated to tICA modes.  We follow here a procedure similar to the one we established for the LE4PD-XYZ modes.

The elements of the tICA eigenvectors ${\Omega}$ can be decomposed into their $x-,\ y-,$ and $z-$projections
\begin{equation}
{\Omega}=\Omega^x \otimes \widehat{x} + \Omega^y \otimes \widehat{y} + \Omega^z \otimes \widehat{z},
    \label{U_decomp}
\end{equation}
where $\widehat{x},\ \widehat{y},$ and $\widehat{z}$ are the unit vectors in the x-, y-, and z-directions, and $\otimes$ denotes the Kronecker product.\cite{Horn1994} This decomposition is useful as it allows for the creation of tIC-dependent free-energy surfaces, which can be compared directly with the LE4PD free energy surfaces (see Section \ref{wwohydro}).

To define a Free-Energy Surface (FES) for each of the tICA mode coordinates, we start by projecting the space coordinates of the fluctuations onto tICA modes using the tICA eigenvectors. 
For the tICA modes, the eigenvector matrix ${\Omega}^T$, which transforms the $\Delta\vec{R}(t)$ into the $\mathbf{z}(t)$ tIC coordinate system, can be decomposed into its contributions from the $x-,y-,$ and $z-$components of $\Delta\vec{R}(t)$, 
\begin{equation}
{\Omega}^T = \Omega^{T,x} \otimes \widehat{x}^T + \Omega^{T,y} \otimes \widehat{y}^T + \Omega^{T,z} \otimes \widehat{z}^T,
    \label{UT_decomp}
\end{equation}
which allows for the decomposition of each tIC $z_a(t)$ into its contributions from the $x-,y-,$ and $z-$components of the input coordinates $\Delta\vec{R}(t)$:
\begin{align}
z_{a,x}(t)=\sum_{i=1}\limits^{N} \left(\Omega^x\right)^{T}_{ai}\Delta{x}_i(t)\\
z_{a,y}(t)=\sum_{i=1}\limits^{N} \left(\Omega^y\right)^{T}_{ai}\Delta{y}_i(t)\\
z_{a,z}(t)=\sum_{i=1}\limits^{N} \left(\Omega^z\right)^{T}_{ai}\Delta{z}_i(t).
\label{xyz}
\end{align}
This decomposition can be used to describe each tIC in a new spherical coordinate system:
\begin{align}
R_a(t)=z_{a,r}(t)=\sqrt{z_{a,x}(t)^2 + z_{a,y}(t)^2 + z_{a,z}(t)^2}\label{rtp1}\\
\theta_a(t)=z_{a,\theta}(t)=\arccos\left(\frac{z_{a,z}(t)}{\vert \mathbf{z}_a(t)\vert}\right)\label{rtp2}\\
\phi_a(t)=z_{a,\phi}(t)=\arctan\left(\frac{z_{a,y}(t)}{z_{a,x}(t)}\right).
\label{rtp3}
\end{align}
With the definitions of $\theta_a(t),\phi_a(t)$ and $R_a(t)$, one can create two-dimensional free-energy surfaces in $(\theta_a, \phi_a)$ by averaging over the radial coordinate $R_a(t)$:
\begin{align}
 F(\theta_a,\phi_a)&=-k_BT\ln\left[P(\theta_a,\phi_a)\right]\notag\\
 &=-k_BT\ln\left[\int P\left(R_a,\theta_a,\phi_a\right)d R_a\right] \ .
 \label{tica_FES}
 \end{align}
The main advantages of constructing the free-energy surfaces in this manner are that 1) each surface is tIC-specific because the dynamics among tICs are largely decoupled (an evaluation of the extent of coupling in tICA modes is reported in Section S1 of the Supplementary Material, and 2) energetic pathways and fluctuations along this surfaces are easy to visualize for each tIC. As with previous LE4PD analyses, a variant of the string method is utilized to find minimum free-energy pathways between energy wells on the surface.\cite{Beyerle2021a,Beyerle2019,E2002} A MSM analysis can provide the time scale associated to each tICA mode.
Note that the tICA can be applied once we have selected a lag time, $\tau_{\text{tICA}}$. In this study, the tICA lag time is $2$ ns; we present the procedure to establish this value in Section \ref{lag time}.

\subsection{Characterization of the first and the second tICA modes: FES, pathways, and conformational transitions}
\label{FEStICA}

As an example of the information inherent in $F(\theta_a,\phi_a)$ for the tICs, Figure \ref{tp1square} and Figure  \ref{tp2square} show the results of the analysis in the $\left(\theta_a, \phi_a\right)$ coordinate space for the two slowest tICA modes extracted from the 1-$\mu$s simulation of ubiquitin. 

Figure \ref{tp1square}a shows the free energy map, $F(\theta_1,\phi_1)$, for the first tICA mode, $z_1(t)$, with a pathway drawn between the two prominent minima on the surface. Figure \ref{tp1square}b displays the fluctuations along the alpha-carbon backbone of ubiquitin when moving along the pathway given in Figure \ref{tp1square}a; the colors of the structures in Figure \ref{tp1square}b correspond to the colors of the images along the pathway in Figure \ref{tp1square}a. Movement along the minimum energy pathway for $z_1(t)$ shows fluctuations in the $50\text{ s}$ loop (blue arrow), the C-terminal tail (black arrow), and the Lys11 loop (red arrow), each of which is a known binding region of ubiquitin to other proteins.\cite{Penengo2006,Komander2009,Komander2012} 

Figure \ref{tp1square}c shows the projection of the most slowly decaying eigenfunction, $\psi_2$, from the MSM transition matrix constructed on this surface starting from the MD trajectory. Note that by assuming different MSM lag times, one obtains different eigenvectors $\psi_2$ and different projections onto the surface. By selecting $\tau_{MSM}=4.0 \  \text{ns}$ we see that the most positive projection of $\psi_2$ lies in the minimum in the bottom half of the surface, and the maximum projection of $\psi_2$ lies in the minimum in the top half of the surface. Thus, with $\tau_{MSM}=4.0 \  \text{ns}$ selected, the spectrum indicates that the slowest process described by the MSM corresponds to transitions between the two minima on the surface, whose fluctuations should be described well by the extracted structures from the pathway given in Figure \ref{tp1square}b.

To test the validity of the $\tau_{MSM}=4.0 \  \text{ns}$ found with this procedure,  Figure \ref{tp1square}d shows the implied timescale of $t_2$, i.e., the timescale of the process described by $\psi_2$, as a function of MSM lag time $\tau_{MSM}$. The vertical dashed line marks the lag time used in the construction of the MSM shown in Figure \ref{tp1square}c. We see that at the selected MSM lag time, the dynamics is markovian, and the $t_2$ time is constant, confirming the validity of the selected  $\tau_{MSM}$.
Thus, for the $\psi_2$ shown in Figure \ref{tp1square}c, the MSM transition matrix, $\mathbf{T}$, is constructed by sampling the trajectory with an interval of $\tau_{MSM} = 4.0$ ns, and the predicted timescale is $t_2\left(\tau_{MSM}=4.0\ \text{ns}\right)=52.6$ ns. In summary, combining the tIC free-energy surface in $\left(\theta_a,\phi_a\right)$ with the Markov state modeling analysis predicts that the timescale of movement between the two minima in Figure \ref{tp1square}a is approximately $53$ ns. The corresponding dynamics along the alpha-carbon backbone during this event are illustrated in Figure \ref{tp1square}b.

Figure \ref{tp2square} illustrates the analogous analysis for the $\left(\theta_a, \phi_a\right)$ surface spanned by the \textit{second-slowest} tIC. Drawing a transition pathway between the two minima on the surface (Figure \ref{tp2square}a) and extracting the structures along that pathway from the MD simulation shows that this tIC describes fluctuations in the Lys11 loop and C-terminal tail regions of ubiquitin (Figure \ref{tp2square}b).\cite{Komander2009, Komander2012} Again, using the decomposition of $\psi_2$ from the MSM on this surface to choose the lag time of the MSM (Figure \ref{tp2square}c), the process of transitioning between the minima on the surface is predicted to occur over a timescale of $6.7$ ns (Figure \ref{tp2square}d). Thus, the $\left(\theta_a, \phi_a\right)$ surface for the second-slowest tIC predicts mainly motion in the tail and Lys11 loop, occurring over a timescale of $6.7$ ns. 

We repeat the evaluation of the $t_2$ times for all the tICs and we present these results for the first ten tICs in comparison with the first ten LE4PD-XYZ modes  in  Tables \ref{timetable1} and \ref{timetable2}, Section \ref{comparison}.
Further information on the timescales associated with the first and second tICA modes and on the amplitude and position of their fluctuations are presented in the Sections \ref{comparison} and \ref{LML}, respectively.

\section{A comparison between the \MakeLowercase{t}ICA and the LE4PD-XYZ slowest fluctuations}
\label{comparison}
In this section we perform a quantitative comparison of the transition times predicted for the slow tICA modes and for the LE4PD-XYZ modes, starting from the same MD trajectory of ubiquitin in solution (for the MD simulation method, see Section S8 in the Supplementary Material).

For both the LE4PD-XYZ and the tICA free energy surfaces we evaluate the MSM times following the procedure presented in Sections  \ref{FES} and  \ref{FEStICA}.
To calculate the transition times, $t_2$, we construct the MSM for each mode and estimate the timescales, $t_2$, using either the mapping of the second MSM eigenvector onto the FES or the markovian criterion (i.e. the Chapman-Kolmogorov [CK] condition) for the mode trajectories (for details on the MSM, see Section S9 in the Supplementary Material). 
Table \ref{timetable1} presents the values of $t_2$ calculated using the second MSM eigenvector $\psi_2$, while Table \ref{timetable2} reports the values of the transition times calculated using the Chapman-Kolmogorov criterion for a markovian process.\cite{Reichl1998,Bowman2013} In both tables, we also report the values of transition times calculated for the internal modes of the isotropic LE4PD equation with hydrodynamic interaction included (the LE4PD theory is briefly summarized in Section S6 in the Supplementary Material). The data for the LE4PD-XYZ approach are reported both with and without hydrodynamic interaction included. From these data we can assess the importance of hydrodynamics in the time scale of the protein's fluctuations. Note that the calculations are performed while including hydrodynamics and also account for residue-dependent friction coefficients.

\begin{table}[h!]
   \centering
   \caption{Comparing the slowest timescales from the isotropic LE4PD, the LE4PD-XYZ (ansiotropic LE4PD), and tICA for the 1-$\mu$s simulation of ubiquitin at the MSM lag time where the spectrum of $\psi_2$ on the free-energy surface is optimized ($\tau_{MSM}$).\cite{Beyerle2019} For the LE4PD-XYZ modes, the table reports data for the approach with (w/ HI) and without hydrodynamic (w/o HI) interaction included. The isotropic LE4PD modes are indexed by internal mode number (see explanation in the text).}
   \label{timetable1}
   \begin{tabular}[c]{ c c c c c}
   &LE4PD & \multicolumn{2}{c}{LE4PD-XYZ} & tICA \\
   & &  w/ HI& w/o HI & \\
   Mode &  $t_2\left(\tau_{MSM}\right)$ , ns & $t_2\left(\tau_{MSM}\right)$, ns & $t_2\left(\tau_{MSM}\right)$ , ns & $t_2\left(\tau_{MSM}\right)$ , ns  \\ \midrule

   1    & 3.9(1.05)&  8.0(3.2) &  6.5(2.8) & 52.6 (4.0)\\ 
   2    & 0.7(0.1)& 3.7(1.1) &  4.6(1.5) & 6.7 (1.6) \\
   3    & 0.9(0.35)& 4.3(2.5) &  4.3(2.0) & 4.8(1.6)\\
   4    & 2.4(0.5)& 6.4(4.0) &  1.0(0.3) & ---(---)\\
   5    & 0.1(0.01)& 4.8(4.0) &  5.5(4.9) & 2.5(1.0)\\
   6    & 11.0(0.9)& 3.3(1.0) &  3.1(2.0) & ---(---)\\
   7    & 0.5(0.25)& 3.6(2.0) &  7.4(3.0) & 2.5(1.6)\\
   8    & 0.4(0.11)& 0.6(0.2) &  ---(---) & 6.1(5.0)\\
   9    & 0.24(0.1)& 0.3(0.1) &  1.3(0.5) & 0.8(0.3)\\
   10   & 0.35(0.3)& 0.4(0.1) &  0.4(0.2) & 7.0(5.0)\\
   \end{tabular}
  \newline
\end{table}

\begin{table}[h!]
   \centering
   \caption{Comparing the slowest timescales from the isotropic LE4PD, the LE4PD-XYZ (ansiotropic LE4PD), and tICA for the 1-$\mu$s simulation of ubiquitin in the long-lag time regime ($\tau_{MSM}$) where the dynamics best satisfy the Chapman-Kolmogorov condition.\cite{Swope2004a} For the LE4PD-XYZ modes, the table reports data for the approach with (w/ HI) and without hydrodynamic (w/o HI) interaction included. The isotropic LE4PD modes are indexed by internal mode number (see explanation in the text).}
   \label{timetable2}
   \begin{tabular}[c]{ c c c c c}
   &LE4PD & \multicolumn{2}{c}{LE4PD-XYZ} & tICA \\
   & &  w/ HI& w/o HI & \\
   Mode &  $t_2\left(\tau_{MSM}\right)$ , ns & $t_2\left(\tau_{MSM}\right)$, ns & $t_2\left(\tau_{MSM}\right)$ , ns & $t_2\left(\tau_{MSM}\right)$ , ns  \\ \midrule

   1    & 5.3(1.8)&  14.6(12.0) &  16.2(12.0) & 54.0 (5.0)\\ 
   2    & 3.3(1.6)& 14.4(10.0) &  16.6(12.0) & 12.6 (5.0) \\
   3    & 1.9(1.2)& 9.6(8.0) &  9.2(8.0) & 10.5 (5.0)\\
   4    & 4.7(1.6)& 7.2(6.0) &  9.5(8.0) & 9.1 (5.0)\\
   5    & 3.6(1.6)& 4.8(4.0) &  7.7(6.0) & 9.3(5.0)\\
   6    & 33.7(25.0)& 4.6(4.0) &  21.5(12.0) & 6.6(5.0)\\
   7    & 1.2(1.0)& 19.9(12.0) &  12.6(10.0) & 5.7(5.0)\\
   8    & 3.0(1.6)& 2.4(2.0) &  4.6(4.0) & 6.1(5.0)\\
   9    & 0.5(0.4)& 4.0(3.5) &  1.8(1.5) & 6.4(5.0)\\
   10   & 0.35(0.3)& 1.3(1.0) &  3.7(3.0) & 7.0(5.0)\\
   \end{tabular}
  \newline
\end{table}

Note that the isotropic LE4PD is an equation of motion in the lab reference system, while the LE4PD-XYZ starts from a body-centered trajectory where translation and rotation have been eliminated.\cite{Eckart1934,Sayvetz2004,Kneller2008,Chevrot2011} Thus, we report the internal motion of LE4PD after the first three rotational modes are discarded. In the tables, all the dashed entries denote free energy surfaces where the extreme projections of $\psi_2$ are never located in minima on the surface and are thus not suited for Markov state modeling in the manner desired here. 

Crossing the energy barriers may slow down differently different modes. Some internal modes may have larger energy barriers than the first mode. This is, in fact, the case for ubiquitin, as one can see from reading the tables. When $t_2$ is calculated using the second MSM eigenvector (Table \ref{timetable1}), the slowest mode for the isotropic LE4PD method, with hydrodynamic interaction included, is mode $6$, while the slowest mode for the anisotropic LE4PD-XYZ model without hydrodynamics is mode $7$. If the markovianity criteria are enforced (Table \ref{timetable2}), the slowest fluctuations are in mode $7$ for the LE4PD-XYZ with hydrodynamics and in mode $6$ for the LE4PD-XYZ without hydrodynamics.  

From the timescales listed in Table \ref{timetable1}, all the LE4PD methods give roughly the same timescales for the slowest motions of the system. The first tICA mode, however, displays dynamics that are five times slower than LE4PD. The first tIC corresponds to the contemporary motion in the three flexible binding regions of ubiquitin, as shown in Figure \ref{tp1square}, and predicts this motion occurs almost ten times slower than the roughly analogous motion predicted by the isotropic LE4PD mode 6 and LE4PD-XYZ mode 7 with hydrodynamics, respectively. However, when the MSM lag time is selected using the CK condition, which does not always coincide with the lag time selected by optimizing the projection of $\psi_2$ from the MSM,\cite{Beyerle2019} the gap between the predicted timescales of the slow LE4PD and tICA modes is reduced, as shown in Table \ref{timetable2}.

Overall, the time scales presented are similar in magnitude, with the tICA modes being generally slower than the LE4PD-XYZ. Please note that the plots to calculate $t_2$ are in all cases on a logarithmic scale and that small changes in the selected $\tau_{MSM}$ can give large differences in the value of $t_2$ (see Figures  \ref{tp1squareLE4PDXYZa}d and \ref{tp1square}d). Except for the first tICA mode, all the other tICA and LE4PD modes do not show an evident markovian nature of the dynamics, and one should take the exact values of $t_2$ with some reservations.

\subsection{Localization of mode-dependent fluctuations detected by tICA and LE4PD-XYZ}
\label{LML}
To compare the dynamics predicted by the slow tICs and LE4PD modes, we calculate the mode-dependent fluctuation profiles as a function of amino acid sequence along the backbone of the protein for the first ten modes of the LE4PD-XYZ theory without hydrodynamic interaction and of the tICA approach. These correspond to the last two columns on the right of Tables \ref{timetable1} and \ref{timetable2}. Local fluctuations are well represented by the local mode lengthscale (LML).\cite{Copperman2016, Beyerle2019} In the anisotropic formalism of LE4PD-XYZ, where $\langle\Delta\vec{R}_i \cdot \Delta\vec{R}_i\rangle = \langle\Delta x_i^2\rangle + \langle\Delta y_i^2\rangle + \langle\Delta z_i^2\rangle$,  the eigenvectors are partitioned into their $x-$, $y-$, and $z-$components, thus isolating the $x-$, $y-$, and $z-$projections of $\text{LML}_{ia}^{2}$ as:
\begin{align}
    \text{LML}_{ia,x}^{2} = \left(Q_{ia}^{x}\right)^2\mu_{a,\text{LE4PD-XYZ}}^{-1}  \label{lmlx}\\
    \text{LML}_{ia,y}^{2} = \left(Q_{ia}^{y}\right)^2\mu_{a,\text{LE4PD-XYZ}}^{-1}  \label{lmly}\\
    \text{LML}_{ia,z}^{2} = \left(Q_{ia}^{z}\right)^2\mu_{a,\text{LE4PD-XYZ}}^{-1} , \label{lmlz}
\end{align}
where $\mu_{a,\text{LE4PD-XYZ}}$ are the eigenvalues of $\mathbf{A}^{\prime}$ \cite{Beyerle2021a}. 


\begin{figure}[htb] 
\center
\includegraphics[width=.8\columnwidth]{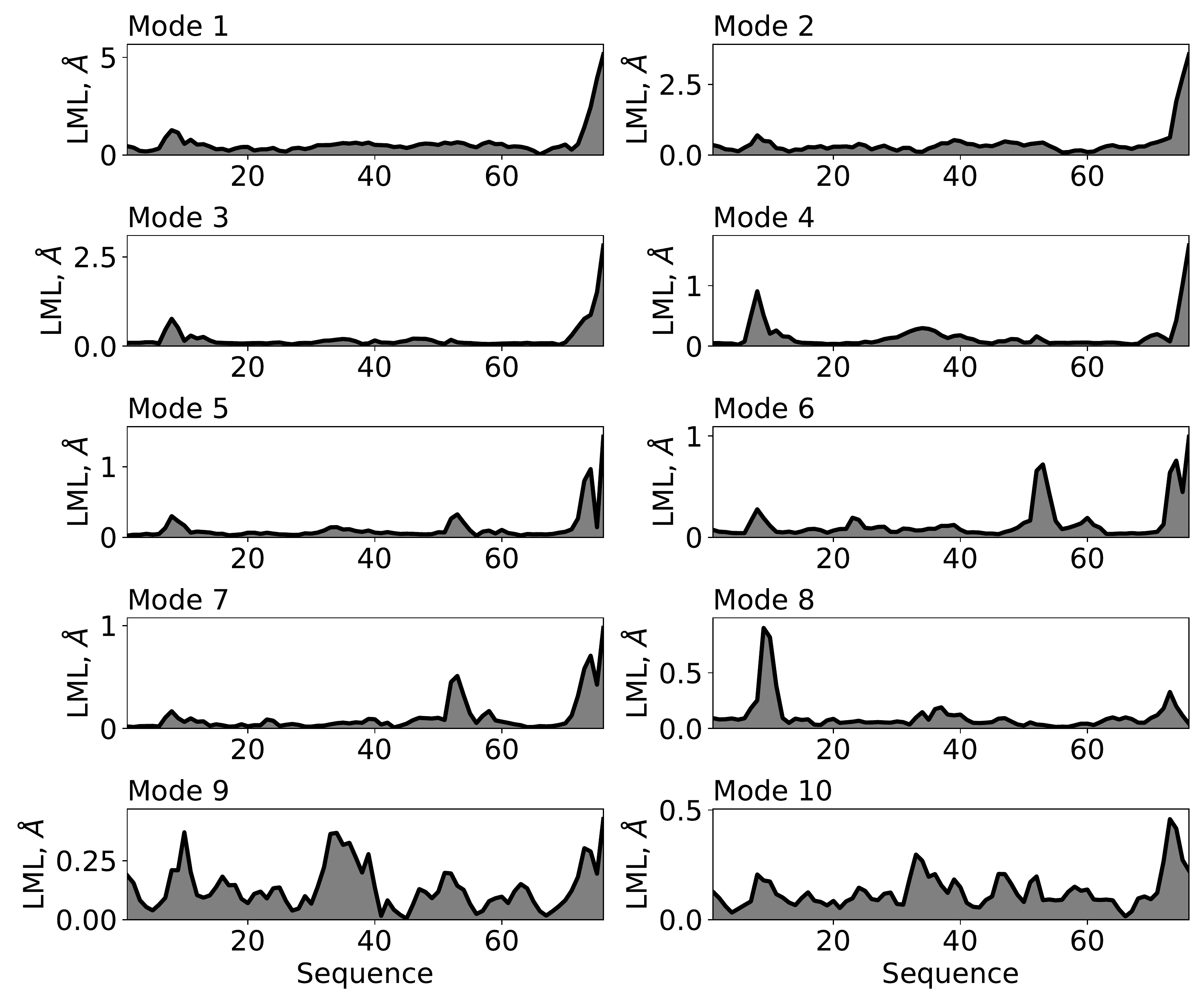}
\caption{Mode-dependent fluctuations or local mode lengthscale (LML) for the ten slowest modes captured from the anisotropic LE4PD-XYZ analysis, without hydrodynamics, of the 1-$\mu$s simulation of ubiquitin. Each panel shows the fluctuations' amplitude as a function of the protein's primary sequence. For example, the first LE4PD-XYZ mode shows fluctuations mostly in the C-terminal tail. One finds the slowest fluctuations corresponding to the 50 s loop  in mode 6.}
\label{msf_1ubq_aniso}
\end{figure}
\begin{figure}[htb] 
\center
\includegraphics[width=.8\columnwidth]{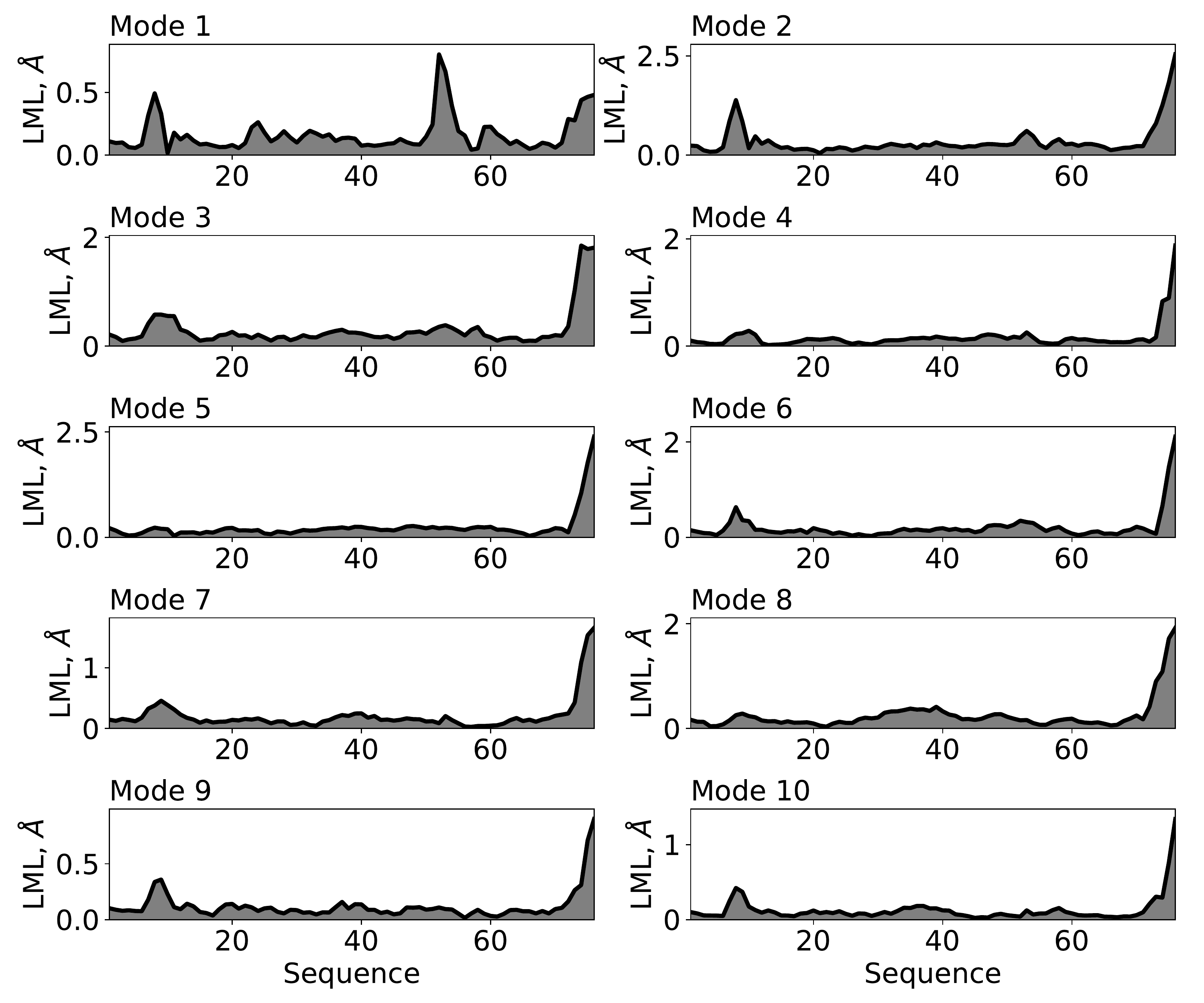}
\caption{Mode-dependent fluctuations or local mode lengthscale (LML) for the ten slowest modes captured from the tICA of the 1-$\mu$s simulation of ubiquitin, with a tICA lag time of 2 ns. Each panel shows the fluctuations' amplitude as a function of the protein's primary sequence. For example, the first tICA mode shows fluctuations in the Lys11 loop, the 50 s loop, and the C-terminal tail.}
\label{msf_1ubq_tica}
\end{figure}

\begin{figure*}[htb] 
\center
\includegraphics[width=1.2\columnwidth]{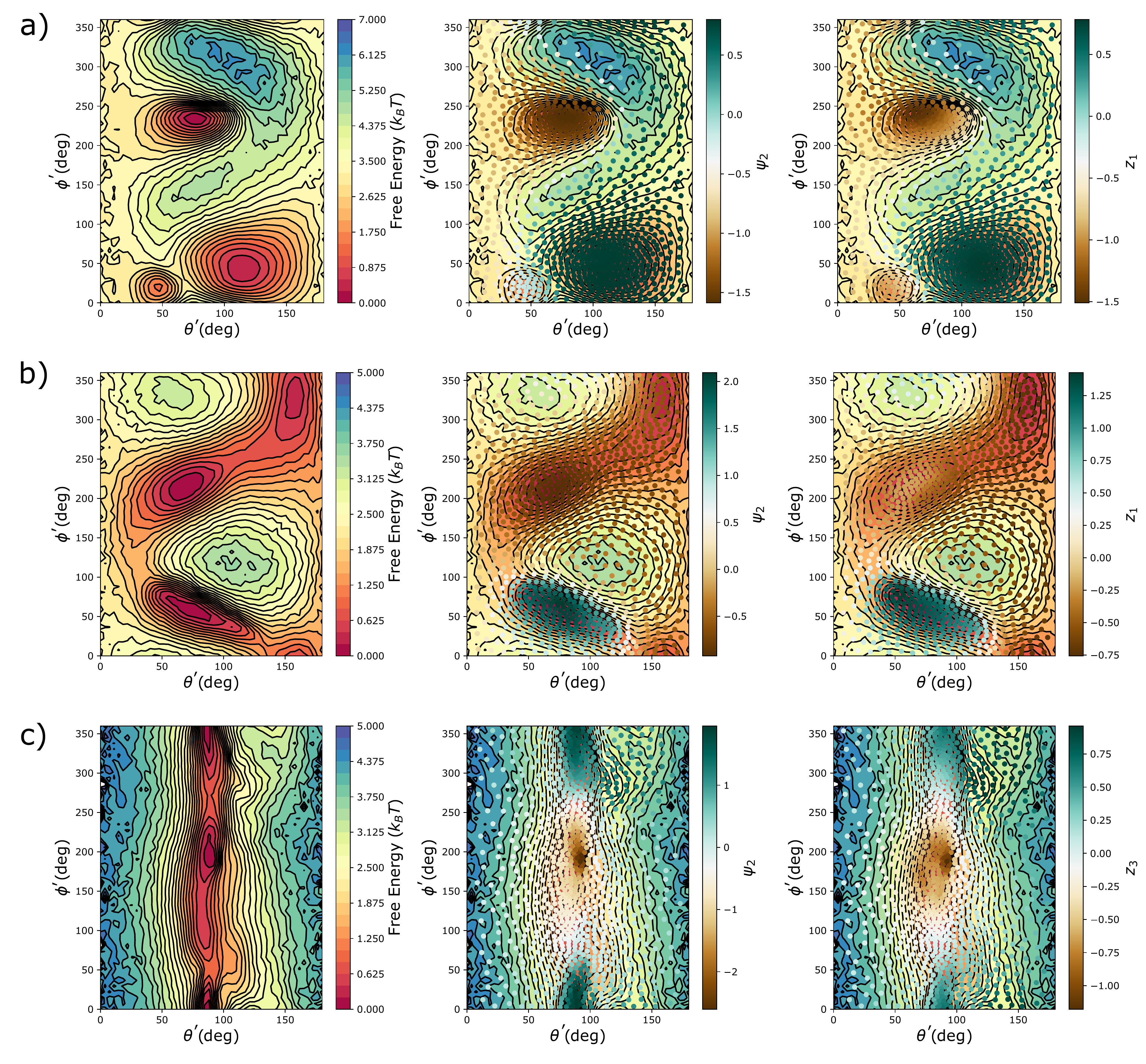}
\caption{a) From left to right: Free-energy surface of isotropic LE4PD internal mode 6 with hydrodynamics from the one-microsecond ubiquitin simulation; projection of $\psi_2$ from the MSM of the trajectory on the $\left(\theta, \phi\right)$ surface; and projection of the first tIC $z_1(t)$ onto the $\left(\theta, \phi\right)$ surface. b) Same as a), but the displayed free-energy surface is for the LE4PD-XYZ mode 7 without hydrodynamics. c) Same as a) and b), except for LE4PD-XYZ mode 5 without hydrodynamics, with the projection in the right-most panel being the third tIC $z_3(t)$ onto the surface.}
\label{z1psi2}
\end{figure*}

For the tICA, fluctuations are derived from the definition of the modes, $z_a(t)=\sum_i\Omega^T_{ai}\Delta R_i(t)$, and the Moore-Penrose generalized inverse\cite{Horn1985} of $\Omega^T, \Omega^{-1T}$ as
\[\Delta R_i(t)=\sum_a\Omega^{-1T}_{ia}z_a(t).\]
The mean-square fluctuations of residue $i$, given by $\langle \Delta x_i(t) \Delta x_i(t)\rangle + \langle \Delta y_i(t) \Delta y_i(t)\rangle + \langle \Delta z_i(t) \Delta z_i(t)\rangle$, can be written in terms of the tICs as
\begin{align}
    &\langle \Delta x_i(t) \Delta x_i(t)\rangle + \langle \Delta y_i(t) \Delta y_i(t)\rangle + \langle \Delta z_i(t) \Delta z_i(t)\rangle\notag\\
    &= \sum_a\sum_b\left(\Omega^{-1T}_{ia,x}\Omega^{-1T}_{ib,x} + \Omega^{-1T}_{ia,y}\Omega^{-1T}_{ib,y} + \Omega^{-1T}_{ia,z}\Omega^{-1T}_{ib,z}\right)\langle z_a(t)z_b(t)  \rangle \notag\\
    &=\sum_a\left(\Omega^{-1T}_{ia,x}\right)^2 + \left(\Omega^{-1T}_{ia,y}\right)^2 + \left(\Omega^{-1T}_{ia,z}\right)^2, \label{msf_tics}
\end{align}
where the definition $\langle z_a(t)z_b(t)\rangle=\delta_{ab}$ is used to obtain Eq. \ref{msf_tics}. The tICA LMLs are reported in Figure \ref{msf_1ubq_tica}

Thus, Figures \ref{msf_1ubq_aniso}, \ref{msf_1ubq_tica} show the mode-dependent fluctuations calculated from the one-microsecond ubiquitin simulation using the anisotropic LE4PD without hydrodynamics, and the tICA, respectively, for the first ten processes of each method.  
For tICA, the slowest tIC describes concerted fluctuations in the tail, Lys11, and $50\text{ s}$ loops of the protein. For the LE4PD-XYZ approach, most of the low-index modes describe fluctuations in the C-terminal tail of the protein. One needs to look at the fourth mode to find slow fluctuations in the Lys11 loop and to the  $6$th and $7$th modes to find slow fluctuations in the $50\text{ s}$ loop. Neither of the LE4PD approaches gives a single mode describing simultaneous motion in the three important regions of the protein. However, we observe that there is good correspondence between the slowest tIC and the anisotropic LE4PD-XYZ mode 7, which is the slowest mode for the anisotropic LE4PD when hydrodynamic effects are neglected.

Because the three slowest processes all appear in the first tICA mode, while they are partitioned in six different LE4PD-XYZ modes, we observe
that the tICA procedure can group the slowest, important, dynamics in a smaller number of modes than the LE4PD, which, instead, partitions the protein's slow dynamics into several leading modes with different time and length scales. When the goal is identifying the slowest fluctuations in one mode, tICA appears to be more efficient than the LE4PD in isolating the slow fluctuations if the lag time, $\tau_{\text{tICA}}$, is selected appropriately. However, suppose the ultimate goal is the accurate analysis of the protein's slow dynamics. In that case, the LE4PD approach has a more desirable outcome because it maintains the information on the fast dynamics and provides better resolution at short times. As shown in Section \ref{TCFS}, the LE4PD can predict the dynamics as measured by time correlation functions with higher accuracy than the tICA modes.

The quantitative comparisons of the mode dependent fluctuations calculated using the isotropic LE4PD with hydrodynamics, the anisotropic LE4PD-XYZ without hydrodynamics, and the tICA are presented in the Supplementary Material, Section S4.

\subsection{Biological interpretation of ubiquitin's fluctuations}
 Following the conformational selection model, a protein in absence of its binding partner samples all the energetically available states, and among those states are some that can bind to the substrates of interest.\cite{Monod1965,Boehr2009,Csermely,Kahler2020} Thus, the residue fluctuations, observed in the tICA and LE4PD analyses, may provide useful information on the time scales and length scales of relevant binding modes. By direct inspection of the LMLs calculated with the LE4PD-XYZ and with tICA, we observe that there are three important regions of slow fluctuation dynamics in ubiquitin.   Fluctuations in the C-terminal tail and in the flexible loop containing Lys11, which are visible in most of the slow modes, are implicated in the covalent association to other proteins, and in covalent binding to lysine in polyubiquitination.\citep{Komander2009, Komander2012,Lv2018} Given that ubiquitin binds covalently to numerous proteins of different sizes and flexibilities, it is perhaps not surprising that these fluctuations cover a wide range of length scales and time scales. The third region of important fluctuations involves the $50\text{ s}$ loop, which is known to participate to the hydrophobic binding to the A20 zinc-finger motif of Ras guanine exchange factor Rabex-5, where the residue Y25 of Rabex-5 forms a hydrogen bond with residue E51 of ubiquitin, which has the largest fluctuations in the $50\text{ s}$ loop.\cite{Penengo2006,Lee2006}
 
Dynamics in the  $50\text{ s}$ loop region of ubiquitin is correlated with the breaking of a hydrogen bond between G53 and E24, which helps maintain the protein's folded structure.\cite{Sidhu2011} Thus, breaking this hydrogen bond serves as a 'gatekeeper' to selecting different conformations. For example, in \cite{Sidhu2011}, the authors found that only 29\% of 155 x-Ray structures examined showed ubiquitin with the same hydrogen bonding pattern between G53 and E24 found in the folded structure from \cite{Vijay-kumar1987}, which is also the starting structure in this study. Furthermore, in \cite{Smith2016} the authors demonstrate that the interchange between hydrogen bonding patterns in the $50\text{ s}$  loop modulate large-scale conformational changes (contraction and expansion) along the entire primary sequence of ubiquitin.  This affects the protein's ability to bind to a set of ubiquitinases known as ubiquitin-specific proteases and marks the $50\text{ s}$  loop as a potential site of allosteric inhibition. Thus, experimental evidence indicates that local conformational changes in the $50\text{ s}$  loop are required for global conformational transitions in ubiquitin.
 

\section{Similarities between the \MakeLowercase{t}ICA  and the LE4PD-XYZ free energy surfaces }
\label{wwohydro}
In this section we quantify the agreement between the energy maps for the slowest modes identified by tICA and LE4PD-XYZ.
Figure \ref{z1psi2} displays in each row the comparison between the LE4PD slowest modes and the tICA slowest mode for the two LE4PD models we study, namely the isotropic LE4PD and the anisotropic LE4PD-XYZ theory. 

In the first column, Figure \ref{z1psi2} shows the FES of the LE4PD projected trajectory, which displays energy minima for the most populated conformations of the protein.  For this FES, the second column of Figure \ref{z1psi2} presents the second eigenvector obtained from the Markov State Model (MSM) analysis of the FES. The superposition of the second MSM eigenvector to the LE4PD energy map indicates which transition represents the slowest fluctuation for the given LE4PD mode.\cite{Reichl1998, Noe2007, Prinz2011} The third column in Figure \ref{z1psi2} shows the comparison between a tICA mode and the LE4PD mode. The superposition is accomplished by projecting the first tIC onto the LE4PD free energy map and testing if the most extreme tICA conformations are the ones that correspond to the minima in the LE4PD FES. To perform this comparison, we assign each conformation in the tICA mode trajectory to the closest MSM microstate in the LE4PD-mode FES surface, using the root mean square distance from each MSM microstate as the assignment metric. Then the tICA mode trajectory populates the FES, giving a projection of the tICA mode that is completely analogous, in both meaning and interpretation, to the projection of an eigenvector $\psi_i$ from the MSM onto the LE4PD FES (see the second column of Figure \ref{z1psi2}). The approach of projecting a tICA mode onto a free-energy surface has been previously applied by Sultan and Pande\cite{Sultan2017} to verify the interpretation for the slowest tIC from a simulation of alanine dipeptide.

When projecting the tICs, $z(t)$, onto the $\left(\theta_a,\phi_a\right)$ surfaces, the average of $z(t)$ within each MSM LE4PD microstate $i$, $S_i$, is calculated as
\[\langle z(t)\rangle_i=\frac{1}{M_i}\sum_{k=1}^{M_i}z(k),\qquad \forall \left(\theta_a(k),\phi_a(k)\right)\in S_i,\]
with $M_i$ the total number of frames the $z(t)$ trajectory resides in the $S_i$ LE4PD microstate over the course of the simulation. This local average of $z(t)$ within each of the discrete states is what is reported in Figure \ref{z1psi2}.

The slowest tIC is the optimal linear approximation to the full-space Markov propagator of the system.\cite{Perez-Hernandez2013} The $\psi_2$ from the MSM on the slowest LE4PD modes are also estimators of the slowest processes of the system; a high similarity between the projected spectra of the slow tICs and $\psi_2$ indicate high similarity between the predicted dynamics from the two models. That is, if the slow dynamics predicted in each approach are consistent with each other, then the spectra of both the slow tICs and $\psi_2$ should predict probability flow between the deep minima on the  $\left(\theta_a,\phi_a\right)$ surfaces of the slowest LE4PD modes. The $\psi_2$ are already parameterized to do so, \cite{Beyerle2019} but the slow tICs are, in principle,  ignorant of the LE4PD $\left(\theta_a,\phi_a\right)$ surface. We use this technique to confirm that the slow LE4PD modes can extract the slow dynamics compatible with tICA modes.\cite{Sittel2018} 

The three rows in Figure  \ref{z1psi2}  represents, from top to bottom, the following calculations. In the first row, the  \emph{isotropic} LE4PD (mode 6) with hydrodynamics agrees with the frist tICA mode. In the second row, the \emph{anisotropic} LE4PD-XYZ (mode 7) without hydrodynamics compares well with the first tICA mode. The third row shows a comparison between the third tICA mode (mode 3) and the fifth mode of the \emph{anisotropic} LE4PD-XYZ without hydrodynamics. It is clear from these results that the slow dynamics detected by tICA and \emph{anisotropic}  LE4PD-XYZ are similar, even if the slow dynamics can be distributed differently in the LE4PD, LE4PD-XYZ, and tICA modes (see Figures \ref{msf_1ubq_aniso} and \ref{msf_1ubq_tica}).

Note that the technique used here of projecting the tICs onto the $\left(\theta_a,\phi_a\right)$ surfaces of the LE4PD modes is analogous to the technique used in \cite{Olsson2017, Keller2012, Noe2011, Chodera2010} to model experimental observables using Markov state models.  Like an experimental observable, the separation of two minima of the $\left(\theta_a,\phi_a\right)$ surfaces into `high $z$' and `low $z$' states indicates that transitions on the $\left(\theta_a,\phi_a\right)$ surface correspond to transitions between a high $z$ state and a low $z$ state, similar to how fluorescence experiments on a protein search for transitions between a high fluorescence state, indicating the protein is sampling conformations where the fluorophores are far apart, and a low fluorescence state, where the protein is sampling conformations where the fluorophors are close together.\cite{Comstock2015,Mazal2019}

In conclusion, Figure \ref{z1psi2} demonstrates that both LE4PD approaches are able to capture the same slow motion as the tICA. The correlation between the time series of $z_1$ and $\psi_2$ from the MSM of the slowest isotropic LE4PD mode is high ($\rho=0.92$), indicating that both $z_1$ and $\psi_2$ are predictive of the slow dynamics in ubiquitin. The correlation coefficient between the time series of $z_1$ and $\psi_2$ from the MSM of the slowest anisotropic LE4PD-XYZ mode is $\rho=0.73$, which is still acceptable. 
The correlation coefficient between the  time series of $z_3$ and the $\psi_2$ for the fifth LE4PD-XYZ mode is $\rho=0.54$.

\section{Testing the \MakeLowercase{t}ICA and LE4PD predictions of time correlation functions against simulations}
\label{TCFS}
The analysis of the amplitude, location, and time scale of the slow fluctuations for ubiquitin with the three methods (tICA, LE4PD, and LE4PD-XYZ) show that they correctly identify the regions in the protein where slow fluctuations occur. However, the slow fluctuations are partitioned in different modes for the two methods. 
To gain further details on the capabilities of the two approaches, we introduce, as the ultimate test of the tICA's and LE4PD's ability to predict with accuracy slow time dynamics, the comparison of their time correlation functions (tcfs) to the tcfs directly calculated from the simulation trajectory.

\begin{figure}[htb] 
\center
\includegraphics[width=.9\columnwidth]{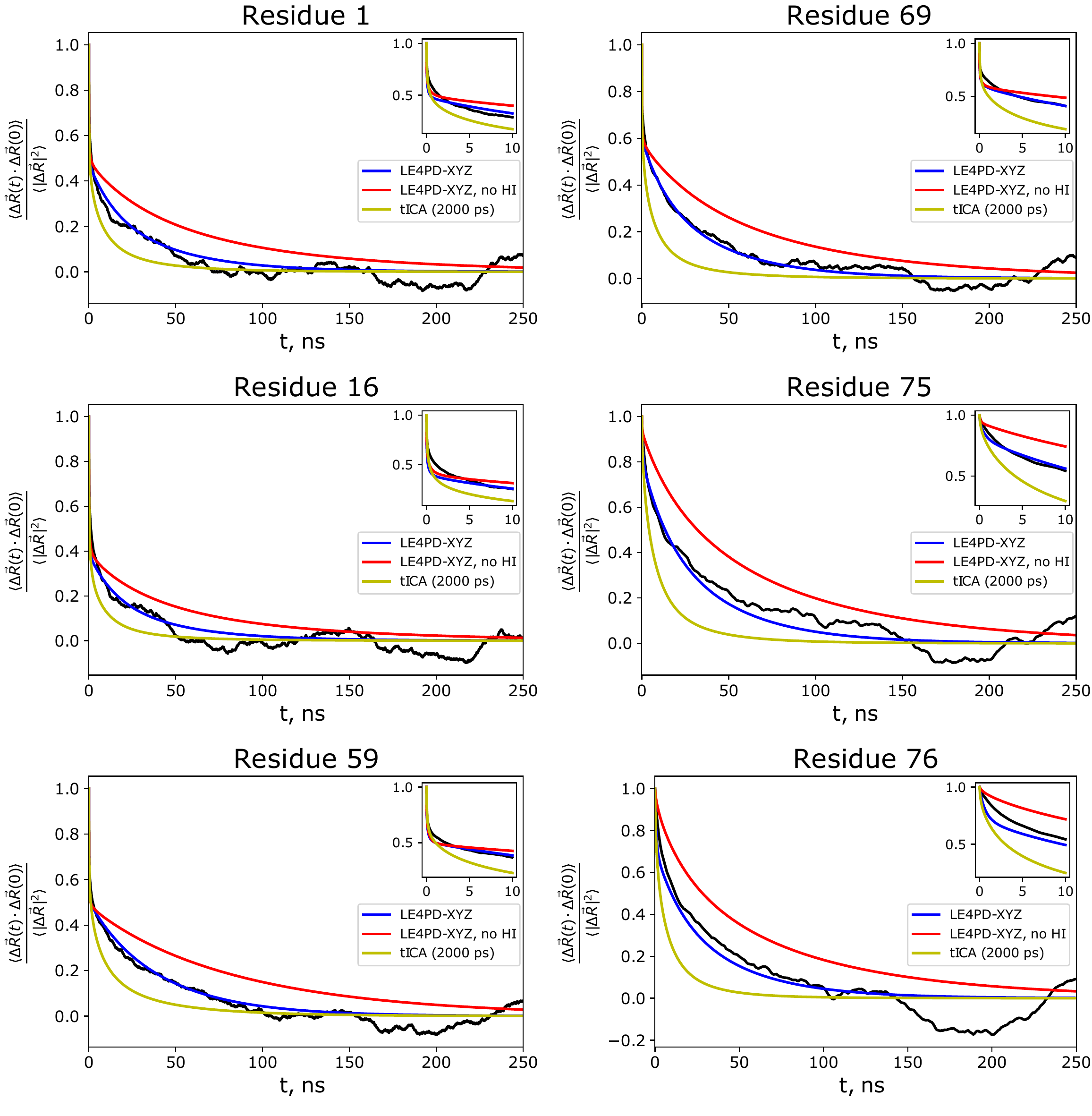}
\caption{Comparison of the time correlation functions (tcfs) for a sampling of residues along the primary sequence of ubiquitin. The black curves in each subplot show the tcf calculated from the simulation trajectory; the blue curves show the tcfs predicted from the LE4PD-XYZ theory with HI, the red curves the tcfs predicted from the LE4PD-XYZ without HI, and the yellow curves show the tcfs predicted from the tICA with a lag time of 2000 ps.}
\label{comp_tcfs}
\end{figure}

The normalized autocorrelation function for the fluctuations of each residue is defined as $C(t)=\frac{\langle \Delta\vec{R}(t) \cdot \Delta\vec{R}(0)\rangle}{\langle \Delta \vec{R}(0) \cdot \Delta\vec{R}(0)\rangle}$. For the LE4PD approaches, the autocorrelation function is calculated by including for each mode the slowing down of the dynamics due to the presence of an energy barrier in the FES.\cite{Beyerle2019,Beyerle2021a} 
Recently, we have shown that neglecting the hydrodynamic interaction modifies the LE4PD-XYZ curves, leading to a (moderately) worse agreement with the simulation data.\cite{Beyerle2021a} Figure \ref{comp_tcfs}  shows the fluctuation decay of the tcfs for residues sampled along the primary sequence of ubiquitin.  The figure compares the LE4PD-XYZ results with hydrodynamic included to the tcfs from the simulations: the agreement is remarkable. It also shows the tcfs for the LE4PD-XYZ without hydrodynamic interactions, which are less in agreement, at least for the residues presented in the figure. 

At a given lag time, $C(t)$ can be written in terms of the tICA eigenspectra by inverting the relationship $z_a(t)=\sum_i\Omega^{T}_{ai}\Delta R_i(t)$ as $\Delta R_i(t)=\sum_a\Omega^{-1T}_{ia}z_a(t)$, and using the (near) independence of the tICs $\langle z_a(t)z_b(0)\rangle\approx\langle z_a(0)z_b(0)\rangle \text{exp}\left[{-t/\tau_a}\right]=\delta_{ab}\text{exp}\left[{-t/\tau_a}\right]$ as
\begin{align}
    C(t)  & = \frac{\langle \Delta\vec{R}(t) \cdot \Delta\vec{R}(0)\rangle}{\langle \Delta \vec{R}(0) \cdot \Delta\vec{R}(0)\rangle} \label{tcfs} \\
      & =  \frac{\sum_{a}\left[\left(\Omega^{-1T}_{ia,x}\right)^2 + \left(\Omega^{-1T}_{ia,y}\right)^2 + \left(\Omega^{-1T}_{ia,z}\right)^2\right]e^{-t/\tau_a}}{\sum_{a}\left[\left(\Omega^{-1T}_{ia,x}\right)^2 + \left(\Omega^{-1T}_{ia,y}\right)^2 + \left(\Omega^{-1T}_{ia,z}\right)^2\right]}. \nonumber
\end{align}
The decay timescales for each tICA mode, $\tau_a$, are calculated empirically by the integration of the autocorrelation function
$\langle z_a(t)z_b(0)\rangle/\langle z_a(0)z_b(0)\rangle=e^{-t/\tau_a}$ obtained from the simulations,\cite{Naritomi2011} and assuming that each mode is represented by a single exponential decay.  This procedure should account for the barriers present along each tICA coordinate in, at least, a coarse manner.\cite{Beyerle2021a,Chan2020} This time, $\tau_a$, is in general different from the inverse of the eigenvalues $\lambda_{IC}$ (Eq. \ref{diagonalization}) because that time does not include the mode-dependent energy barrier. If one adopted the inverse of the eigenvalues $\lambda_{IC}$ as the timescale of decay, the tcfs calculated from tICA would display an even faster and more unphysical decay than the one observed when including mode-dependent energy barrier for tICA (see Fig. \ref{comp_tcfs}).  Once a lag time is selected, we build the matrix $\mathbf{C}^r(\tau_{\text{tICA}})$ (Eq. \ref{diagonalization}) and, by diagonalization, we derive the eigenvectors and eigenvalues that enter Eq. \ref{tcfs}. 

The time correlation functions calculated from the tICs (Eq. \ref{tcfs}) are directly compared to the one from the simulation trajectory in Figure \ref{comp_tcfs}.  For each residue shown, and for most residues across the primary sequence of ubiquitin, the tcfs predicted from the LE4PD-XYZ with hydrodynamics are in better agreement with the simulated tcfs than those predicted from the tICA or the LE4PD-XYZ without hydrodynamics. 

\begin{figure}[htp]
    \centering
    \includegraphics[width=.8\columnwidth]{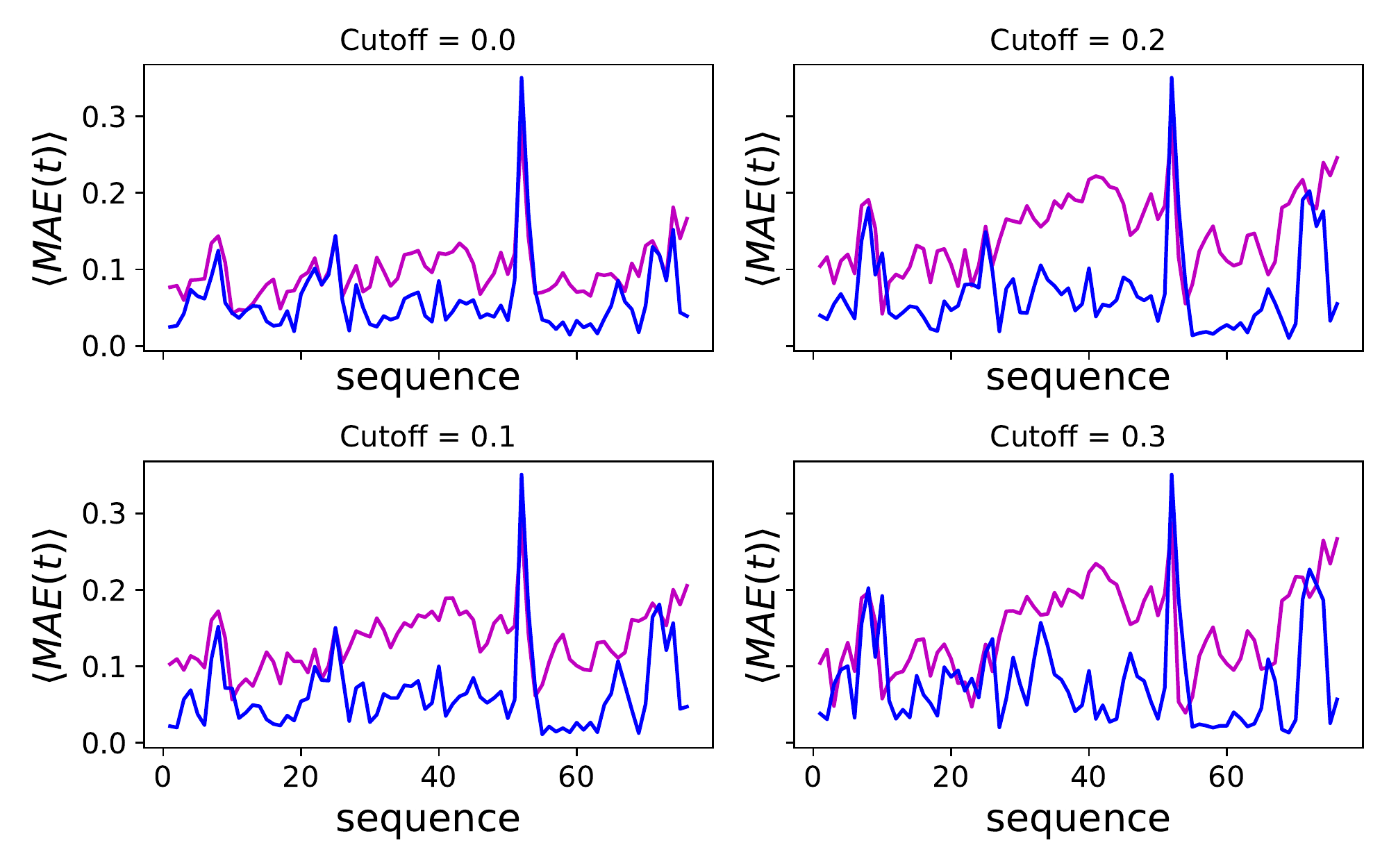}
    \caption{Mean absolute error ($\langle$MAE(t)$\rangle$) between the simulated autocorrelation function (acf), C(t), and the predicted autocorrelation functions from the tICA (purple), and from the LE4PD-XYZ with hydrodynamics (blue). The error is reported for all amino acids as a function of the primary sequence of the protein ubiquitin and at increasing acf cut-off. The error is found by calculating the right-hand side of Eq. \ref{rmse} for all values of $t$ until C(t) reaches the given cut-off value for the first time $t^{\prime}$. On average, across all residues and for all cut-off values, the anisotropic LE4PD with hydrodynamics out-performs the tICA predictions.
    }
    \label{error}
\end{figure}

The accuracy of the two approaches is quantified in Figure \ref{error}, which shows the mean absolute error ($\langle$MAE(t)$\rangle$) between the simulated and predicted C(t) for all the residues in ubiquitin. This metric of quantifying the distance between the `true' (simulated, C(t)) and `estimated' (predicted, $\widehat{\text{C}}$(t)) is defined as
\begin{equation}
\langle \text{MAE(t)}\rangle = \frac{1}{M_0}\sum\limits_{t=1}\limits^{M_0}\left[\vert \text{C}(t) - \widehat{\text{C}}\text{(t)}\vert\right],
    \label{rmse}
\end{equation}
which is the average distance between the two autocorrelation functions (acf). $M_0$ is the number of frames before an acf cutoff time, which is the time at which the acf first decays to a specified value. For example, using a cutoff of C(t) = 0.0 calculates Eq. \ref{rmse} over all points of the acf until the acf attains a value of 0.0 for the first time.

Figure \ref{error} shows that for most aminoacids the error metric is lower for the anisotropic LE4PD with hydrodynamics compared to the tICA estimator of C(t). The figure shows data for four different cutoff times, indicating that the result is robust and is not affected by the choice of the cutoff time.  Thus, using the anisotropic LE4PD with hydrodynamics gives a better prediction, on average, of the C(t) autocorrelation function compared to tICA (lagtime 2.0 ns) across all the residues in ubiquitin.

\begin{figure}[htp] 
\includegraphics[width=.8\columnwidth]{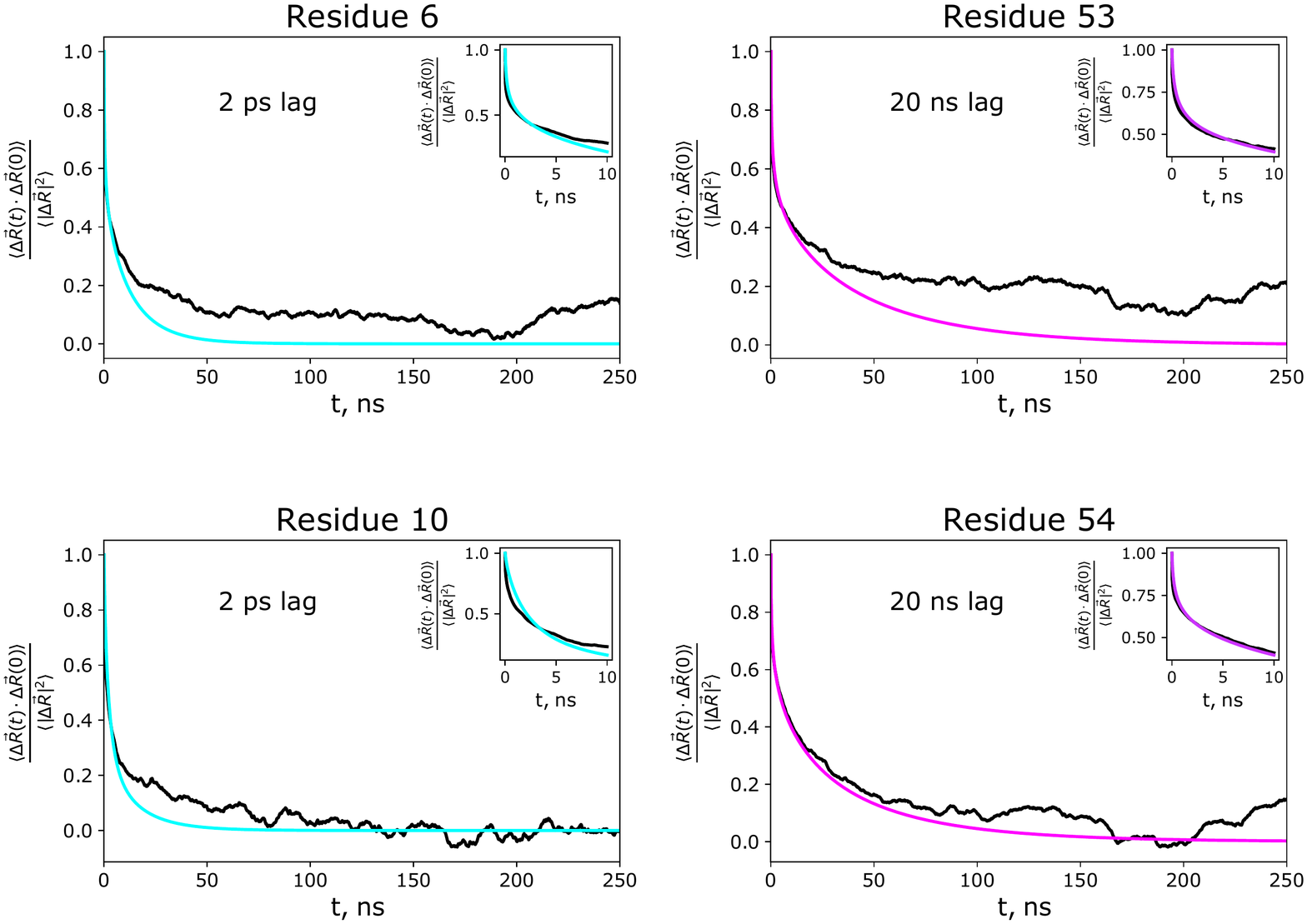}
\caption{Left column: two residues in the Lys11 loop of ubiquitin whose tcfs from the simulation (black) are well approximated at timescales less than 10 ns by the tICs predicted using a lag time of 2 ps (cyan). Right column: two residues in the $50\text{ s}$ loop of ubiquitin whose tcfs from the simulation (black) are well approximated at timescales less than 10 ns by the tICs predicted using a lag time of 20 ns (magenta).}
\label{tica_tcf_alt}
\end{figure}

One may assume that the disagreement of tICA with the simulated tcfs observed in Figure \ref{comp_tcfs} is related to the choice of the lag time and that choosing either a longer or shorter tICA lag time may give a better agreement in the tcf of specific bonds. This is, in fact, the case. Figure \ref{tica_tcf_alt} shows how using a shorter lag time ($2$ ps) yields tcfs in good agreement with residues' tcfs in the highly flexible Lys11 loop, especially at timescales less than $10$ ns. Similarly, using a longer lag time ($20$ ns) gives tICs that agree well with the simulated tcfs of several residues in the $50$s loop, where the slowest fluctuations of the protein occur. This analysis supports the heterogeneity of ubiquitin dynamics, and the concept that there may be a different optimal lag time for each different tICA mode,  since one can locally optimize the residues' relaxation in different regions by varying the tICA lag time.

\section{The optimum \MakeLowercase{t}ICA lag time corresponds to the time that samples the highest barrier in the free energy surface}
\label{lag time}

\begin{figure*}[htb] 
\center
\includegraphics[width=1.8\columnwidth]{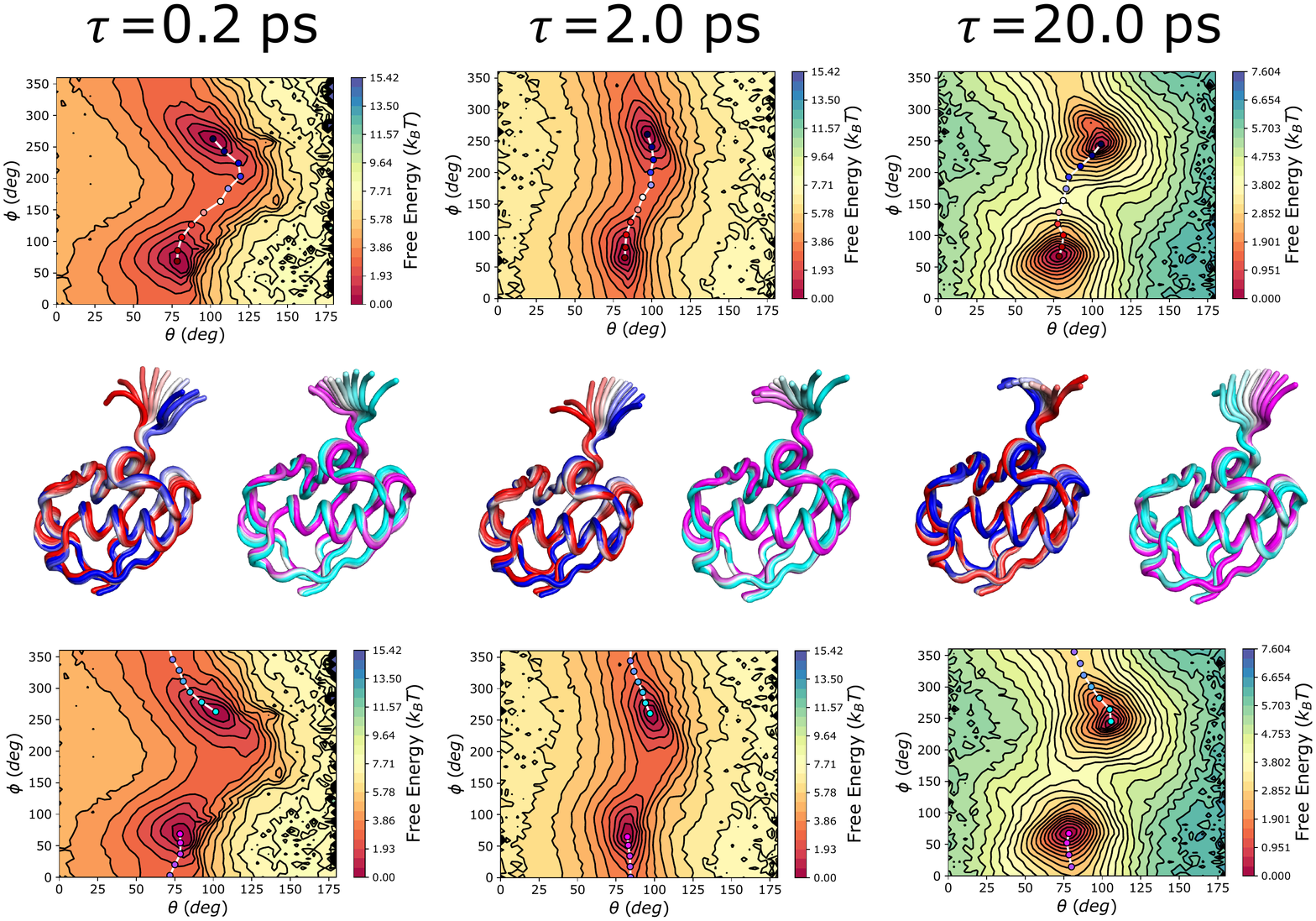}
\caption{Effect of changing the tICA lag time on the first tICA mode free energy surface (FES) and the associated fluctuations. Note that each FES has two possible pathways to transition between the two energy minima, depicted in the panels above and below the protein fluctuations pictures. In the protein cartoon, the configurations on the left (blue-white-red) represent the path in the top FES (blue-white-red path). In contrast, the configurations on the right (pink-yellow-light blue) represent the path in the bottom FES (pink-yellow-light blue path). As one increases the lag time, the FES  detects different internal energy barriers.}
\label{due}
\end{figure*}

\begin{figure*}[htb] 
\center
\includegraphics[width=1.8\columnwidth]{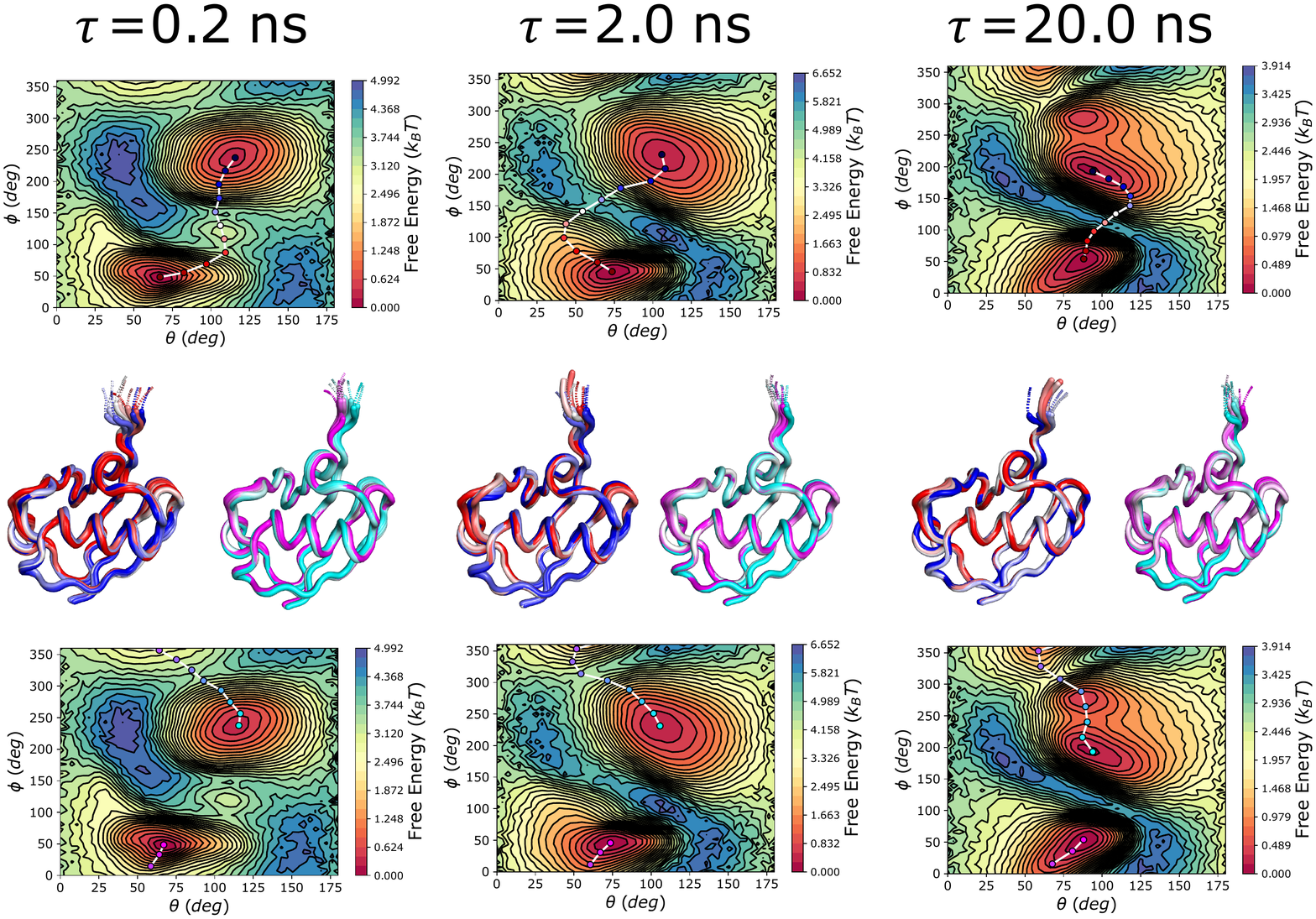}
\caption{Effect of changing the tICA lag time on the first tICA mode free energy surface (FES) and the associated fluctuations. Note that each FES has two possible pathways to transition between the two energy minima, depicted in the panels above and below the protein fluctuations pictures. In the protein cartoon, the configurations on the left (blue-white-red) represent the path in the top FES (blue-white-red path). In contrast, the configurations on the right (pink-yellow-light blue) represent the path in the bottom FES (pink-yellow-light blue path). As one increases the lag time, the FES  detects different internal energy barriers. While the system crosses the barrier, it samples fluctuations in the tail and in the important loops, but as the lag time increases, the predicted motion moves from the C-terminal tail and Lys11 loop into the $50\text{ s}$ loop.}
\label{lag timeFES}
\end{figure*}


The tICA approach is general and applies to any time-dependent set of coordinates and any lag time, $\tau_{\text{tICA}}$. After selecting the input coordinates to the tICA, which in this study are the coordinates of the fluctuations away from the average structure calculated over the MD trajectory, $\Delta \mathbf{R}$, there remains a single adjustable parameter: the observation lag time or $\tau_{\text{tICA}}$. This time parameter is used to construct the time-lagged covariance matrix (see Eq. \ref{gee}).  The tICA modes illustrate the dynamics taking place over a timescale longer than $\tau_{\text{tICA}}$, while dynamical phenomena that are faster  are averaged out and cannot be detected. Thus, only selecting the proper lag time can lead to the correct sampling of the dynamical phenomena that one desires to study.\cite{Perez-Hernandez2013, Schwantes2013,Naritomi2011} 

The selection of  $\tau_{\text{tICA}}$ is usually accomplished by performing the MSM analysis of multi-dimensional free-energy maps at different lag times. This step is followed by testing the results \textit{a posteriori} to verify which lag time leads to the longest possible timescale and to the most efficient separation of those timescales. Here, we propose a procedure to select \textit{a priori} the optimal tICA lag time. However, note that the traditional \textit{a posteriori} verification of the optimal tICA lag time agrees with the $2$ ns value used in our calculations (see Section S5 in the Supplementary Material).

We start here from the one-dimensional mode-dependent free energy surface constructed as described in Section \ref{FESTICAa} and select the optimum $\tau_{\text{tICA}}$ lag time as the one that samples the highest energy barrier in the FES. To illustrate how the choice of $\tau_{\text{tICA}}$ affects the tICA modes, Figures \ref{due} and  \ref{lag timeFES} show how an increase of the lag time modifies the dynamics that the first tICA mode samples. The energy landscape displays two deep minima and two possible paths that connect the two minima at all the lag times studied. 
Both figure \ref{due} and figure \ref{lag timeFES} display the two pathways in the two FES at the top and the bottom of each panel, respectively. The protein configurations that populate the two pathways are shown in the middle of the panel: the configurations on the left (blue-white-red) represent the path in the top FES (blue-white-red path). In contrast, the configurations on the right (pink-yellow-light blue) represent the path in the bottom FES (pink-yellow-light blue path). While at all the time lags studied, the dynamics of the protein involves mostly fluctuations in the C-terminal tail, at increasing lag time, the fluctuations in the tails become less pronounced, and new fluctuations start to appear in the Lys11 loop and in the $50\text{ s}$ loop.

\begin{figure}[htp] 
\includegraphics[width=0.8\columnwidth]{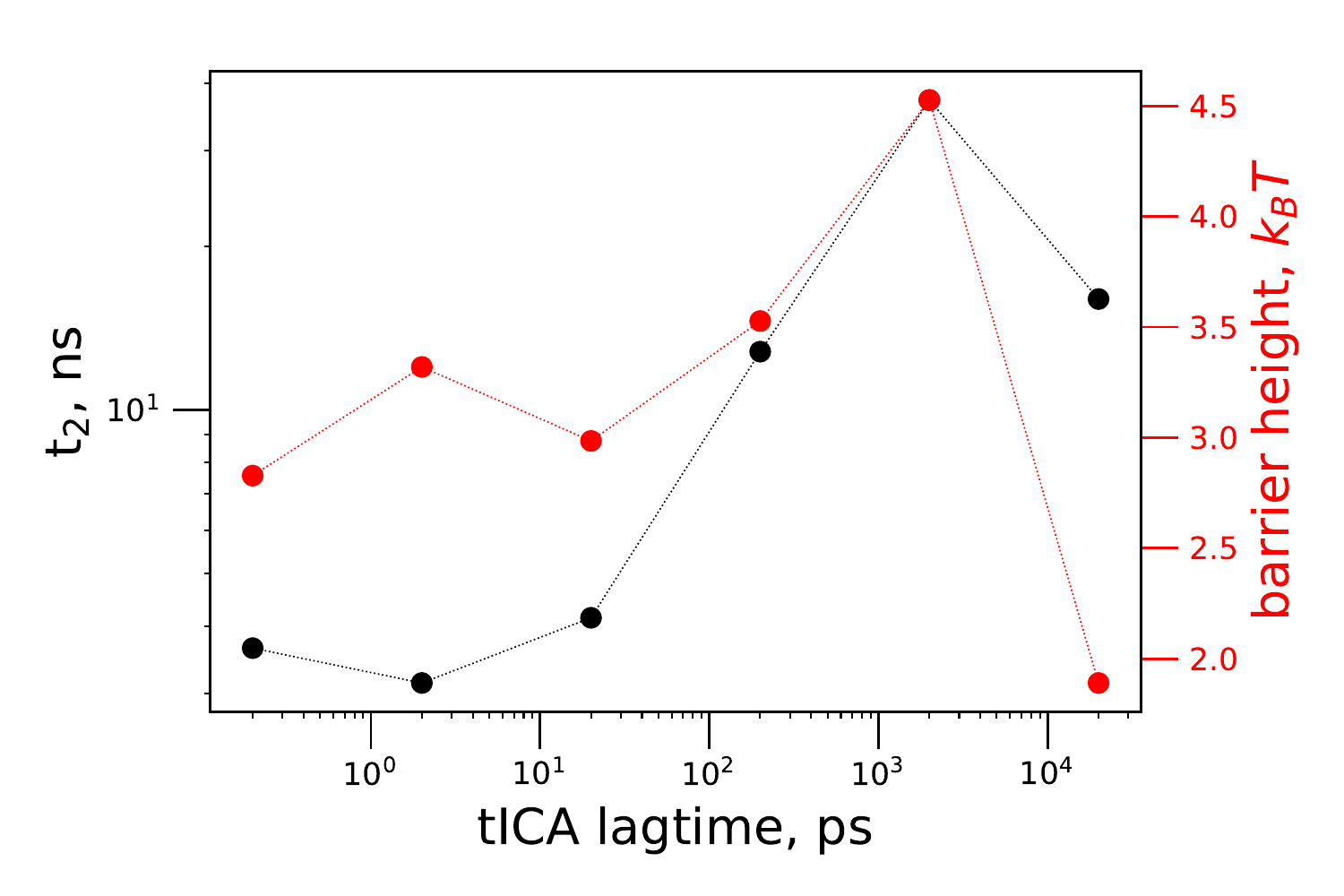}
\caption{Correlation between the barrier surmounted by the red-white-blue pathway between minima in Figures \ref{due} and \ref{lag timeFES} (red markers) and the t$_2$ timescale of the MSM constructed on the surface (black markers), as a function of tICA lag time. The correlation coefficient between the two sets of data, $\rho$, is 0.56. Dotted lines between markers are a guide to the eye.}
\label{tp1_t2_bar_corr}
\end{figure}

The mode-dependent FES look qualitatively similar at  $\tau_{\text{tICA}}$ smaller than $20$ ps, and at $\tau_{\text{tICA}}$ larger than $2$ ns. If we report the barrier height of the red-white-blue pathway between minima in Figures \ref{due} and \ref{lag timeFES} as a function of the lag time (see Figure \ref{tp1_t2_bar_corr}), we observe that when the FES is calculated at increasing $\tau_{\text{tICA}}$ the barrier height  increases until $\tau_{\text{tICA}}\approx2.0$ ns, when it starts decreasing. 
Figure \ref{tp1_t2_bar_corr} also reports the calculated Markov State Model (MSM) time, $t_2$, which is given by the projection of  the second MSM eigenvector as described in Section \ref{LE4PD}. $t_2$ is the time needed by the system to cross the barrier and shows a nice correlation with the barrier height for increasing tICA lag times. This analysis agrees with the concept that the best tICA lag time is the one that leads to the slowest dynamics, and hence the longest timescales, in a MSM analysis.

Intuitively, the non-homogeneity of ubiquitin's dynamics when changing the tICA lag time, observed in Figure \ref{tp1_t2_bar_corr},  seems associated with the well-known hierarchical energy landscape of proteins in the folded state.\cite{WU2008,Zhuravlev2010} At short lag times the tICA is sampling faster dynamics than at large lag times. Fast fluctuations cross small barriers along the pathway while sampling the energy landscape. As the tICA lag time is increased, the analysis picks up slower fluctuations, with a corresponding increase in predicted timescales and barrier heights. The fall-off in t$_2$ (and in barrier heights) at longer tICA lag times is likely due to a loss of statistics as the lag time is made large and the system makes direct `hops' between deep minima, thus avoiding the sampling of the barriers. Because the $t_2$ from the MSM is being reported as the timescale of the slowest processes found by the tICA, and at both long and short $\tau_{\text{tICA}}$ there are no large barriers sampled, the tICA coordinates, which are unit-free and do not encode lengthscales, return a similar quadratic or barrier-free surface to the MSM analysis. 

To conclude this section we observe that  for each mode, our procedure identifies the optimal mode-dependent tICA lag time using the height of the energy barriers.  Thus, different tICA modes are likely to have different optimal lag time, so that the definition of an optimal lag time can be not unique. We adopt for the optimal lag time the one measured in the slowest tICA mode.

\section{A comparison between 1D and 2D MAPS of \MakeLowercase{t}ICA modes}
\label{2D}
In the last section, we compare the outcome of this study with the results of the analysis performed using the conventional procedure, which combines a two-dimensional, or multidimensional, tICA free energy map with an MSM analysis of the kinetics.\cite{Schwantes2013, Sittel2018, Perez-Hernandez2013, Wehmeyer2018, Noe2016, Razavi2015, Wang2017} \cite{Schwantes2013}
In what has become a fairly typical workflow for the analysis of MD simulations of biomolecules using Markov State Models, the MD trajectory is projected onto not just one mode but a number $n$ of the slowest tICA modes. This procedure reduces the high dimensionality of the original free energy landscape by identifying the slowest dominant modes. One then performs an MSM analysis on the reduced subspace to parse the slowest dynamics and corresponding timescales of the system.\cite{Schwantes2013, Perez-Hernandez2013, Perez-Hernandez2016, Klus2018, Scherer2015, Husic2016, Wang2017} Usually, one selects the two slowest modes, but in some cases one considers instead more than two tICA modes: the latter procedure may lead to even slower measured kinetic timescales. \cite{Schwantes2013} This is because the transitions among the selected modes can become even less probable, while statistical insufficiencies in the necessarily finite simulation data can also play a role. Note that if the tICA modes were fully uncoupled, as we would like them to be, there would not be transitions between the modes (or the time for the transition would be infinite).  Transitions between tICA slow modes are rare, and represent the slowest dynamics in the MD trajectory. See also Section S1 in the Supplementary Material for an analysis of the independence of the tICA modes.

Here, we report the results of this type of `traditional' tICA-MSM approach for the 1-$\mu$s ubiquitin simulation. We build the MSMs on the space spanned by the first two tICs calculated from the ubiquitin simulation at a tICA lag time of $\tau_{\text{tICA}}=2$ ns, the same as that used in the $\left(\theta_a, \phi_a\right)$ surfaces presented in Section \ref{FEStICA}. Then we compare the results of this model to the analysis performed on the single-mode projections, namely the FES of a tIC, and the LE4PD-XYZ mode-dependent FES, presented earlier in this paper.

\begin{figure}[htb] 
\includegraphics[width=.9\columnwidth]{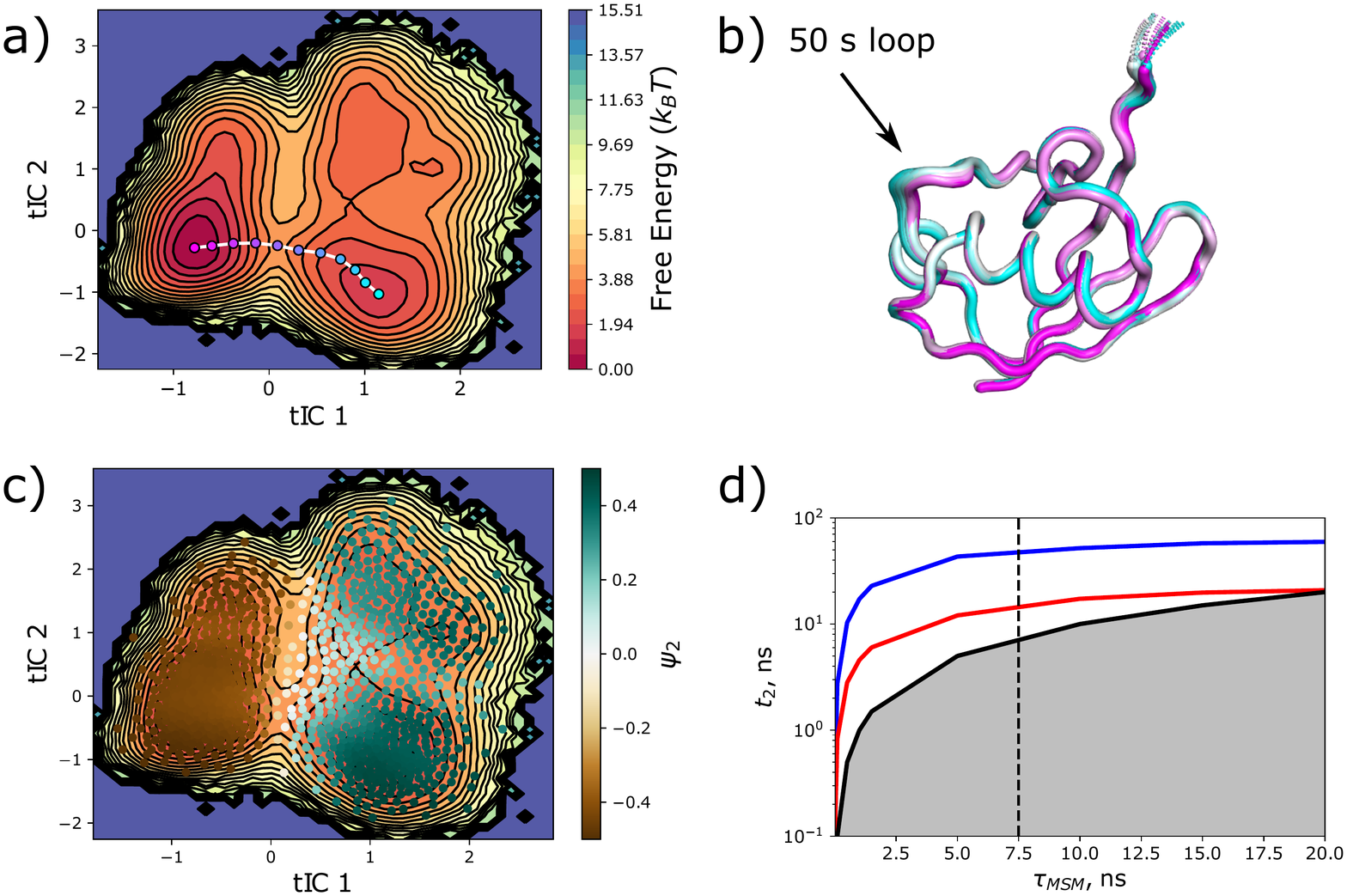}
\caption{Results for the MSM of the two slowest tICs. a) Free-energy surface for the first two tICs. b) Structures of ubiquitin from the trajectory along the free-energy surface given in a). The colors of the structures correspond to the given colored marker along the transition pathway. c) projection of $\psi_2$ onto the discrete states of the MSM; colors correspond to the scaled-and-shifted value of $\psi_2$ at that discrete state, $\psi_2 = \frac{\psi_2-\min(\psi_2)}{\max(\psi_2)-\min(\psi_2)}-0.5$. d): the two slowest implied timescales,  $t_2$ (blue curve) and $t_3$ (red curve), of the MSM as a function of MSM lag time, which is completely unrelated to the lag time used in the prior tICA step. The black vertical line demarcates the lag time selected when constructing the MSM, 7.5 ns.}
\label{square1}
\end{figure}

\begin{figure}[htb] 
\includegraphics[width=.9\columnwidth]{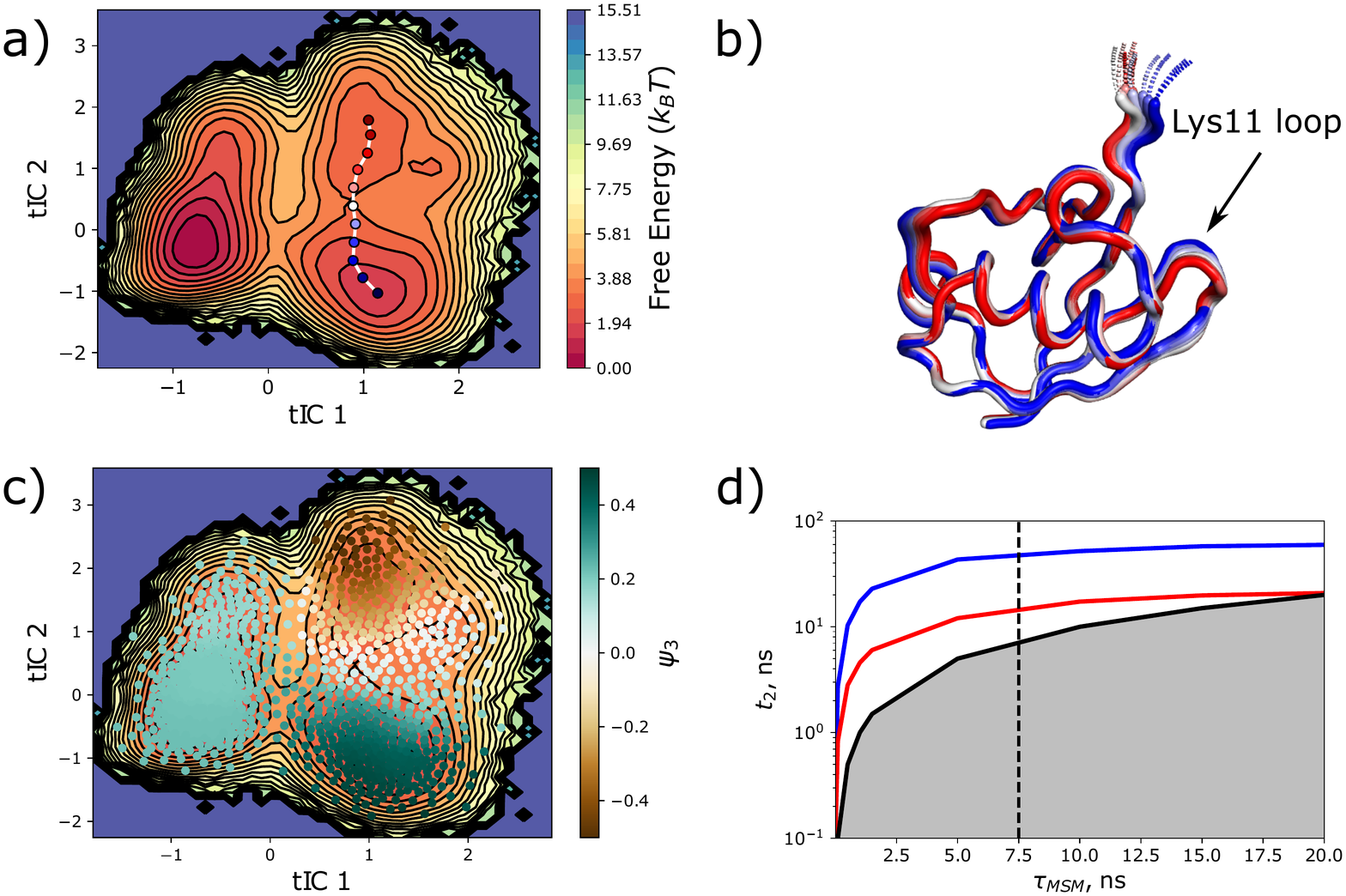}
\caption{Same as Figure \ref{square1}, except examining the second slowest process of the MSM, which is described by $\psi_3$ in c), where $\psi_3$ is scaled and shifted in the same manner as $\psi_2$ is in Figure \ref{square1}. }
\label{square2}
\end{figure}

Figure \ref{square1} shows that the free-energy surface spanned by the first two tICs has two `lobes' having two minima each. When a MSM is constructed on this surface, it predicts that the slowest motion spanned in this two-dimensional space is a transition between the two lobes, i.e. the transition between the two tICA modes, as can be seen by an examination of the spectrum of $\psi_2$ projected onto the free-energy surface (Figure \ref{square1}, top left panel). The MSM predicts that the transition between the two lobes occurs over a timescale of approximately 200 ns. Tracing a pathway between the two deepest minima in each lobe, using the same method we utilized for the $\left(\theta_a, \phi_a\right)$ surfaces shows that inter-lobe transitions correspond to dynamics in the $50\text{s}$ loop of ubiquitin. The second-slowest relaxation processes on the surface spanned by the first two tICs correspond to movement between the intra-lobe minima on the right-hand lobe (Figure \ref{square2}). The MSM predicts that this transition occurs over a timescale of $\sim$70 ns and that the transition causes motions in the Lys11 of ubiquitin. Note that the same slow fluctuations are identified by the single mode analysis of tICA and by the LE4PD models.

Thus, although the timescales predicted using the space spanned by the first 2 tICs are slightly slower than using the $\left(\theta_a, \phi_a\right)$ surfaces for the first 2 tICs individually ($\sim$40 versus 24.0 and $\sim$20 versus 10.7, respectively) the timescales are still within a factor of 2 in both cases, and the qualitative dynamics are predicted to be similar from both methods (comparing Figures \ref{tp1square} and \ref{tp2square} with Figures \ref{square1} and \ref{square2}). 
Thus, the single-mode analysis of tICA, the two-mode analysis of tICA, and the single-mode analysis of LE4PD and LE4PD-XYZ give consistent results when identifying, in one simulation trajectory, the local fluctuations for the slowest processes. 

We report in Section S7 of the Supplementary Material the analysis of the two-dimensional map of the two slowest LE4PD-XYZ modes. Analysis of this map confirms the previous calculations for the single-mode LE4PD-XYZ analysis. Because the LE4PD modes are slightly more coupled than the tICA modes, the energy barriers between the two slowest modes are smaller than in the tICA case.

\section{Discussion and conclusions}
\label{conclusions}
Atomistic MD simulations of proteins have been shown to describe with accuracy relevant biological processes. However, the leading behavior that guides the properties in biological systems, including the slow cooperative dynamics involved in protein binding and function, is often hidden by the broad spectrum of phenomena and the multitude of atomistic details displayed in simulations. Many studies have begun to rely on machine learning techniques to distill the essential leading kinetic information from MD trajectories. The most straightforward and widely used analytical tool in machine learning  is the time-lagged independent analysis or tICA, where the covariance matrix samples fluctuations at a given lag time, $\tau_{tICA}$. One can obtain a careful analysis of the kinetics of  slow fluctuations by combining the tICA with the MSM analysis. \cite{Perez-Hernandez2013, Schwantes2013} Similarly to tICA, the LE4PD and LE4PD-XYZ approaches accurately model  nonlinear protein motions by partitioning the MD dynamics into quasi-independent diffusive mode, while including the mode-dependent free-energy surfaces. \cite{Copperman2014, Copperman2015, Copperman2016, Beyerle2019, Copperman2017,Beyerle2019} The LE4PD is an isotropic representation of protein dynamics in a lab reference frame, analogous to the celebrated Rouse-Zimm equation for synthetic polymer dynamics,\cite{Doi1988,Bird1987} extended to consider the protein's hydrophobic core and the sequence-dependent hydrodynamic interaction. The LE4PD-XYZ is the anisotropic version of the previous equation of motion in the protein's center-of-mass reference system, where rotation and translation are removed.  The LE4PD-XYZ represents the anisotropic fluctuating dynamics of the proteins around its average structure.

Among deep learning methods, more sophisticated approaches than tICA have been proposed to model nonlinear effects in protein motions, such as kernel tICA \cite{Schwantes2015}, state-free reversible VAMPnets \cite{Chen2019a}, and time-lagged autoencoders (TAEs).\cite{Wehmeyer2018, Chen2019b} However, the use of such deep learning approaches to modeling nonlinearities in dynamics often comes with an increased computational cost, paired with a loss of physical intuition for the system under study. Thus, the tICA coordinates are considered, in general, the optimal linear approximation to the order parameters for relevant slow processes in proteins' dynamics.\cite{Perez-Hernandez2013, Sultan2018, Naritomi2011, Noe2015, Scherer2015} However, the tICA modes and other slow processes identified by machine learning lack information on their physical origin, having no associated equation of motion.

In this study, we compare tICA predictions with both the isotropic LE4PD  and with the anisotropic LE4PD, or LE4PD-XYZ. To do so, we associate to each tICA mode a free energy landscape obtained by the eigenvector projection of the simulation trajectory onto the tIC modes. This representation is convenient because it allows one to analyze the tICA's predictions based on the time evolution of fluctuations onto the mode-dependent free energy landscape. Because this representation depends on a single mode, it is free from the need to decide the number of modes to consider when building a free energy map, as one usually does.

LE4PD and  tICA agree in identifying the regions in the protein's primary sequence that undergo slow dynamics. 
There are three regions of slow dynamics for ubiquitin, namely the $50\text{ s}$ loop, the Lys 11 loop, and the C-terminal tail. All of them are known to be involved in ubiquitin's multiple biological functions. \cite{Komander2009, Komander2012, Penengo2006, Lv2018} Because the primary sequence of ubiquitin is highly conserved in the family of proteins with a similar function, we expect the processes identified in our study to be kinetically and thermodynamically robust, and that similar mechanisms are likely to guide the binding of other proteins that perform the same function.
 
Both tICA and LE4PD consistently identify the leading slow modes. However, while tICA tends to collect all the slow processes in the first or a few modes, the LE4PD provides a more detailed picture of the time- and length-dependence of the slow dynamics, which are partitioned into a larger number of modes. Thus, if one aims at identifying the slowest fluctuations in one mode, tICA may be more efficient than the LE4PD, if $\tau_{tICA}$ is opportunely selected. A detail to note is that when the tICs are sorted in descending order of decorrelation time (i.e., in the order of the tICA eigenvalues), the relative timescale may change after the free energy barriers are accounted for through the Markov state modeling, so that the slowest tICA mode could be different from the first tICA mode.\cite{Copperman2014, Beyerle2019, Beyerle2021a} 

In general, the tICA captures the slow fluctuations that occur at a timescale longer than the given tICA lag time, while faster dynamics are averaged out. The LE4PD method, instead, which is based on the solution of a ``bead-and-spring" model of macromolecular dynamics, provides detailed information on the dynamics at the different length scales. It follows that the LE4PD is accurate in reproducing the time decay of amino acid fluctuations at all timescales when the dynamics is represented by the time correlation functions calculated from the simulation trajectory (Figures \ref{comp_tcfs} and \ref{error}). The similar calculation performed using tICA modes is, with a few exceptions, much less accurate (Figures \ref{comp_tcfs}, and \ref{tica_tcf_alt}).

The tICA's lack of accuracy in the description of the time dependence of the fluctuation decorrelation as described by the simulated tcfs is not surprising because the tICA averages out the information at times shorter than the lag time. Setting a lag time for tICA affects the modality of sampling the dynamics in the free energy landscape.  These considerations lead us to perform a study on how the tICA lag time affects the properties analized by tICA, namely the slow fluctuations and the calculation of the time correlation functions. For example, if the lag time is too short or too long, the tICA cannot properly sample the free energy barriers. Thus, we propose and test several methods to evaluate \textit{a priori} an optimal lag time. The optimal $\tau_{tICA}$ calculated with the different methods is fairly consistent.

We also observed an almost quantitative agreement between the time correlation functions directly calculated from the simulation and the ones obtained by solving the LE4PD-XYZ equation when hydrodynamics is included. This result confirms the importance of hydrodynamics in the Langevin dynamics of proteins in solution, which is not surprising given that the Langevin is an equation of motion in the protein's reduced coordinates, where the effect of the solvent enters through friction, random forces, and hydrodynamic interactions. Thus, the hydrodynamic forces that enter the LE4PD equations result from the projection of the forces due to the solvent and the protein's atomistic fast degrees of freedom onto the reduced coordinates of the alpha carbons. Finally, hydrodynamics is more important for modes that are local, while large-scale fluctuations and slow modes are less affected.   

In conclusion, if a rapid identification of the leading slow dynamics is required, the tICA analysis is a practical and valuable strategy to collect that information. However, suppose the time propagation of the slow leading dynamics is of interest. In that case, the LE4PD-XYZ with hydrodynamics provides a more accurate representation of the slow processes based on its superior ability to reproduce the protein's dynamics at all times.

Note that different tICA modes are likely to have different optimal lag time. Thus, in principle, one cannot find one lag time that is optimal for the whole multiscale dynamics of the protein. If fact, in general, one observes that the barrier height tends to decrease with increasing mode number, as the mode-dependent dynamics become increasingly more local and less cooperative.\cite{Beyerle2019} Because defining the lag time in tICA implies that dynamics at shorter lag time are not detected, the dynamics on the more local modes may be not be correctly represented.

\section{Supplementary  Material}
The Supplementary  Material presents in Section S1 a study of the 
independence of the tICA and LE4PD modes, while Section S2 shows the stability of the tICA modes as a function of the selected lag time. To most efficiently identify the LE4PD-XYZ modes that best overlap to tICA modes, Section S3 in the Supplementary Material presents some additional methods to the ones described in Section \ref{wwohydro} of the Main document, including the projection of the slowest tICA mode onto the LE4PD-XYZ modes with and without hydrodynamics interactions. Section S4 presents the quantitative comparison of the mode-dependent fluctuations for the different LE4PD methods and tICA. A selection of alternative methods to calculate the optimal tICA lag time is summarized in Section S5. Section S6 presents a brief overview of the isotropic LE4PD: the calculations for this model compare in several sections of the Main Document together with the anisotropic LE4PD-XYZ's data. Section S7 shows the two-dimensional free energy maps for the two slowest modes of the anisotropic LE4PD-XYZ. Finally, the Supplementary Material document concludes in Section S8 with a presentation of the Molecular Dynamics methodology we used to simulate ubiquitin and the post-processing of the trajectory, followed by Section S9 with a brief overview of the Markov State Model method.

\section{Data Availability}
The codes used to perform the isotropic LE4PD and the anisotropic LE4PD-XYZ analyses described here are available on GitHub (https://github.com/GuenzaLab). The processed MD trajectory and $\left(\theta_a(t),\phi_a(t),\vert\vec{\xi}_a(t)\vert\right)$ trajectories for the first 10 LE4PD-XYZ modes both with and without hydrodynamics included are available on Zenodo (https://doi.org/10.5281/zenodo.4312224).

\section{Acknowledgements}
E.R.B. was supported by the John Keana Graduate Student Fellowship from the University of Oregon and by the National Science Foundation through grant
CHE-1665466. The computational work was performed on the
supercomputer Comet at the San Diego Supercomputer Center, with the support of XSEDE\cite{Towns2014} allocation TG-CHE100082 (XSEDE is a program supported by the National Science Foundation under Grant No. ACI-1548562).

\providecommand{\noopsort}[1]{}\providecommand{\singleletter}[1]{#1}%

\end{document}


\title{Supplemental Material for ``Identifying the leading dynamics of ubiquitin: a comparison between the tICA and the LE4PD slow fluctuations in amino acids' position"}
\date{\today}

\author{E. R. Beyerle}
 \affiliation{Institute for Fundamental Science and Department of Chemistry and Biochemistry, University of Oregon, Eugene, Oregon 97403, USA}
\author{M. G. Guenza}%
 \email{mguenza@uoregon.edu.}
\affiliation{Institute for Fundamental Science and Department of Chemistry and Biochemistry, University of Oregon, Eugene, Oregon 97403, USA}

\maketitle

\section{Evaluation of the degree of independence of the Slow LE4PD-XYZ Modes and of the \MakeLowercase{t}ICs}
\label{Indipendence}
As mentioned in the Main Text, the analysis of protein dynamics is facilitated by selecting coordinates that are independent so that the dynamics can be partitioned into trajectories that can be examined one at a time. In this section, we calculate the degree of correlation of the tICA and LE4PD-XYZ modes following the analysis presented in \cite{Naritomi2011} where the authors compare the principal components and tICs from a rigid body trajectory of the LAO protein. 
Both LE4PD-XYZ and tICA modes are built to have a minimum degree of correlation, even if they are not fully independent. The LE4PD-XYZ modes are designed to be statistically independent at zero lag time, but they may show statistical dependence at non-zero lag times. The degree of correlation is measured by calculating the normalized cross-correlation $C_{ab}(\tau)=\langle\xi_a(\tau)\xi_b(0)\rangle / \sqrt{\mu_a\mu_b}$ function between LE4PD-XYZ modes $a$ and $b$ at varying time lags $\tau$, where $\mu_a=\langle \xi_a^2\rangle$ is the mean-square fluctuation of mode $a$. The modes are independent when the cross-correlation is equal to zero. 

The tICs have an extra measure of statistical independence in that they are required to be independent at both zero lag time and the lag time at which the tICA is constructed (which is 2 ns here).\cite{Hyvarinen2001} Since the tICs have the guaranteed independence at the parameterized lag time, $\tau_{tICA}$, it might be anticipated that they will also be more independent than the LE4PD-XYZ modes at all lag times, as measured by the analogous cross-correlation functions. 

\begin{figure}[htb] 
\includegraphics[width=.9\columnwidth]{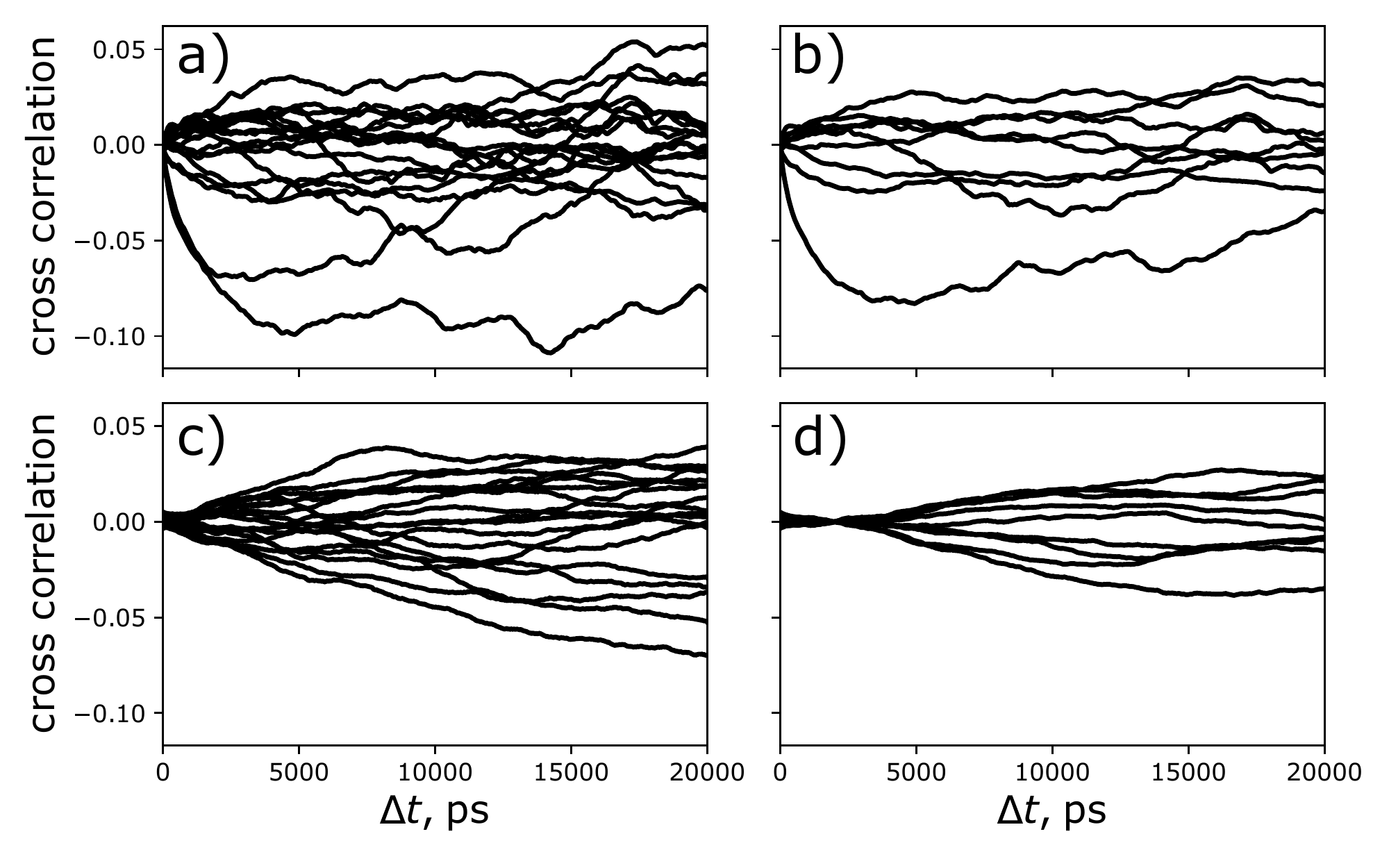}
\caption{a) Unsymmetrized and b) symmetrized cross-correlation functions for the 10 slowest LE4PD-XYZ modes. Panel c) shows the unsymmetrized  cross-correlation of the 10 slowest tICs and panel d) the symmetrized cross-correlation of the 10 slowest tICs.}
\label{crosscorr}
\end{figure}

Figure \ref{crosscorr} shows the cross-correlation functions for the 10 slowest ($a,b\in\{1,2,\ldots,10\}$) LE4PD-XYZ modes (top) and tICs (bottom) using lag times varying from 0 to 20 ns. The cross-correlations for both sets of dynamical variables are similar in magnitude to those observed for the principal components and tICs in \cite{Naritomi2011}. We also observe that, on average, the slowest LE4PD-XYZ modes show a larger magnitude of cross-correlation than the slowest tICs, both with (imposing that $C_{ab}(\tau) = C_{ba}(\tau)$) and without symmetrization, indicating that the tICs are more statistically independent than the LE4PD-XYZ modes (however, note the small correlation values for both approaches). 
When symmetrized, the slow tICs (Figure \ref{crosscorr}d) are on average more independent than the slow LE4PD-XYZ modes (Figure \ref{crosscorr}b).

\section{Effect of the \MakeLowercase{t}ICA lag time on the stability of the slow tICA modes}
\label{stability}
Figure \ref{crosscorr} shows that the tICs are more statistically independent than the LE4PD-XYZ modes because they are independent not just at zero lag time, but also at the tICA lag time, $\tau_{tICA}$, which here is $2$ ns. Thus, we want to investigate if the eigenvectors calculated at the tICA lag time still diagonalize the tICA matrix at a different lag time, or  how much the slow tICs depend on the chosen tICA lag time given that the tICA modes are defined as $\mathbf{z}(t)={\Omega}^T\Delta\mathbf{R}(t)$.  It has been shown that, depending on the system under study, the behavior of the tICs may \cite{Schwantes2013} or may not \cite{Naritomi2011, Sittel2018} be sensitive to a change in the tICA lag time. When the tICA lag time is modified, the dynamics predicted by the slowest tICs can change, sometimes significantly. 

For the protein ubiquitin examined here, we evaluate the stability of the slowest tICs by measuring the normalized self-overlap between two tICA eigenvectors calculated at two different time lags, $\tau_1$ and $\tau_2$. The overlap function between two tICA modes, specifically mode $a$ calculated with lag time $\tau_1$ and mode $b$ calculated with lag time $\tau_2$, is $O_{ab}\left(\tau_1,\tau_2\right) = \sum_{i,j}\Omega_{ai}\left(\tau_1\right)^TC_{ij}(0)\Omega_{jb}\left(\tau_2\right)$. If the overlap is close to one, it means that the same eigenvectors approximately diagonalize the matrix at those different lag times and the the tICs derived at $\tau_1$ and $\tau_2$ describe similar dynamics, which are robust to the lag time parameter.

\begin{figure}[htb] 
\includegraphics[width=.5\columnwidth]{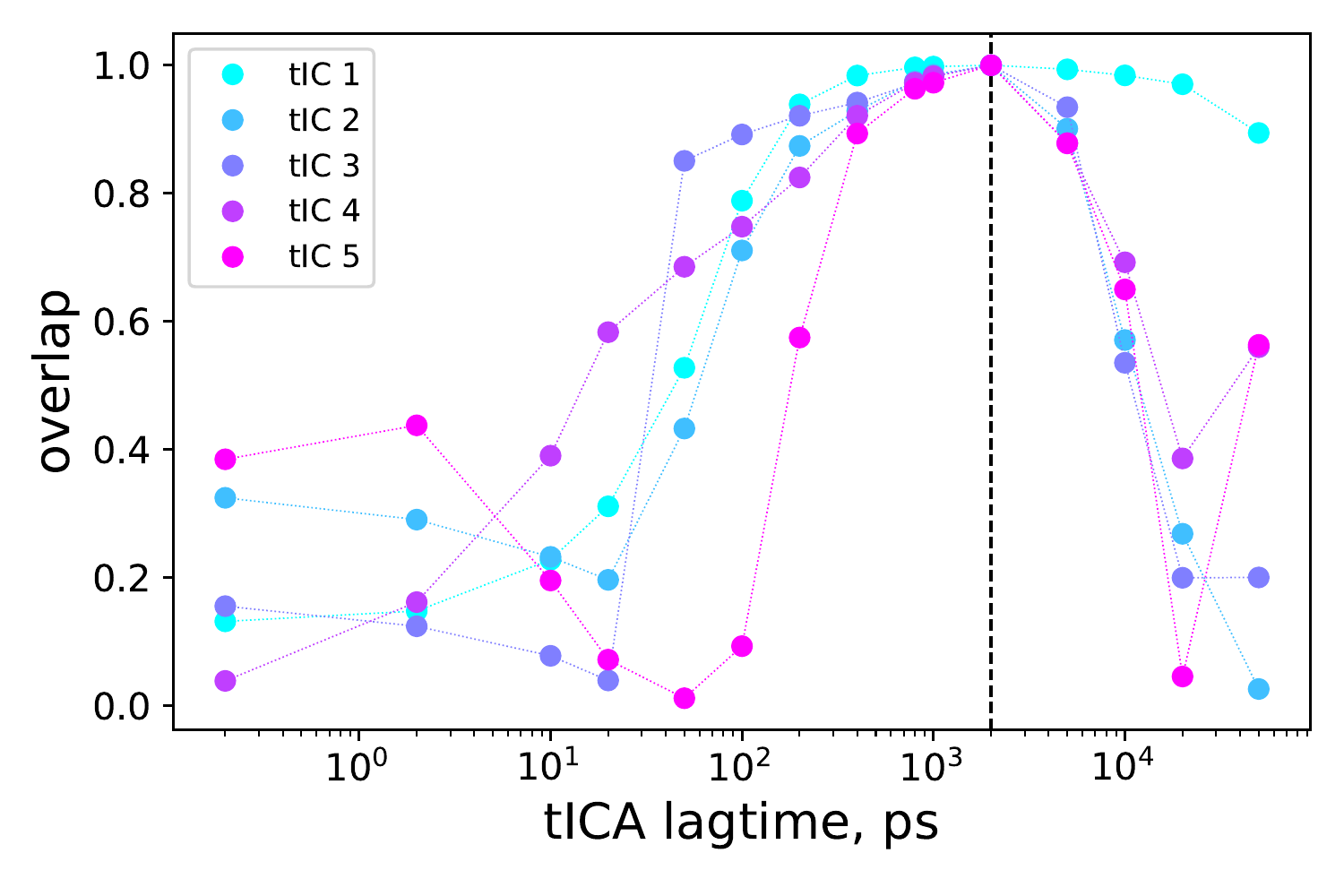}
\caption{Self-self overlap $O_{aa}\left(\tau_1,\tau_2\right)$ for the slowest five tICs at varying tICA lag times $\tau_1$ and $\tau_2$ calculated from the ubiquitin simulation. The vertical, dashed line gives the reference lag time $\tau_1=2000$ ps.}
\label{olap_1ubq}
\end{figure}

Figure \ref{olap_1ubq} shows the self-self overlap, $O_{aa}\left(\tau_1,\tau_2\right)$ for the five slowest tICs of the ubiquitin simulation. Overlaps are calculated with respect to the reference tICA lag time of $\tau_1=2$ ns. For the three slowest tICs, there is about one-half an order of magnitude of stability on either side of $\tau_1$ before the overlap falls sharply. For the fourth and fifth slowest tICs, there is less stability surrounding the reference lag time. These results indicate that the optimum lag time for the slow modes is less optimal for the fast modes.

\section{Methods to find the LE4PD-XYZ modes that best overlap with the slowest \MakeLowercase{t}ICA mode}
\label{overlapsection}
While the LE4PD-XYZ and tICA modes are ordered following the magnitude of their eigenvalues, i.e., from the slowest to the fastest modes, the eigenvalues are obtained from the linear approximation of the forces and do not account for anharmonicity and non-linear dynamics. Once we rescale the modes by using their internal energy barriers, we find that the slowest LE4PD-XYZ mode is not the first mode as expected. Thus, to effectively compare the predictions for the slow dynamics of LE4PD-XYZ and tICA, one needs first to identify the slowest modes and then compare them. This section presents two methods that identify the LE4PD-XYZ mode whose dynamics have the highest resemblance with the slowest tICA mode.

\subsection{Direct calculation of the mode-mode overlap function}
In this section we calculate the correlation or overlap between a given tICA, $z_a(t)$, and a LE4PD-XYZ mode, $\xi_b(t)$. The overlap between the slow tICs and LE4PD-XYZ modes without hydrodynamics can be determined analytically using the $\mathbf{V}$ matrix introduced in the Main Text. Since $\mathbf{z}(t)=\mathbf{V}^T\widehat{\xi}(t)$ and $\langle\mathbf{z}(t)\mathbf{z}(t)^T\rangle=\mathbf{I}$, one finds
\begin{align}
&\langle z_a(t)\xi_b(t)\rangle=\lambda^{\frac{1}{2}}_b V_{ba}.
    \label{Aoverlap}
\end{align}
So, $V_{ba}$ measures the scaled overlap of $z_a(t)$ and $\xi_b(t)$, with the scaling given by the square root of the eigenvalue corresponding to $\xi_b(t),\ \lambda_b^{\frac{1}{2}}$. 
Likewise, the overlap of the tICs and the LE4PD-XYZ modes with hydrodynamics, $\xi^{\text{HA}}$ are found using the overlap matrix defined in \cite{Beyerle2021a} and Eq. (\ref{Aoverlap}). Starting from the definition $\xi_a(t)=\sum_iQ^{T}_{ai}\Delta R_i(t)$ and given that $\xi_b^{HA}(t)=\sum_i(Q^{HA})^{T}_{bi}\Delta R_i(t)$, we have
\begin{align}
\xi_a(t)=\sum_{i,b}Q^{T}_{ai}Q^{\text{HA}}_{ib}\xi^{\text{HA}}_b(t)=\sum_bO_{ab}\xi^{\text{HA}}_b(t) \ .
\end{align}
Thus, 
\begin{equation}
\langle z_a(t)\xi^{\text{HA}}_b(t)\rangle = \sum_dO^{-1}_{bd}\langle z_a(t)\xi_d(t)\rangle=\sum_dO^{-1}_{bd}V^{T}_{da},
\label{overlapHA}
\end{equation}
so that $\sum_dO^{-1}_{bd}{V^T}_{da}$ measures the overlap between tIC $a, z_a(t)$, and LE4PD-XYZ mode $b$ with hydrodynamics, $\xi_b^{\text{HA}}$.

\begin{figure}[htb] 
\includegraphics[width=.5\columnwidth]{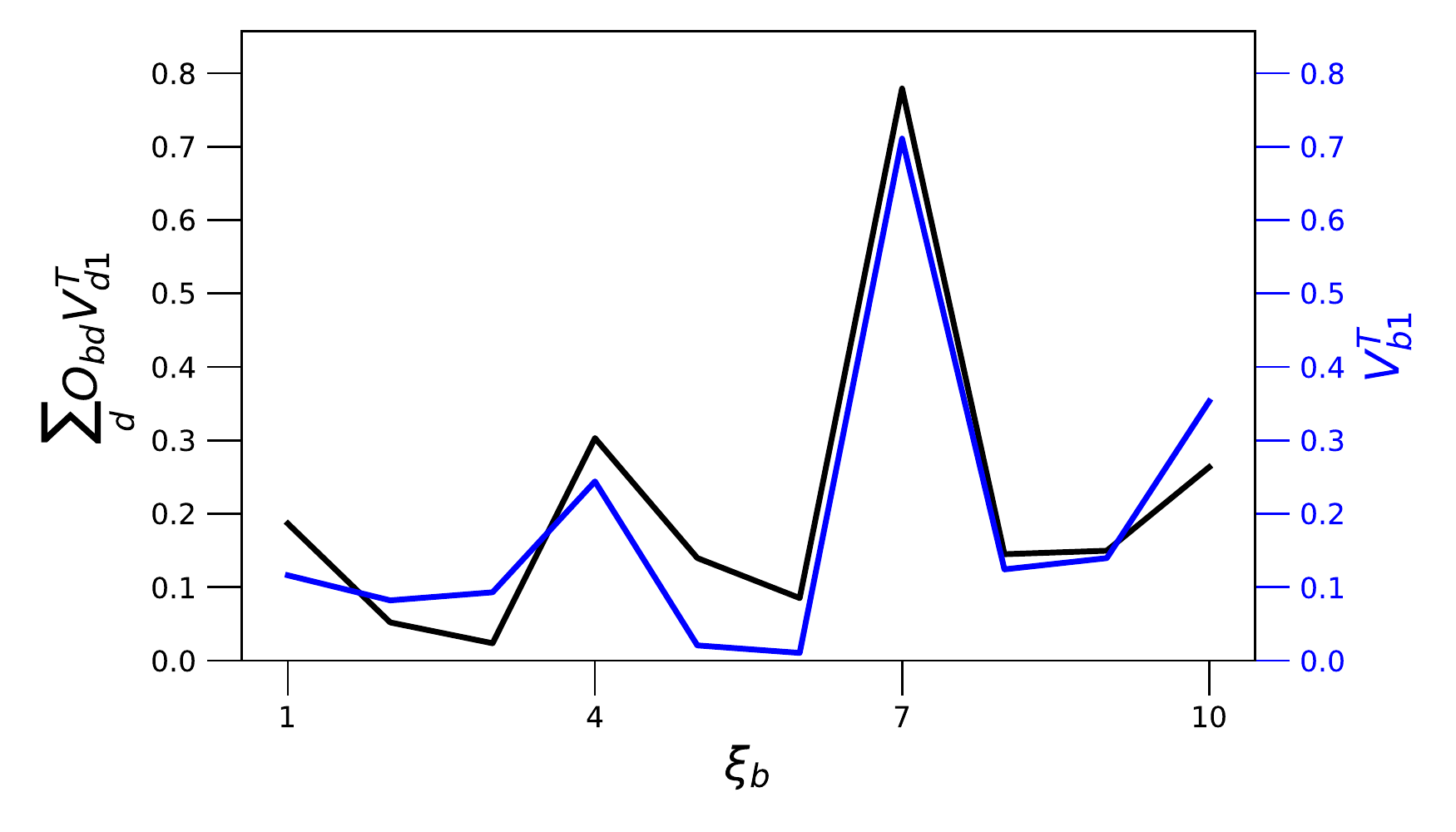}
\caption{Absolute value of the overlap between $z_1(t)$ and the first 10 LE4PD-XYZ modes with (black) and without (blue) hydrodynamics; the x-axis is generic and $\xi_b$ refers to mode $b$ from either approach.}
\label{overlap_fig}
\end{figure}

Figure \ref{overlap_fig} shows the unscaled overlap of $z_1(t)$ with the first 10 LE4PD-XYZ modes $\xi_b(t)$ both with (black curve in Figure \ref{overlap_fig}) and without hydrodynamics (blue curve in Figure \ref{overlap_fig}); the overlap of $z_1(t)$ with higher index LE4PD-XYZ modes is not significant and is not shown. It is clear from Figure \ref{overlap_fig} that the LE4PD-XYZ mode 7, both with and without HI, overlaps most strongly with the $z_1(t)$, while the overlap of $z_1(t)$ with the other LE4PD-XYZ modes is much lower. Thus, both the projection of $z_t(t)$ on the $\left(\theta_a, \phi_a\right)$ surface (presented in the Main Text) and the strength of the overlap between $z_1(t)$ and $\xi_7(t)$, which is the slowest mode from the LE4PD-XYZ analysis, provide evidence that this LE4PD-XYZ mode best corresponds to the slowest tIC (even if the agreement is not $100\%$). 

\begin{figure}
    \centering
    \begin{minipage}{0.45\textwidth}
        \centering
        \includegraphics[width=.9\textwidth]{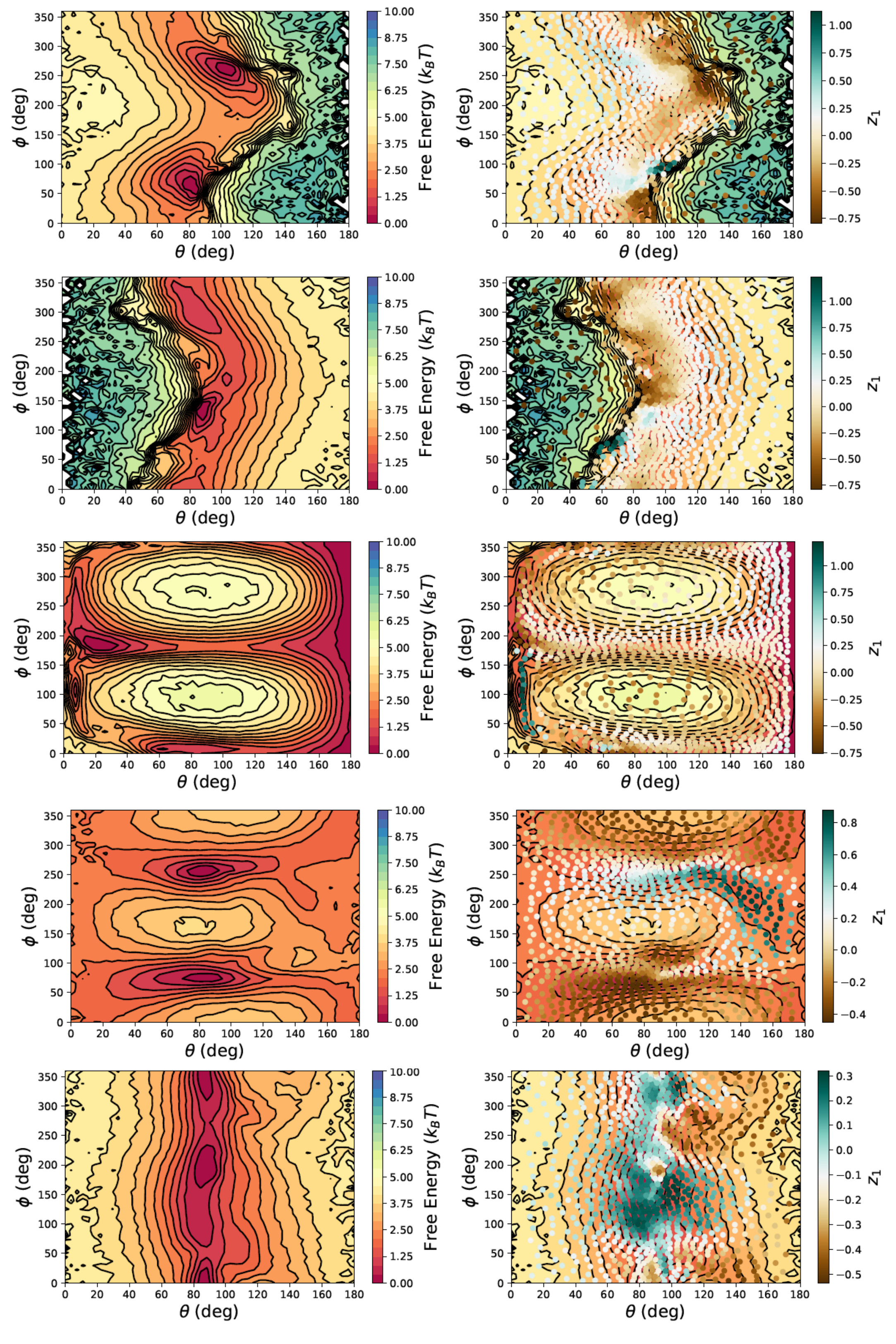}
        \includegraphics[width=.9\textwidth]{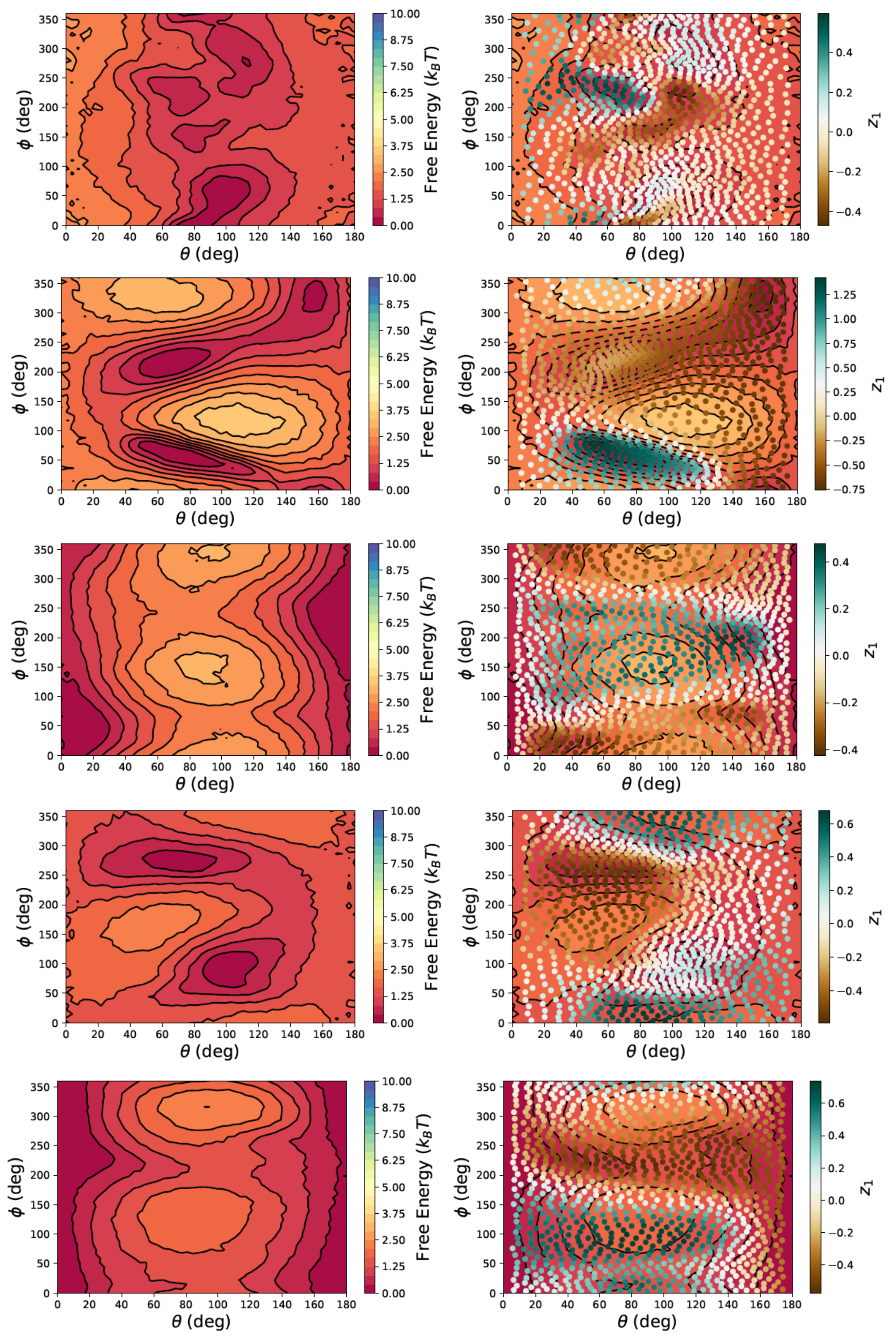}
        \caption{Projection of $z_1$ onto the ten slowest LE4PD-XYZ modes without HI. From the top panel to the bottom panel the LE4PD-XYZ mode number goes from one to ten. }
        \label{gallery1}
    \end{minipage}\hfill
    \begin{minipage}{0.45\textwidth}
        \includegraphics[width=.9\textwidth]{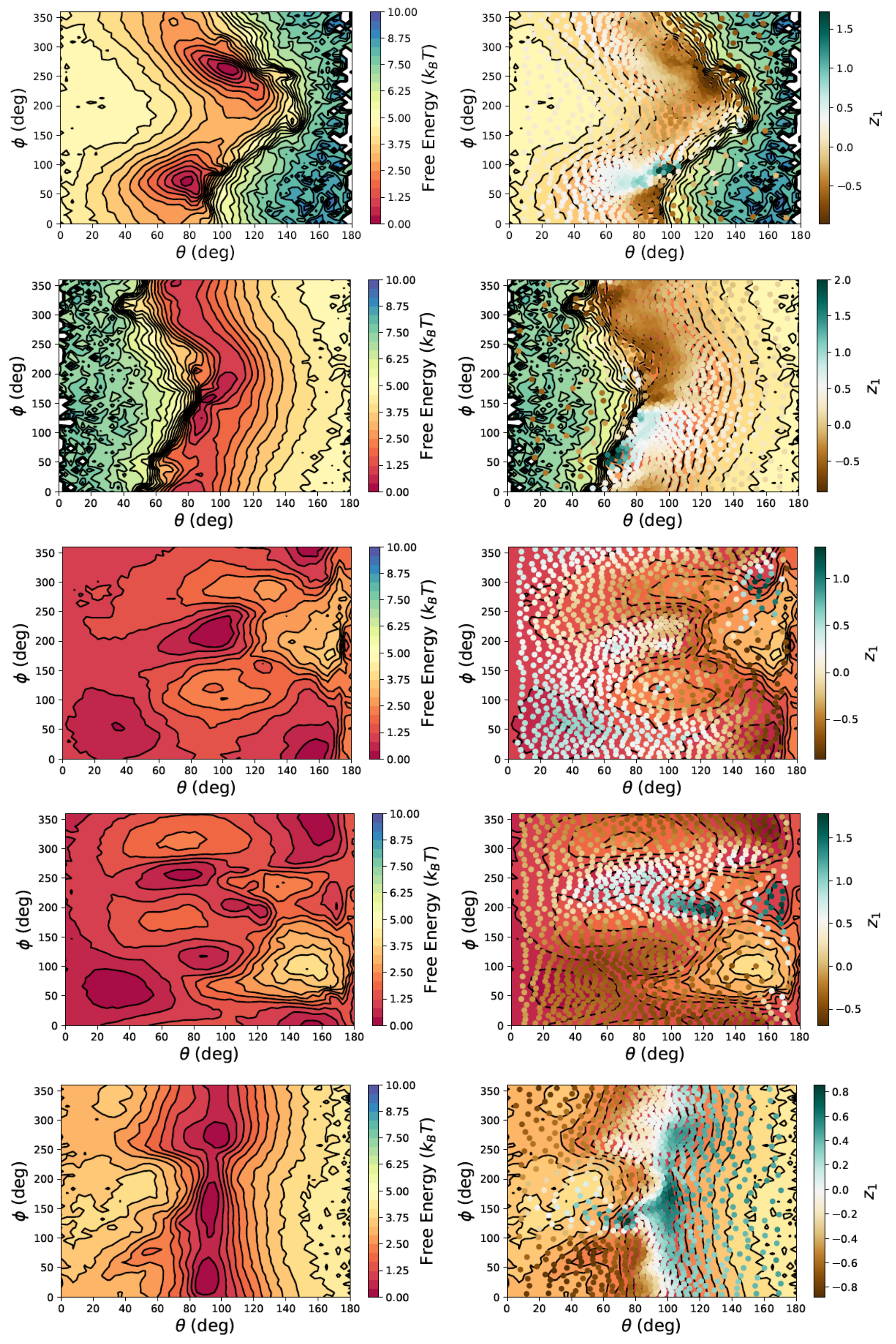}
        \includegraphics[width=.9\textwidth]{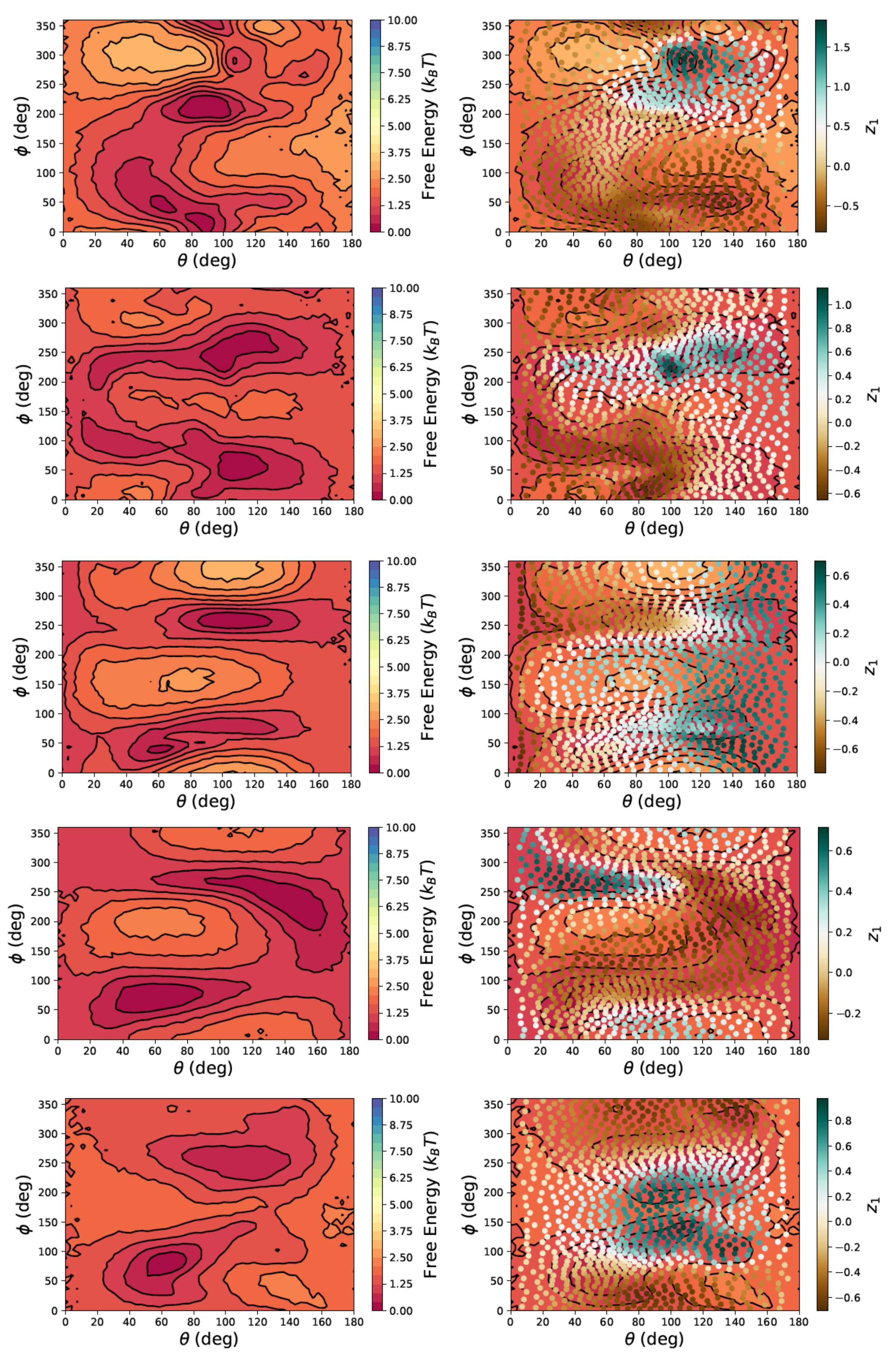}
        \caption{Projection of $z_1$ onto the ten slowest LE4PD-XYZ modes with HI. From the top panel to the bottom panel the LE4PD-XYZ mode number goes from one to ten. }
        \label{gallery2}
    \end{minipage}
\end{figure}

\subsection{Projections of the first tICA mode onto the first ten LE4PD-XYZ free-energy surfaces with and without hydrodynamic interaction}
To quantify the correspondence between the tICA first mode and the ten slowest modes in LE4PD-XYZ, we project the first tIC onto the energy surface of the different LE4PD-XYZ modes, with and without hydrodynamic interaction, following the procedure explained in the Main Text.
In the Main Text, the slowest tIC $z_1$ is projected onto the energy surfaces of the slowest LE4PD-XYZ mode (mode $7$) to show the correspondence between $z_1$ and the slowest eigenfunction from the MSM on that surface, $\psi_2$. Here, Figures \ref{gallery1} and \ref{gallery2} show the slowest tIC, $z_1$, projected onto the first ten LE4PD-XYZ both without and with HI, respectively. In general, there is no correspondence between $z_1$ and any features of the landscape, with the exception of LE4PD-XYZ mode 7 without HI. When a pattern emerges, it is significant and links the interpretation of $z_1$ and the given LE4PD-XYZ mode. That is, when the pattern of the extrema of the slowest tICs projects onto the minima of the slow LE4PD or LE4PD-XYZ energy surfaces, this indicates that the slow dynamics predicted by both the tICA and LE4PD-type analysis are similar.

\section{Quantitative comparison of the mode-dependent fluctuations calculated with the isotropic LE4PD with hydrodynamics, the anisotropic LE4PD-XYZ without hydrodynamics, and the tICA}
\label{quantitativeLML}
\begin{figure}
    \centering
    \begin{subfigure}[t]{0.45\textwidth}
        \centering
        \includegraphics[width=.9\textwidth]{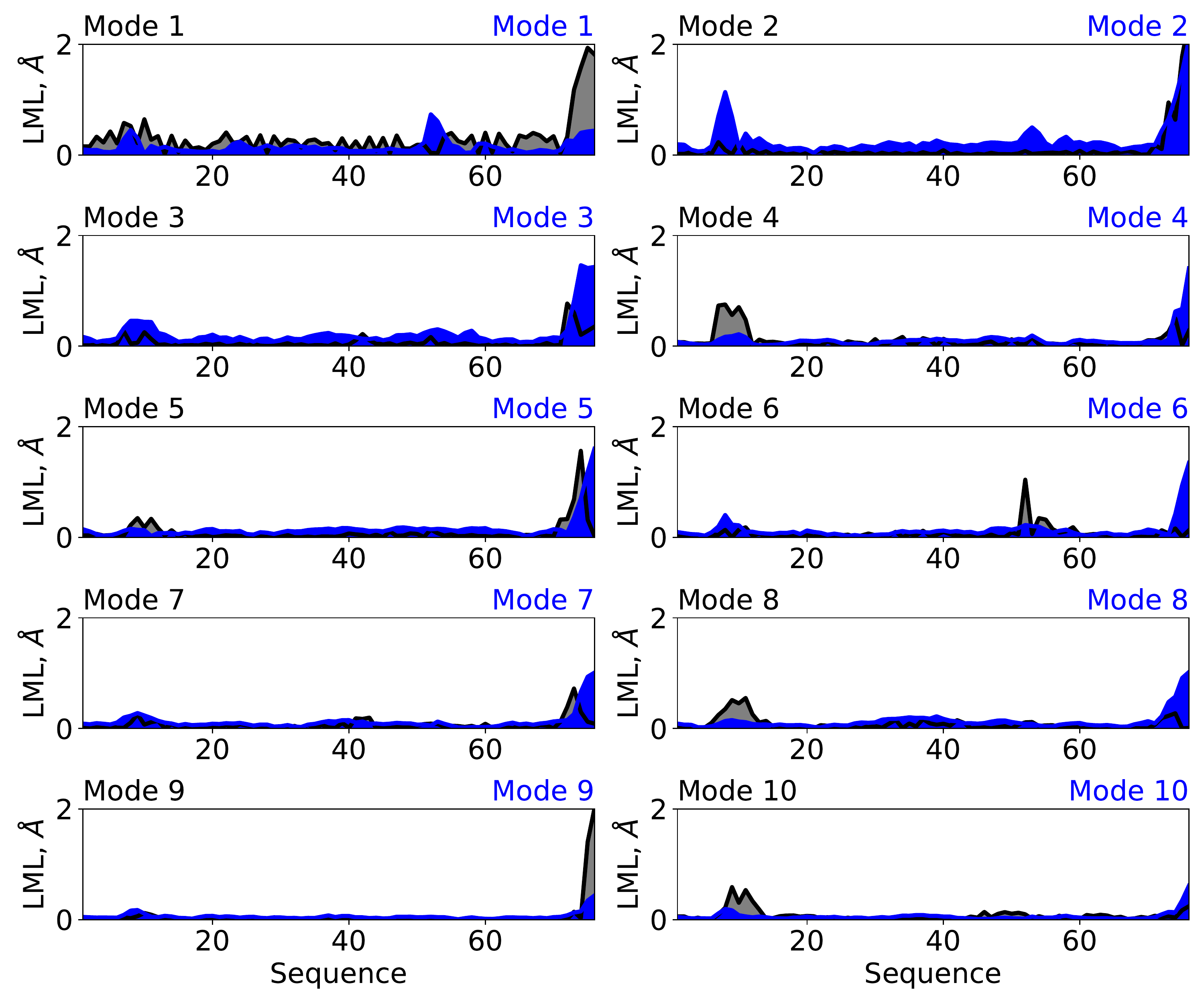}
        \caption{Mode fluctuations}
        \label{LML1}
    \end{subfigure}\hfill
    \begin{subfigure}[t]{0.55\textwidth}
        \includegraphics[width=.9\textwidth]{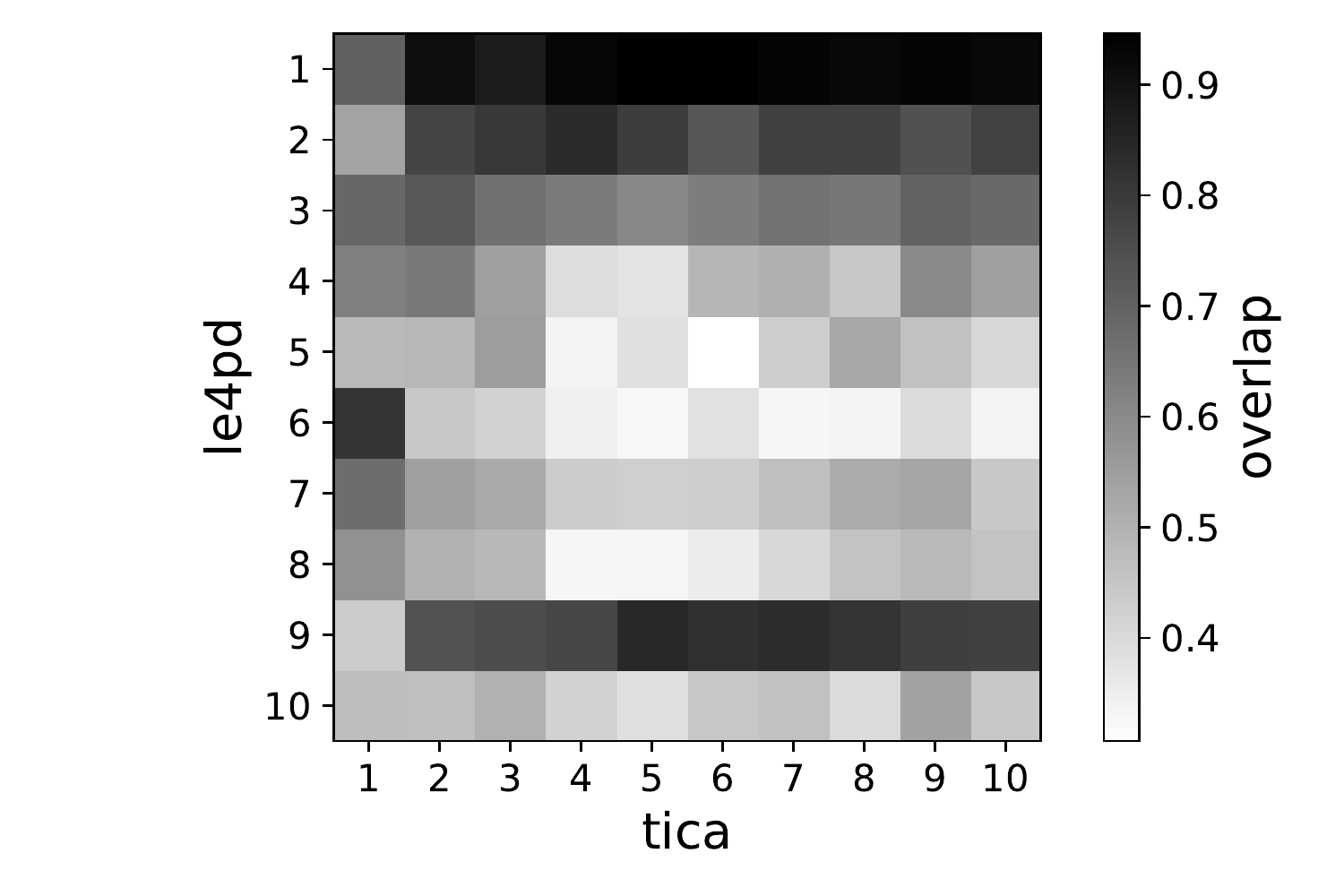}
        \caption{Mode overlaps}
        \label{lml_mat1}
    \end{subfigure}
\caption{a) Quantitative comparison of the first ten isotropic LE4PD mode fluctuations (black) with the first ten tICA modes (blue). b) Overlap of the normalized first ten local mode fluctuations for the isotropic LE4PD and the tICA. The best correlation between the first tICA mode is with mode 6 of the isotropic LE4PD.}
\label{pippo1}
\end{figure}

In this Section we compare in Figure \ref{LML1} the mode dependent fluctuations (LML) predicted by the isotropic LE4PD and the tICA as well as the anisotropic LE4PD-XYZ without the hydrodynamic interaction included and the tICA; to identify cross-correlations between these modes, we report in Figure \ref{lml_mat1} the overlap between the normalized fluctuations of the first ten modes of the isotropic LE4PD and tICA. The overlap function is defined as

\begin{equation}
\text{Overlap}_{a,b}= \sum_{i=1}^N \Big( \frac{LML^{(LE4PD)}_{ia}}{\sum_{j=1}^N LML^{(LE4PD)}_{ja}} \times  \frac{LML^{(tICA)}_{ib}}{\sum_{k=1}^N LML^{(tICA)}_{kb}} \Big)\ .
\end{equation} 

\begin{figure}
    \centering
    \begin{subfigure}[t]{0.45\textwidth}
        \centering
        \includegraphics[width=.9\textwidth]{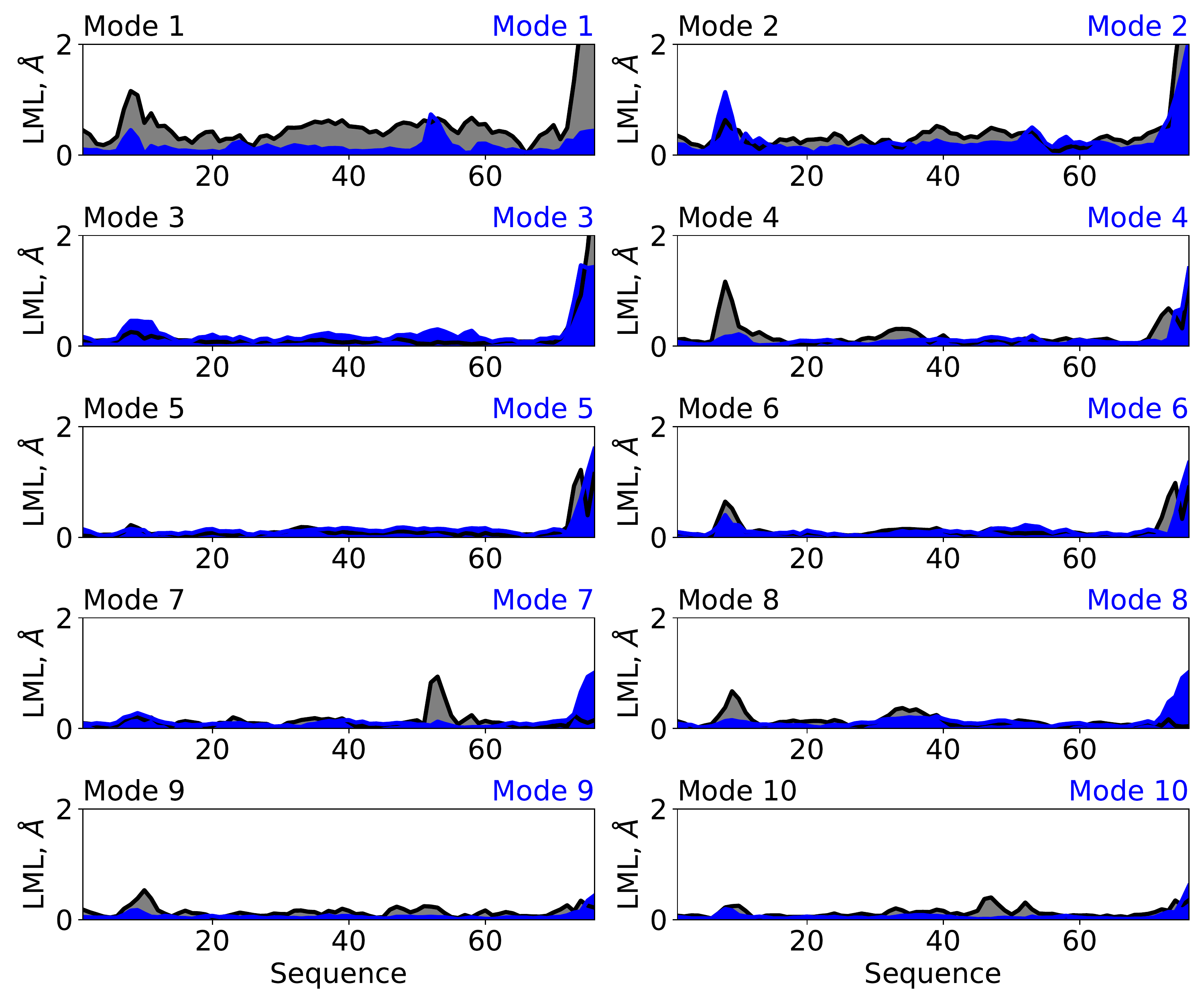}
        \caption{Mode fluctuations}
        \label{LML2}
    \end{subfigure}\hfill
    \begin{subfigure}[t]{0.55\textwidth}
        \includegraphics[width=.9\textwidth]{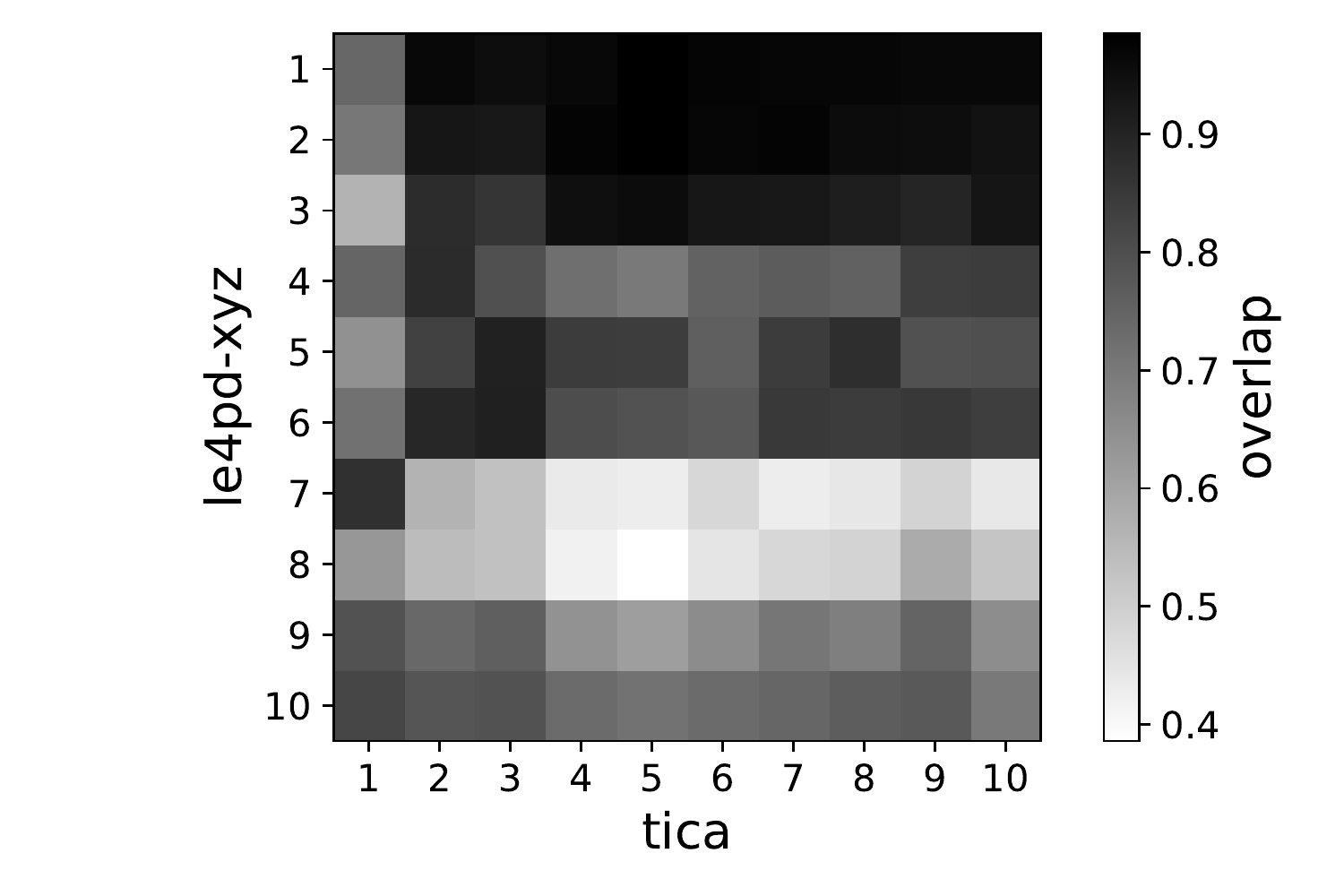}
        \caption{Mode overlaps}
        \label{lml_mat2}
    \end{subfigure}
\caption{a) Quantitative comparison of the first ten anisotropic LE4PD-XYZ mode fluctuations (black) with the first ten tICA modes (blue). b) Overlap of the normalized first ten local mode fluctuations for the anisotropic LE4PD and the tICA. The best correlation between the first tICA mode is with mode 7 of the anisotropic LE4PD or LE4PD-XYZ.}
\end{figure}

Figure \ref{LML2} displays the fluctuations for the anisotropic LE4PD model without the hydrodynamic interaction included and the tICA fluctuations. Figure \ref{lml_mat2} reports the overlap between normalized fluctuations in the first ten modes of the two approaches.  Consistently with our previous findings, the first tICA mode best overlaps with the fluctuations in mode seven of the LE4PD-XYZ approach.

It is immediately apparent that the fluctuations in the C-terminal tail and Lys11 loop are much larger than those in the 50 s loop, although they occur on faster timescales. Second, it is clear how the tICA funnels the dynamics of the Lys11 and 50 s loops into the first three modes while the LE4PD analyses tend to allocate the first few modes to dynamics in the C-terminal tail, although the first mode in both the isotropic and anisotropic LE4PD analyses show large `background' fluctuations along the entire primary sequence of ubiquitin that are comparable to the amplitude of the fluctuations selected by tICA for the 50 s and Lys11 loops in the first tICA mode. Since tICA modes 2 through 10 describe the dynamics in the tail the first LE4PD mode and the first three LE4PD-XYZ mode's all overlap strongly with tICA modes 2 through 10, as shown in Figures \ref{lml_mat1} and \ref{lml_mat2}, respectively.

The overlap between each of the first ten tICA modes and the first ten internal modes from the isotropic LE4PD and the LE4PD-XYZ method (without hydrodynamics) and are plotted as matrices in  Figures \ref{lml_mat1} and \ref{lml_mat2}, respectively. These matrices  show that the overlap between the first tICA mode and the seventh LE4PD-XYZ mode and the sixth isotropic LE4PD mode, respectively, is large; all three modes, as shown extensively in the main text, describe the slow fluctuations in ubiquitin's 50 s loop.

\section{Alternative Methods for Determining the optimal \MakeLowercase{t}ICA Lag Time}
\label{optimaltICA}
The Main Text presents a method that identifies the optimal tICA lag time, $\tau_{tICA}$, by finding the time that maximizes the barrier height of the slowest tICA mode. In this section, we propose three additional methods for selecting the tICA lag time, all of which show that a tICA lag time of 2.0 ns is an acceptable choice.
\begin{figure}[htb] 
\includegraphics[width=.9\columnwidth]{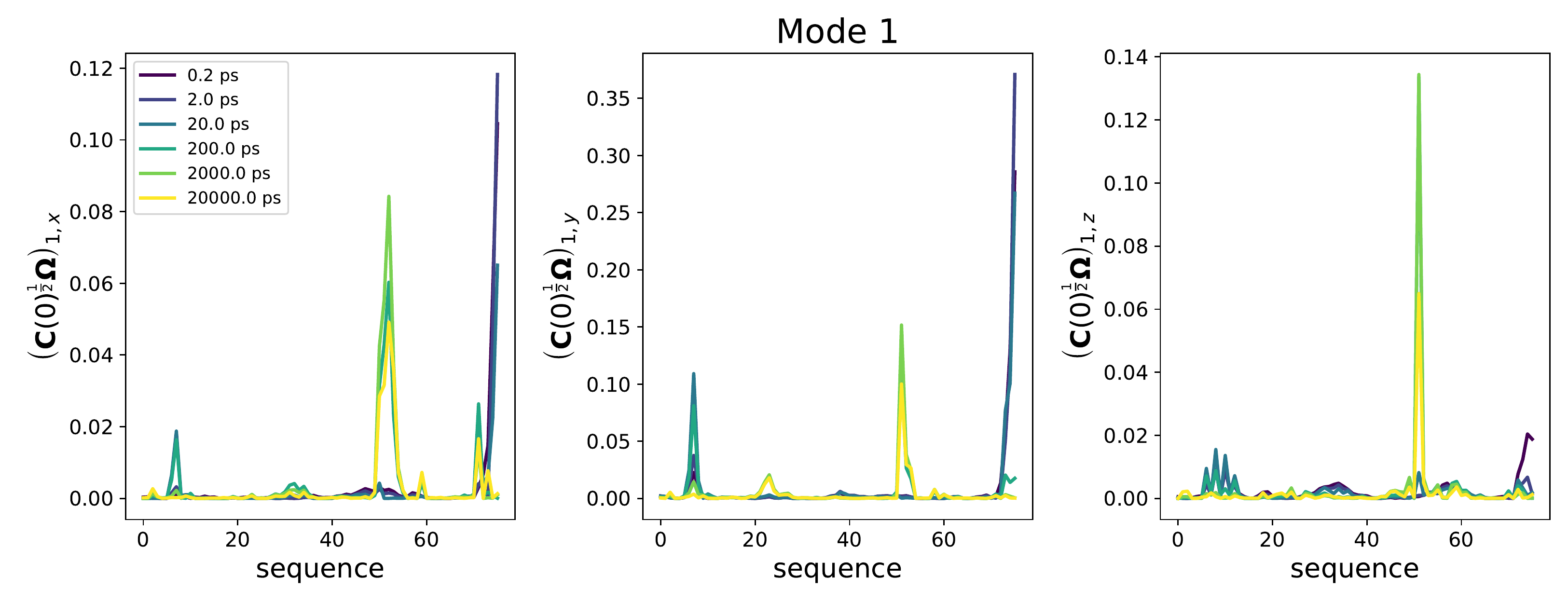}
\caption{Elements of first eigenvector of $\mathbf{C}(0)^{\frac{1}{2}}\mathbf{\Omega}$, $(\mathbf{C}(0)^{\frac{1}{2}}\mathbf{\Omega})_1$,  as the lag time $\tau_{tICA}$ of the tICA increased. This eigenvector corresponds to the first tIC, $\mathbf{z}_1(t)$. The three plots show, from left to right, $(\mathbf{C}(0)^{\frac{1}{2}}\mathbf{\Omega})_{1,x},\ (\mathbf{C}(0)^{\frac{1}{2}}\mathbf{\Omega})_{1,y},$ and $(\mathbf{C}(0)^{\frac{1}{2}}\mathbf{\Omega})_{1,z}$. The x-axis, `sequence', corresponds to the residue location along the primary sequence of ubiquitin.}
\label{pick_lag}
\end{figure}

\subsection{Using the tICA eigenvectors to select the tICA lag time}
An independent method to identify the optimal lag time starts from the analysis of the tICA eigenvectors. Figure \ref{pick_lag} shows how the symmetrized eigenvector $\mathbf{C}(0)^{\frac{1}{2}}\mathbf{\Omega}$ corresponding to the slowest tIC, $z_1$, changes as the lag time is increased. At short lag times ($\tau_{tICA}\leq$  2 ps), the slowest tIC describes fluctuations in the C-terminal tail (sequence numbers 71-76) and the Lys11 loop (sequence numbers 7-11) of ubiquitin. At higher lag times ($\tau_{tICA}\ge$ 200 ps) the structure of the eigenvector of the slowest tIC shifts to describe motion mainly in the $50\text{ s}$ loop (sequence number 48-51) of ubiquitin, which is where the LE4PD analysis also found the slowest motion of the protein.\cite{Beyerle2019} The observed slow dynamics is in agreement with the slow dynamics observed by NMR experiments \cite{Sidhu2011} and millisecond-length simulations.\cite{Piana2013} At all lag times larger than 200 ps tested, there was no significant changes to the structure of $\mathbf{C}(0)^{\frac{1}{2}}\mathbf{\Omega}_1$, which is consistent with the VAMP scores given in Section \ref{VAMP}. However, as the tICA lag time continues to rise, the eigenvalues of $\mathbf{C}(\tau_{tICA})$ and hence the timescales of the predicted dynamics continues to change, which can be correlated with changing barrier heights on the associated single-mode tICA energy surfaces, as shown in the Main Text. Using the structure of $\mathbf{C}(0)^{\frac{1}{2}}\mathbf{\Omega}_1$ alone, it appears that adopting any lag time from 200 ps to 20.0 ns is acceptable, and a more specific criterion is needed to refine the lag time selection further.

\subsection{Choosing the optimal tICA lagtimes from the VAMP-2 score of the 2D tICA maps}
\label{VAMP}
In the usual procedure, the selection of a tICA lag time, $\tau_{tICA}$, for the calculation of the time-lagged covariance matrix $\mathbf{C}^{r}(\tau_{tICA})$, is evaluated \emph{a posteriori} by constructing a Markov sate model (MSM) in the space spanned by the first two (the two slowest, based on the tICA lag time and eigenvalue) tICs and evaluating the so-called VAMP-2 score.\cite{Wu2017} This is the procedure that we adopt in this section.

\begin{figure}[htb] 
\includegraphics[width=.5\columnwidth]{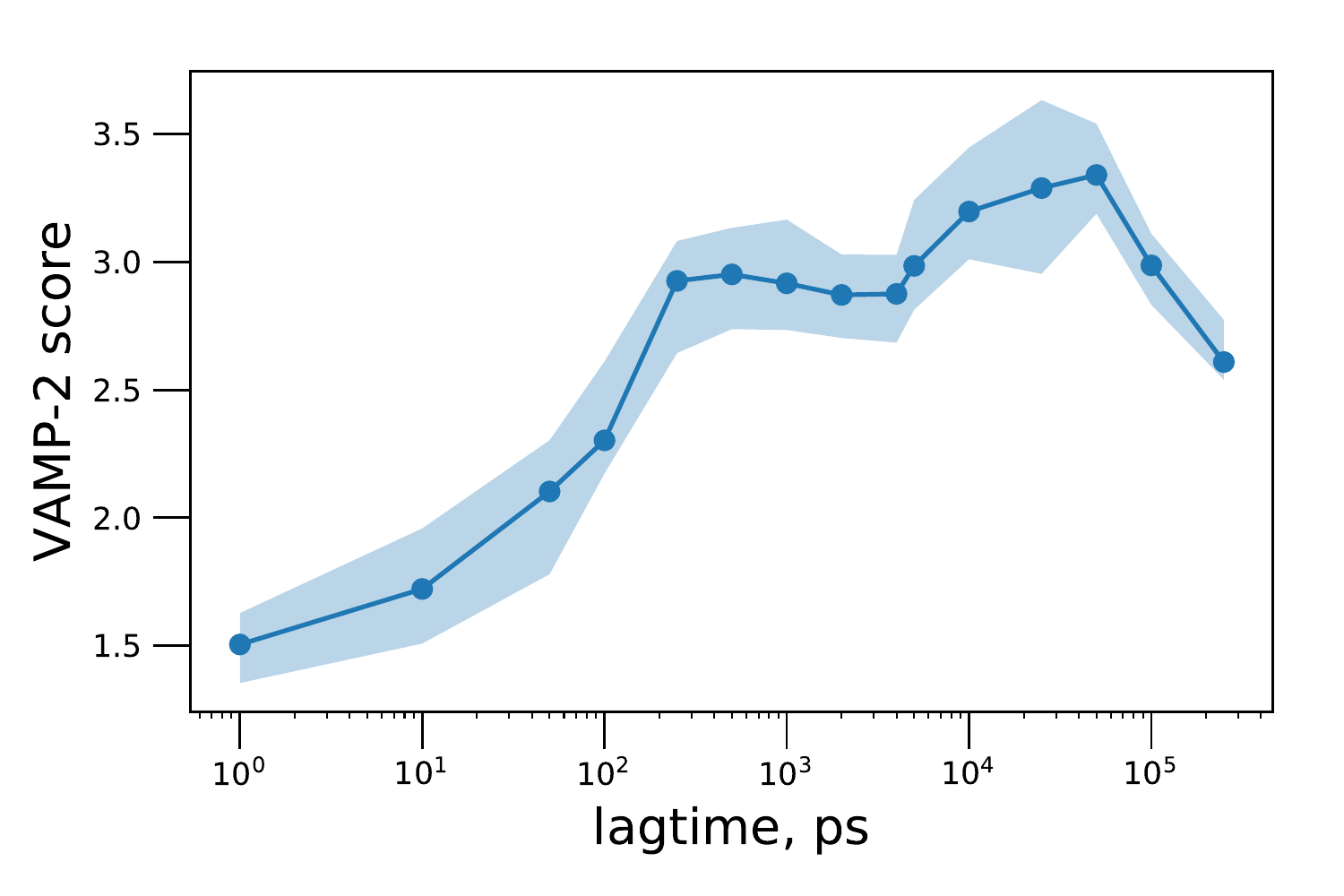}
\caption{VAMP-2 scores as a function of tICA lag time for the ubiquitin simulation. The VAMP-2 score is calculated using cross-validation, as described in the Main Text. MSMs were constructed in the reduced space of the first two tICs, and the tICs were found from a tICA performed at the indicated lag time. From the plot, the VAMP-2 score is maximized around $\tau_{tICA}=7.5\times10^{4}$ ps, but the value of the VAMP-2 score does not change significantly between $\tau_{tICA} = 250$ ps and $\tau_{tICA}=7.5\times10^{4}$. The markers show the average VAMP-2 score for the MSM at a given tICA lag time, and the shaded regions indicate the 90\% confidence intervals. The lines connecting the markers are a guide to the eye.}
\label{VAMP_1ubq}
\end{figure}

 By the variational theorem of conformational dynamics \cite{Perez-Hernandez2013} the transition matrix that leads to the slowest motions is the one that best approximates the `true' (continuous) transition matrix. The VAMP-2 score evaluates the sum of the eigenvalues of the transition matrix to find the $\tau_{tICA}$ that has the slowest transition times and the largest VAMP score. 
 
 \begin{figure}[htb] 
\includegraphics[width=.5\columnwidth]{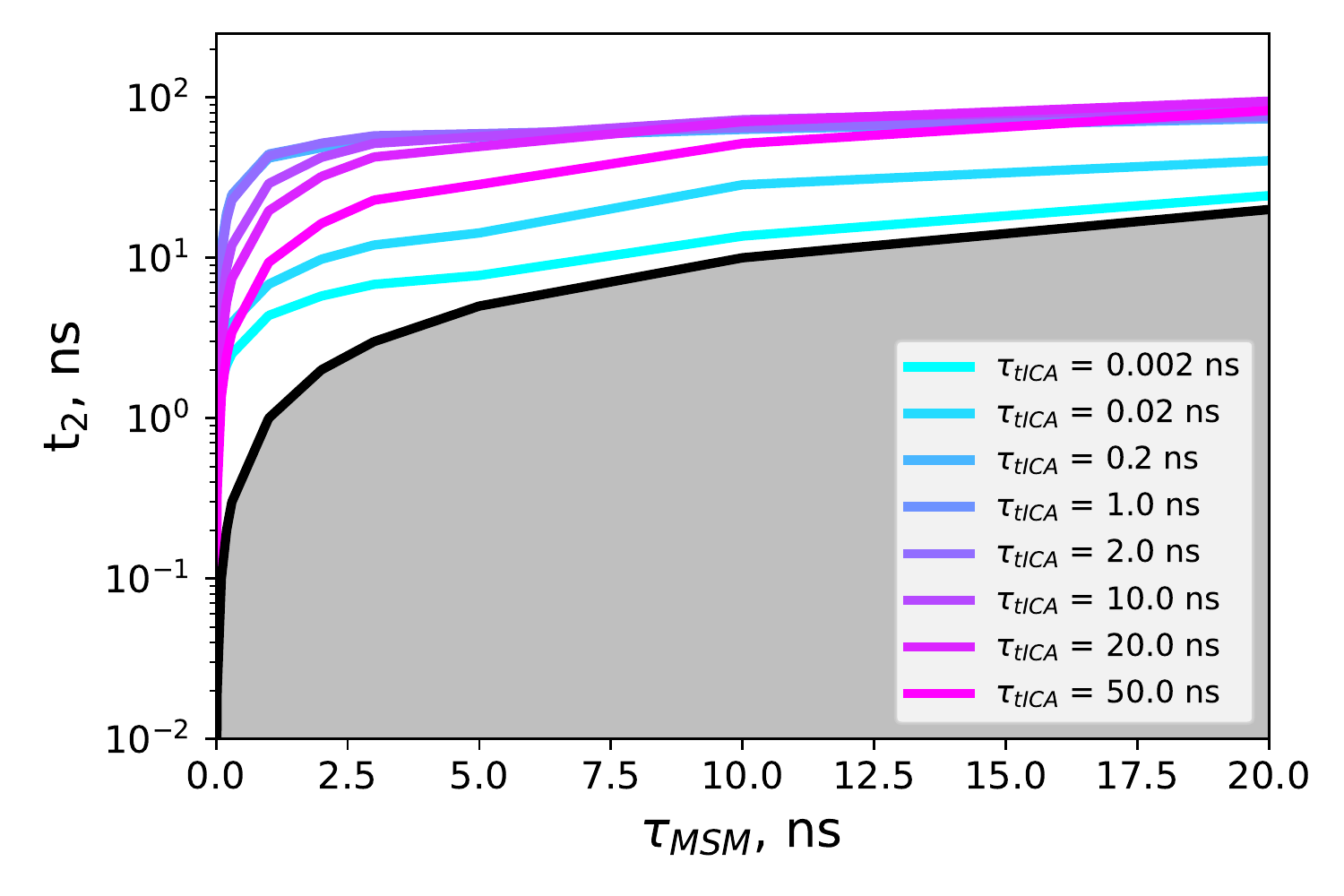}
\caption{Implied timescales for MSMs constructed using the first two tICA coordinates from tICA at each of the indicated tICA lag times.}
\label{comp_ITS}
\end{figure}

Following the traditional procedure, we first identify the MSM lag time at which the two slowest processes become markovian (see the section on Markov State Model, Section \ref{MSM}).\cite{Swope2004a, Scherer2019} After fixing the MSM lag time (here $\tau_{MSM}=5$ ns for all $\tau_{tICA} \ge 100 $ ps, as shown in Figure \ref{comp_ITS}), the VAMP score becomes a function of the tICA lag time alone, and the largest VAMP score identifies the optimal tICA lag time. Figure \ref{VAMP_1ubq} shows the VAMP-2 scores calculated using cross-validation: the trajectory is broken into a set of 10 sub-trajectories, and the MSM is constructed for one of those sub-trajectories and then tested on the remaining sub-trajectories. Figure \ref{VAMP_1ubq} shows that the score is relatively constant between lag times of 250 ps and 75 ns, implying that any value in that range is effective for a tICA analysis of the trajectory, including the value of $\tau_{tICA}=2$ ns that we selected.

To test the VAMP results, we calculate the MSM implied timescale $t_2$ at increasing $\tau_{tICA}$ for the surface defined using the trajectories of the first two tICs, which are the slowest modes. Figure \ref{comp_ITS} displays these implied timescales, and shows that $t_2$ is maximized for the tICA lag time $\tau_{tICA}=1$  or $2$ ns, in agreement with our previous calculations.

For short tICA lag times, $\tau_{tICA} < 0.2$ ns, the implied timescale does not converge at any MSM lag time, and the dynamics are not markovian (see Section \ref{MSM} for further details). At the tICA lag times of $0.2 \text{ ns } \le \tau_{tICA} \le 2$ ns, the implied timescales plots appear to converge to $t_2 \approx 40$ ns. For longer tICA lag times, the system does not approach markovian behavior at any MSM lag time. This result is in agreement with the findings reported in the Main Text and with the VAMP data in Figure \ref{VAMP_1ubq}. The only major difference between Figure \ref{comp_ITS} and Figure \ref{VAMP_1ubq} is that the VAMP analysis shows that tICA lag times between 2 and 50 ns are also optimal. However, some level of disagreement has to be expected between the two methods because the VAMP is maximizing the 10 slowest processes in the MSM while the calculations in Figure \ref{comp_ITS} focuses on the single slowest MSM mode.

\begin{figure}[htb] 
\includegraphics[width=.8\columnwidth]{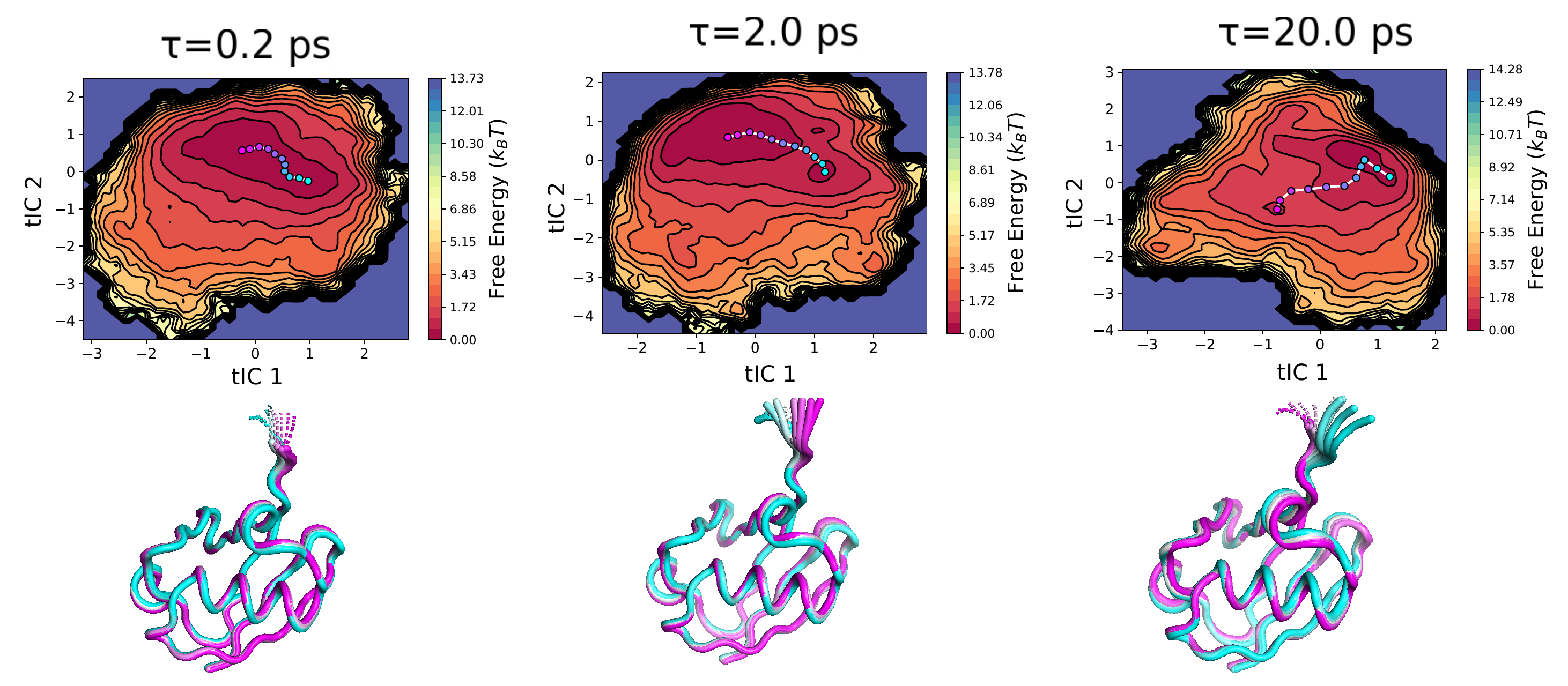}
\includegraphics[width=.8\columnwidth]{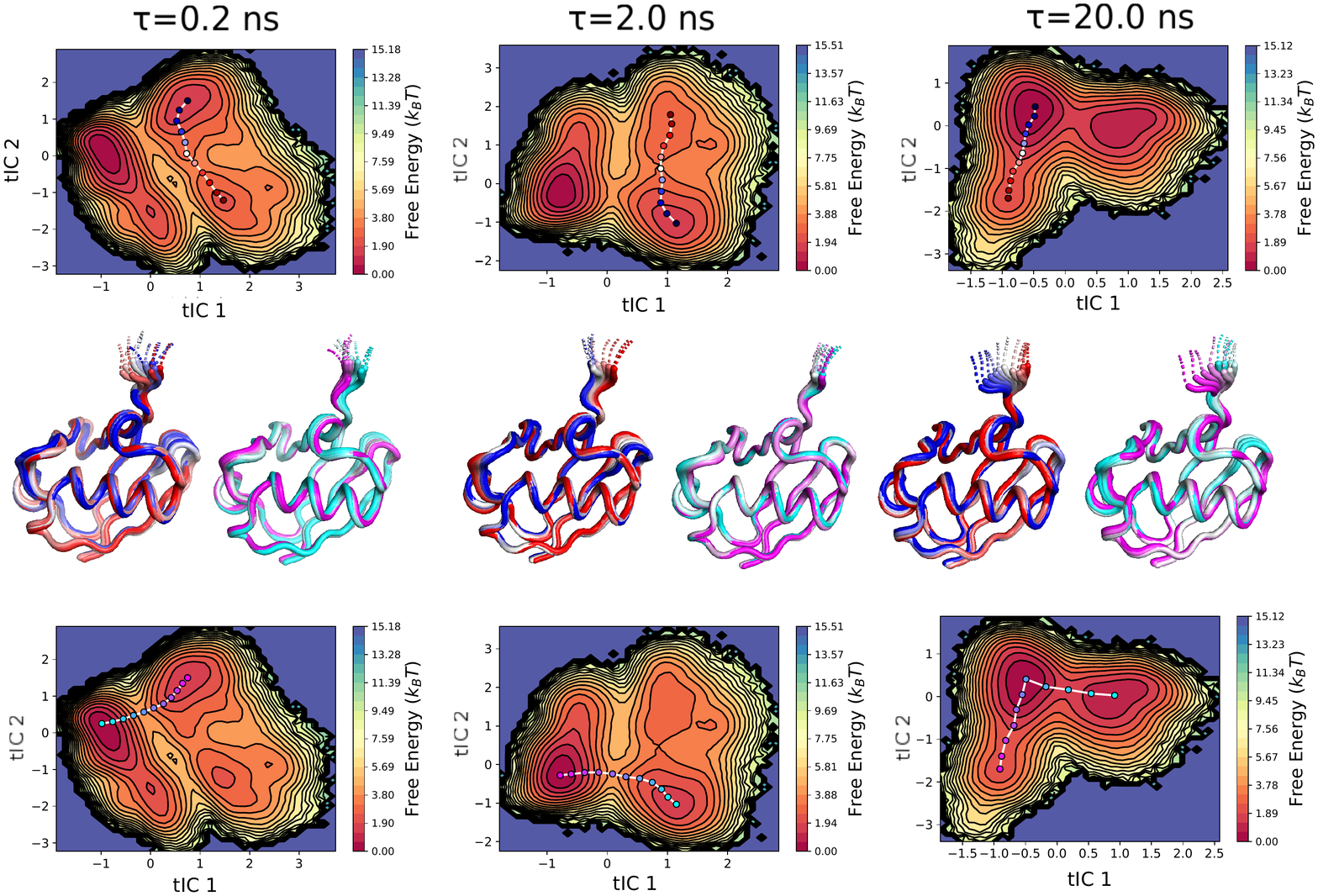}
\caption{Effect of changing the tICA lag time on the resulting tIC 1 - tIC 2 FESs and associated dynamics. As the lag time is increased, the predicted motion of the slowest tIC moves from the C-terminal tail and Lys11 loop into the 50 s loop. Concurrently, the barrier between the two minima on the surface rises until $\tau_{tICA}=2.0$ ns, when the barrier between minima starts to decrease. This decrease in the barrier between minima coincides with the loss of markovian behavior at lag times above 2.0 ns seen in Figure S10. Only a single pathway for $\tau_{tICA}\le 20.0$ ps is drawn because there is no second minimum on the surface.}
\label{tica_change_time}
\end{figure}

\subsection{Selecting the optimal tICA lag time from the analysis of the non-homogeneous dynamics in 2D free energy plots}

To confirm the tICA lag time selection, we started from the 2D tICA FES, and we explored the dynamics on the tIC 1 - tIC 2  energy surface while varying the lag time and examining the trajectories between minima. This analysis is analogous to the one we completed on the FES for one tICA mode in the Main Text. Such an analysis of the 2D FES shows how both the FESs and the associated mode-dependent dynamics change as the lag time of the tICA is adjusted. 

\begin{figure}[htb] 
\includegraphics[width=.5\columnwidth]{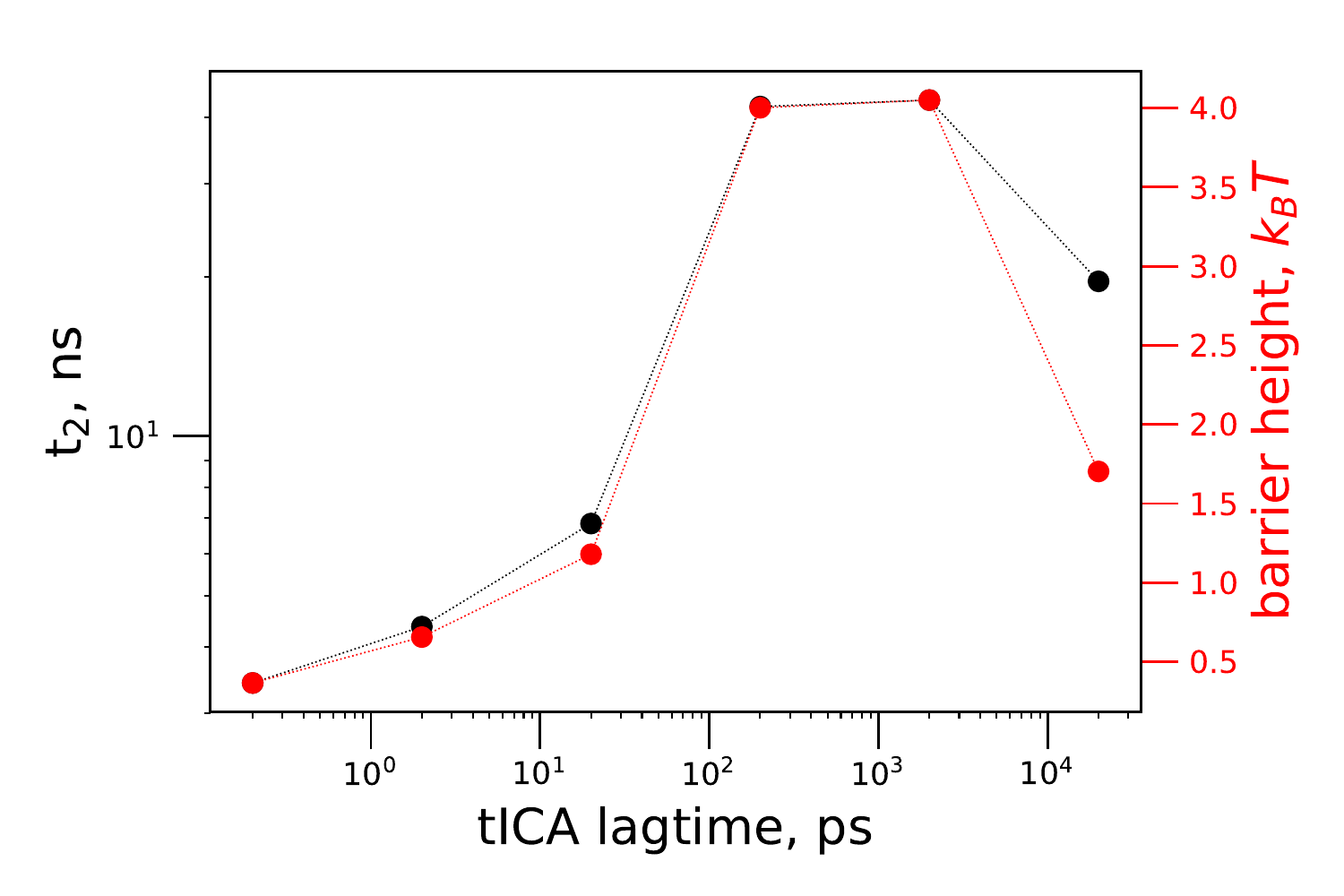}
\caption{Correlation between the barrier surmounted by the red-white-blue pathway between minima in Figure \ref{tica_change_time} (red markers) and the t$_2$ timescale of the MSM constructed on the surface (black markers), as a function of tICA lag time. The correlation coefficient $\rho$ is 0.99. Dotted lines between markers are a guide to the eye.}
\label{bar_t2_corr}
\end{figure}

Figure \ref{tica_change_time} compares the FESs and dynamics along the pathways between minima on these surfaces when the tICA lag time is increased from $\tau_{tICA}=0.2$ ps to $\tau_{tICA}=20.0$ ns. When $\tau_{tICA} \le$ 20.0 ps, the two slowest tICs describe mainly dynamics in the C-terminal tail of the protein (Figure \ref{tica_change_time}). However, once $\tau_{tICA}$ rises above .2 ns, the motion shifts to a combination of fluctuations in the Lys11 and $50\text{ s}$ loops. This shift in the foci of the dynamics coincides with an increase in the barrier between the two minima on the surface and the attainment of markovian dynamics for the transitions between minima (Figure \ref{bar_t2_corr}). Once the lag time rises above 2.0 ns, the barriers between minima begin to decrease, and the dynamics on the surface become again non-markovian.

This result suggests two conclusions. First, the barrier to conformational change in the Lys11 and $50\text{ s}$ loops are larger than those in the C-terminal tail, because the free-energy barrier rises when the dynamics of the two slowest tICs shifts from short lag times (where the tICs describe the dynamics in the protein's tail) to longer lag times (where the tICs describe dynamics in the Lys11 and $50\text{ s}$ loops). That the energy barriers in the tails are smaller than in the loop is not surprising, given the intrinsically disordered nature of the tail region. 

Second, at least for these surfaces, there is a strong correlation between the observed barrier height and how markovian is the dynamics (see Figure \ref{bar_t2_corr}). Again, this result is probably not surprising since large barriers between conformational states are required to `erase' the intra-state memory and generate markovian dynamics among states.\cite{Deuflhard2000, Deuflhard2005} These results are in qualitative agreement with the analysis performed on the single mode tICA free energy surfaces, presented in the Main Text.

\section{A brief overview of the Isotropic LE4PD}
\label{LE4PDisotropic}
The isotropic LE4PD projects the MD trajectory of a protein onto the slow coordinates of the alpha-carbon of each residue, $\vec{R}(t)$. It models the time evolution of these coordinates using an overdamped Langevin equation, where the residues interact through the potential of mean force, defined by the matrix $U_{jk}=\langle\vec{l}_i\cdot \vec{l}_j\rangle/ \langle |\vec{l}_i|\rangle \langle |\vec{l}_j| \rangle$.  Here $\vec{l}_i=\vec{R}_{i+1}-\vec{R}_{i}$ is the bond vector between residue $i$ and residue $i+1$ along the protein's primary sequence and the bracket defines the statistical average over all the trajectory's conformations.  The dynamics is guided by the intramolecular potential of mean force (matrix $\mathbf{A}$, which defines the potential of mean force in the set of $\vec{R}$ coordinates) and hydrodynamic interactions, as well as the random forces generated by the collisions with the surrounding solvent. Thus, the propagation in time of the protein's dynamics follows a Langevin equation that in the $\alpha$-carbon coordinates reads:
\begin{equation}
\frac{d\vec{R_i}(t)}{dt} = -\frac{3k_BT}{l^2\overline{\zeta}}\sum\limits_{j=1}\limits^{N}\sum\limits_{k=1}\limits^{N}H_{ij}A_{jk}\vec{R}_k(t) + \frac{\vec{F}_i(t)}{\overline{\zeta}},
    \label{le4pd}
\end{equation}
where $k_B$ is the Boltzmann constant, $T$ is the temperature of the protein-solvent system, $l^2$ is the mean-square bond length between alpha-carbons, $\overline{\zeta}$ is the average amino-acid friction coefficient, and $H_{ij}$ describes the hydrodynamic interaction between residues $i$ and $j$.  $\vec{F}_i(t)$ is a random force modelling the effect of solvent collisions with the protein, and obeys the following fluctuation-dissipation theorem: $\langle\vec{F}_i(t) \cdot \vec{F}_j(t)\rangle=6\overline{\zeta}k_BT\delta_{ij}$. The transformation from bead to bond coordinates effectively removes the global center-of-mass translation.

The LE4PD takes into account hydrodynamic effects and the chemical specificity of each residue in semiflexibility and friction coefficient. Diagonalizing the LE4PD leads to a Langevin equation of motion in a set of quasi-linearly independent, diffusive normal modes. Eq. (\ref{le4pd}) is solved using the eigenvalue decomposition of the $\mathbf{HA}$ matrix product, $\mathbf{Q}^{-1}\mathbf{HAQ}=\Lambda$,
\begin{equation}
    \frac{d\vec{\xi}_a(t)}{dt}=-\frac{3k_BT}{l^2\overline{\zeta}}\lambda_a\vec{\xi}_a(t) + \frac{\vec{F}_a(t)}{\overline{\zeta}}, \label{le4pdxi}
\end{equation}
with $\vec{\xi}_a(t)=\sum_i\left(\mathbf{Q}^{-1}\right)_{ai}\vec{R}_i(t)$ the $a^{th}$ LE4PD mode, and $\vec{F}_a(t)$ the random force vector transformed into the normal mode coordinates. The equation of motion, Eq. \ref{le4pd}, can be written as a function of the bond vector coordinates, $\vec{l}$, thus uncoupling the center-of-mass translation from the internal dynamics of proteins. The two approaches yield equivalent information; however, for all the isotropic LE4PD results presented here, our analysis starts from the bond vector basis, $\vec{l}$. In the LE4PD formalism in bond coordinates, the first three modes represents the rotational dynamics of the protein, while modes with index higher than three describe the internal dynamics of the protein.\cite{Copperman2014} Since, in this study, we are interested only in describing the internal dynamics of a protein, we ignore the three isotropic LE4PD rotational modes, and, when referring to isotropic LE4PD mode $a$, we implicitly mean isotropic LE4PD \emph{internal} mode $a$.

\begin{figure}[htb] 
\includegraphics[width=.8\columnwidth]{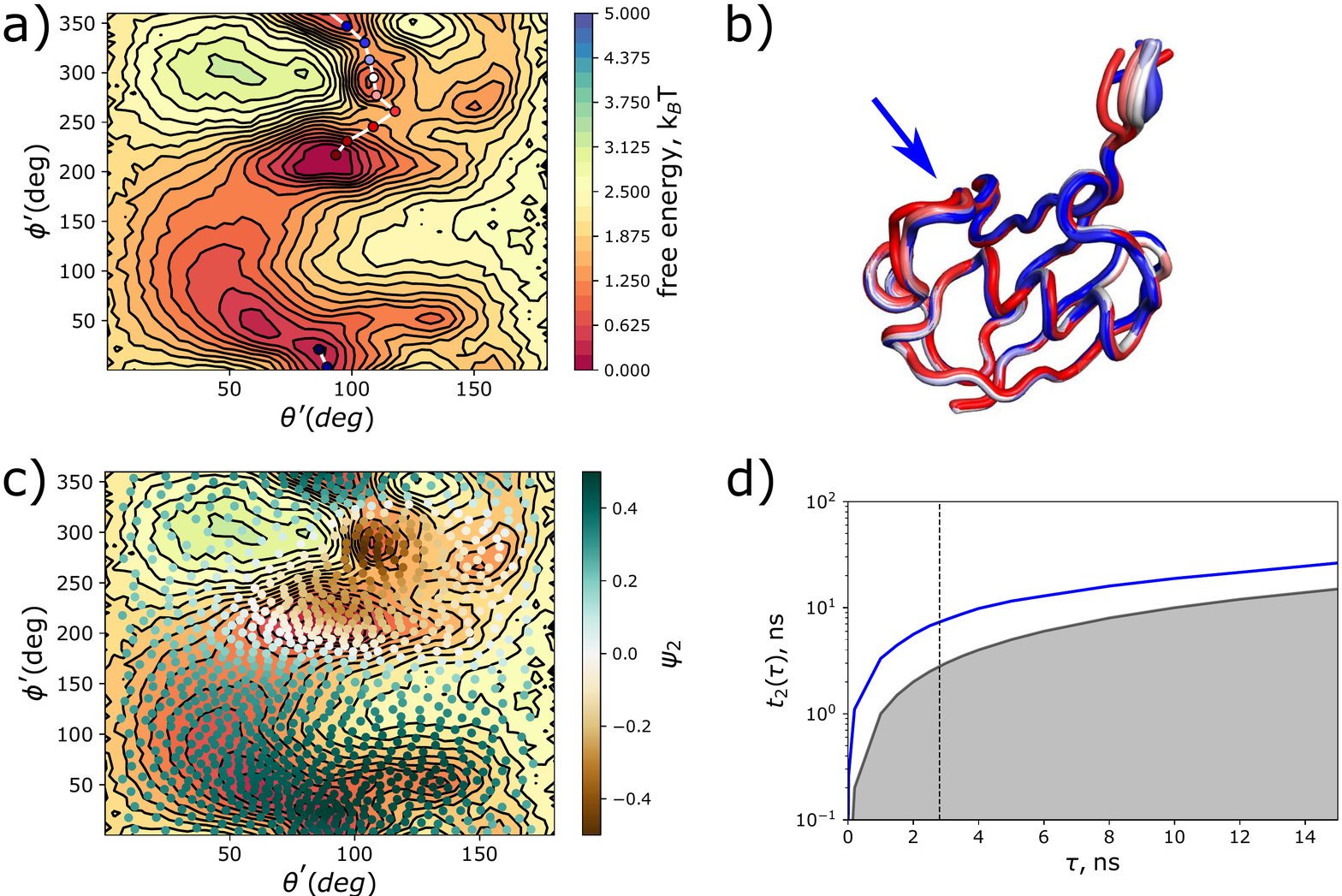}
\caption{Analysis of the sixth  LE4PD-XYZ mode free energy map, with the hydrodynamic interaction included. Panel a) shows the free-energy landscape of the sixth LE4PD-XYZ mode in the two spherical coordinate reference system. The pathway of crossing the energy barrier between the two minima is identified with a rubber band, using the string method.\cite{Beyerle2019} Panel b) shows ubiquitin's conformations that correspond to the pathway identified in panel a) with the red conformation identifying the energy minimum at the top of the map, and the blue conformation corresponding to the energy minimum at the bottom of the map. The arrow points to the region of the $50\text{ s}$ loop that shows the slowest fluctuations. The second eigenvector resulting from the diagonalization of the transition matrix defined in the Markov State Model (MSM) procedure for this mode identifies the two minima in the FES. The projection of $\psi_2$ onto the discrete states of the MSM has colors that correspond to the scaled-and-shifted value of $\psi_2$ at that discrete state, $\psi_2 = \frac{\psi_2-\min(\psi_2)}{\max(\psi_2)-\min(\psi_2)}-0.5$. Panel d) shows how the transition time for the second MSM eigenvector changes when we select a different lag time in the calculation of the MSM transition matrix.  The black, vertical line demarcates the lag time corresponding to the second MSM eigenvector mapping the two minima, as reported in panel c).}
\label{tp1squareLE4PDXYZ}
\end{figure}

\subsection{Building a free energy map in isotropic coordinates and measuring fluctuation timescales}

To each mode is associated a free energy map, which describes mode-dependent local fluctuations of the aminoacids at specific locations along the protein's primary sequence (see Figure \ref{tp1squareLE4PDXYZ}). The maps are constructed as follows: each isotropic LE4PD mode is a linear transformation of the amino acid position vectors, \[\vec{R}_i(t)=\left(R_{i,x}(t),R_{i,y}(t),R_{i,z}(t)\right)^T,\] through the eigenvector matrix $\mathbf{Q}^{-1}$, giving a mode vector with $x-,y-,$ and $z-$components:
\[\vec{\xi}_a(t)=\left(\xi_{a,x}(t),\xi_{a,y}(t), \xi_{a,z}(t)\right)^T.\] For each LE4PD mode one can construct a free-energy surface in spherical coordinates, using the $x-,y-,$ and $z-$components of $\vec{\xi}_a(t)$ as
\begin{align}
\theta_a(t) &= \arccos\left(\frac{\xi_{a,z}(t)}{\vert\vec{\xi}_a(t)\vert}\right) \label{theta}\\
\phi_a(t) &= \arctan\left(\frac{\xi_{a,y}(t)}{\vert\xi_{a,x}\vert}\right) \label{phi}\\
    F(\theta_a,\phi_a)&=-k_BT\ln\left[P(\theta_a,\phi_a)\right]  \ .
    \label{fes}
\end{align}
In Eq. (\ref{fes}), the dependence on the radial coordinate $\vert \vec{\xi}_a(t)\vert$ is averaged over to obtain the joint probability used in the definition of $F\left(\theta_a, \phi_a\right)$: 
\[P(\theta_a,\phi_a)=\int P\left(\vert \vec{\xi}_a\vert,\theta_a,\phi_a\right)d\vert\vec{\xi}_a\vert.\] 
To each mode is associated an energy map with a complex energy landscape where fluctuations have defined pathways, with characteristic amplitudes and timescales. From the linear combination of all the modes one can reconstruct the overall dynamics of the protein and its time correlation functions.\cite{Copperman2014,Copperman2015,Copperman2016,Beyerle2019, Copperman2017} Among these  LE4PD modes one can identify and separate the slow, important motions of the protein based on their timescale. However the information of each mode is retained during the whole process. 

Figure \ref{tp1squareLE4PDXYZ} displays the analysis of the slowest mode, i.e., the sixth mode, for the LE4PD-XYZ with hydrodynamic interaction. A similar study is reported in Figure 1 in the Main Text for the seventh mode of the LE4PD-XYZ  \emph{without} hydrodynamics. The free-energy surfaces look different, but the conformational changes along the minimum energy pathway, and the timescales $t_2$ measured by the Markov state model are almost identical between the two approaches, indicating that the slowest dynamics of ubiquitin is almost insensitive to the presence of hydrodynamics. This result is in agreement with what has been observed previously for the LE4PD-XYZ analysis of this system. \cite{Beyerle2021a}.

\section{Two-Dimensional Energy Maps for the Slow LE4PD-XYZ Modes}
\label{2DLE4PD}
\begin{figure}[htb] 
\includegraphics[width=.5\columnwidth]{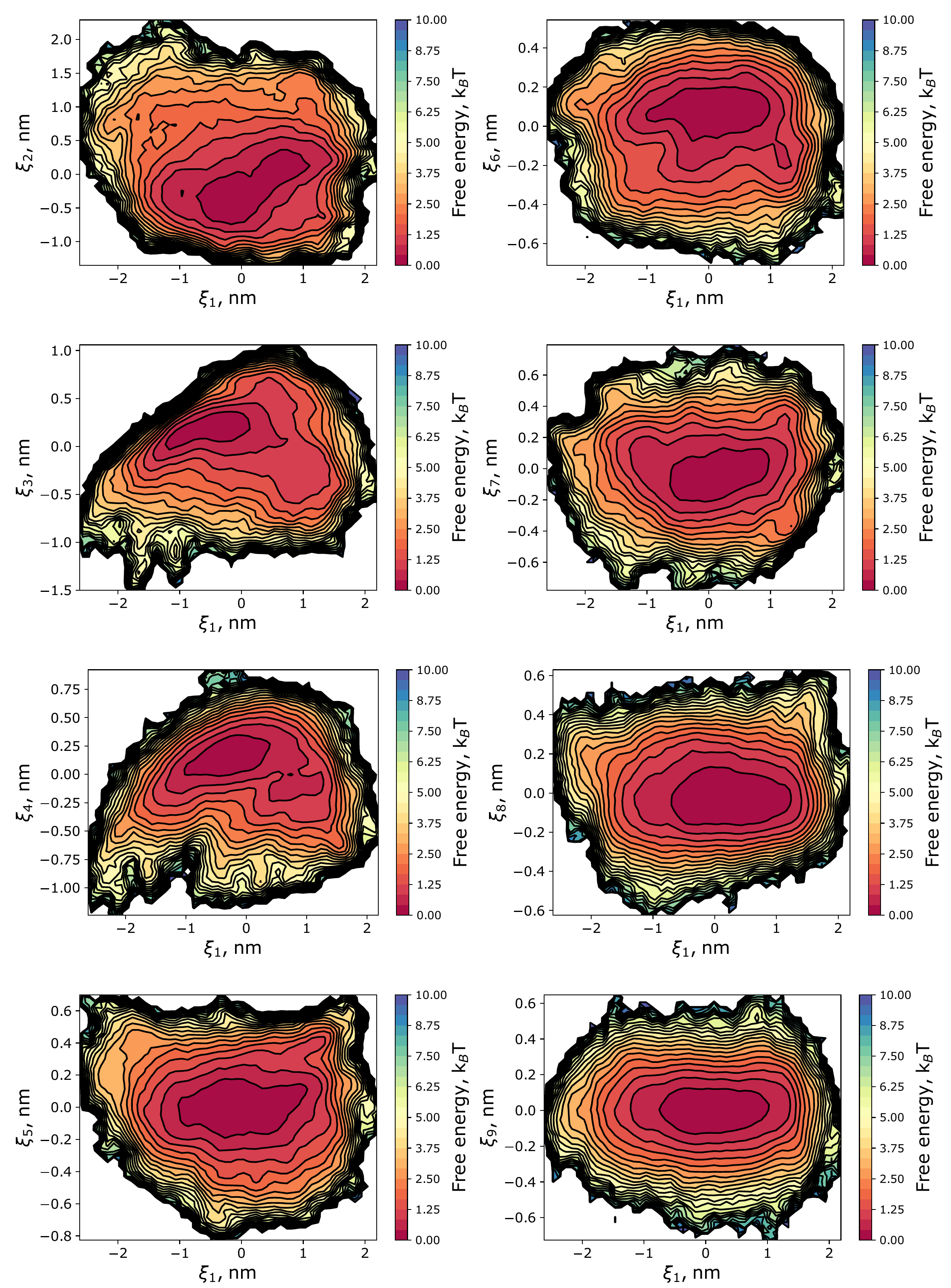}
\caption{Two-dimensional free-energy surfaces of the first LE4PD-XYZ mode with the next eight LE4PD-XYZ modes. All the LE4PD-XYZ modes shown here include the effects of hydrodynamic interactions. Similar surfaces (single minimum) surfaces are obtained if the LE4PD-XYZ without hydrodynamics or the isotropic LE4PD are used.}
\label{2Dfes}
\end{figure}

When analyzing the dynamics of proteins using a set of collective coordinates, the overall dimensionality of the system is first reduced to the space spanned by the slowest $n$ modes or collective coordinates. Then, free-energy surfaces in this reduced space are generated, and the dynamics quantified using a Markov state model or other techniques.\cite{Sittel2018, Bowman2013} In the Main text and in previous studies \cite{Copperman2014, Copperman2016, Beyerle2019, Beyerle2021a} we have used the $\left(\theta, \phi\right)$ coordinates of the LE4PD modes projected into the spherical coordinate system of either the lab \cite{Copperman2014, Copperman2016, Beyerle2019} or body-fixed \cite{Beyerle2021a} frames. Here, we analyze in Figure \ref{2Dfes} the two-dimensional free-energy surfaces for the first LE4PD-XYZ mode with the next eight LE4PD-XYZ modes. Although there are some features on the surfaces shown in Figure \ref{2Dfes}, the surfaces exclusively possess a single well, unlike typical two-dimensional tICA surfaces, which display multiple minima (see Figures \ref{tica_change_time}). 

Similar results (i.e. surfaces with a single minimum) are found when composing free-energy surfaces from the first PCA mode and the next eight PCA modes (data not shown). These observations indicate the usefulness of using the single-mode ($\theta,\phi$) coordinates of the LE4PD-XYZ to extract barriers and conformational changes along each mode coordinate, especially in the case (as presented here) where there are not obvious barriers along the two-dimensional mode coordinate systems.


\section{Equilibrium Molecular Dynamics Simulation of Ubiquitin}
\label{MD}
The MD simulations of ubiquitin were generated using GROMACS version 5.0.4,\cite{Abraham2015} and the AMBER99SB-ILDN atomistic force field,\cite{Lindorff-Larsen2010} on the Comet supercomputer at the San Diego Supercomputing Center. The starting structure was taken from the Protein Databank, PDB ID: 1UBQ.\cite{Vijay-kumar1987} We solvated the protein with spc/e water and minimized the energy using the steepest descent algorithm. We added Na$^+$ and Cl$^-$ ions until the ion concentration was 45 mM, with the concentration of ions selected to match that used in nuclear magnetic resonance experiments of ubiquitin.\cite{Tjandra1995} We subjected the protein-solvent system to two rounds of equilibration: first, a 50-ps equilibration in the NVT ensemble at 300 K, with the temperature-controlled using a Nos\'{e}-Hoover thermostat; then, a 450-ps NPT equilibration at 300 K, with the same thermostat and a Berendsen barostat set to 1 bar. 

Following the NPT equilibration, we performed a 10-ns `burnout' simulation at 300 K with the Nos\'{e}-Hoover thermostat again used to maintain the temperature. We used the last frame of this burnout run as the initial configuration for the 1 $\mu$s production run, which utilized the same simulation parameters as the burnout simulation. Based on a manual inspection of the root-mean-squared deviation (RMSD) of the alpha-carbons from this first frame, the entire trajectory was deemed to fluctuate around an equilibrium value,\cite{Beyerle2019} and the entire 1-$\mu$s of trajectory was used for the subsequent LE4PD and MSM analysis. We used the LINCS algorithm\cite{Hess1997} to constrain all hydrogen-to-heavy-atom bonds in the system and adopted an integration timestep of 2 fs during both the equilibration and MD simulation. We saved the trajectory to file every 100 integration steps (every 0.2 ps), obtaining a total of $\frac{10^6\ \text{ps}}{0.2\ \text{ps}/\text{frame}}=5\times10^6$ frames for analysis.

The MD simulation protocol is the same as that given in the Supplemental Material of \cite{Beyerle2019}. However, the post-processing steps are different. Before performing the tICA, the `raw' MD trajectory is processed to remove the rigid-body rotational and translational motions. First, the reference frame, the first frame in the MD simulation, is centered at the origin of the simulation box. Then, all subsequent frames are centered on this reference structure, and all frames where the protein is broken across the periodic boundaries are made whole. Finally, the rotational motion is removed by fitting each frame in the trajectory to the first, centered frame of the trajectory.

\section{Markov State Modeling}
\label{MSM}
This study adopts the Markov State Model approach to evaluate each normal mode's fluctuations' timescale for both the LE4PD and the tICA. Given the number of resources available describing the theory and application of Markov state models (MSMs) to the analysis of protein dynamics,\cite{Noe2007,Pande2010,Bowman2013,Schwantes2013,Beauchamp2012,Scherer2015,Meng2016,Husic2018} we only present a brief overview of the method and describe the parameter we use in our MSM calculations.  To construct an MSM from an MD simulation, one first identifies a subset of the degrees of freedom or important coordinates. Then one constructs MSM in the state space of these essential collective coordinates. Here, these collective coordinates are the isotropic or anisotropic LE4PD modes or the tICs. Second, a sample space of a small number of these important coordinates is discretized by assigning frames from the trajectory to an appropriate volume of the sample space. 
Third, the transitions among these discrete volumes of the sample space are counted to build a transition matrix $\mathbf{T}$, with the elements $T_{ij}$ defining the conditional probability of transitioning from discrete state $i$ to discrete state $j$. 

The MSM transition matrix is parameterized by a lag time $\tau_{MSM}$, such that the eigenvalues and eigenvectors of $\mathbf{T}=\mathbf{T}\left(\tau_{MSM}\right)$ are generally functions of the MSM lag time.

The eigenvalues of the MSM transition matrix are ordering by descending value of its eigenvalues $\Lambda^{MSM}\left(\tau_{MSM}\right)$, with the first eigenvalue $\lambda^{MSM}_1\left(\tau_{MSM}\right)=1$ and all other eigenvalues of modulus strictly less than 1. The first eigenprocess from the transition operator describes the stationary distribution, and all other eigenprocesses describe dynamic (i.e. decaying) processes of varying timescale. The timescale for the $i^{th}$ process, $t_i$, is given by
\[t_i=-\frac{\tau_{MSM}}{\ln\left(\lambda^{\text{MSM}}_i\right)}.\]
Furthermore, the spectrum of the corresponding right eigenfunction of $\mathbf{T}(\left(\tau_{MSM}\right)$, $\psi_i$, details the dynamics described on the sample space over the timescale given by $t_i$.\cite{Berezhkovskii2004, Buchete2008, Peters2017}

In markovian processes, the eigenvalue decomposition of $\mathbf{T}\left(\tau_{MSM}\right)$ is independent of $\tau_{MSM}$ because they obey the Chapman-Kolmogorov condition.\cite{Reichl1998,VANKAMPEN} In markovian processes the transition matrix sampled at a multiple, \textit{n}, of the lag time $\tau_{MSM}$, is equal to the transition matrix at lag time $\tau_{MSM}$ to the $n$ power: $\mathbf{T}(n\tau_{MSM})=\mathbf{T}(\tau_{MSM})^n$, which implies that the eigenvalues fulfill the property that $\lambda(n\tau_{MSM})=\lambda(\tau_{MSM})^n$. It follows that the timescale of a transition becomes independent of the time used to sample the simulation trajectory. In fact, $t_i(n\tau_{MSM})=\frac{n\tau_{MSM}}{ln{\lambda_i(n\tau_{MSM})}}=\frac{n\tau_{MSM}}{nln{\lambda_i(\tau_{MSM})}}=t_i(\tau_{MSM})$, which is the Chapman-Kolmogorov condition. 

For the MSMs presented here, all steps are performed using the PyEMMA package (http://emma-project.org).\cite{Scherer2019} For all free-energy surfaces, the state space was broken into 1000 discrete states using the k-means++ algorithm,\cite{Arthur2007} which we found previously to be acceptably optimal for ubiquitin.\cite{Beyerle2019} The transition matrix between discrete states is estimated using the reversible estimator given in \cite{Trendelkamp-Schroer2015}. The lag times for the MSMs on the $\left(\theta_a, \phi_a\right)$ surfaces are selected using the spectrum of $\psi_2$.\cite{Beyerle2021a,Beyerle2019}. Briefly, the spectrum of the second right eigenfunction $\psi_2$ of the MSM transition matrix is examined, and the largest MSM lag time such that the maximum and minimum projection of $\psi_2$ reside in the minima on the given free-energy surface is selected for construction of the MSM to be analyzed.
\providecommand{\noopsort}[1]{}\providecommand{\singleletter}[1]{#1}%